\documentclass{aa}  
\usepackage{graphicx}
\usepackage[colorlinks=true,citecolor=blue]{hyperref}
\usepackage{newtxtext,newtxmath}
\usepackage[T1]{fontenc}
\usepackage{amsmath}
\usepackage{amssymb}
\usepackage{color}
\usepackage{xcolor}
\usepackage{pdflscape}
\usepackage{natbib,twoopt}
\usepackage{longtable}
\usepackage{multirow}
\usepackage{lscape}
\usepackage{tablefootnote}
\newcommand{\ha}{H$\alpha$}
\newcommand{\Msun}{M$_\odot$}
\newcommand{\orcid}[1]{\href{https://orcid.org/#1}{\includegraphics[width=10pt]{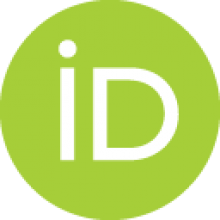}}}

\begin{document} 

\title{Statistical Analysis of Early Spectra in Type II and IIb Supernovae}

\author{Maider Gonz\'alez-Ba\~nuelos, 
\inst{1,2}\thanks{E-mail: mgonzalez@ice.csic.es}\orcid{0009-0006-6238-3598}
\and
Claudia P. Guti\'errez,
\inst{2,1}\orcid{0000-0003-2375-2064}
\and
Llu\'is Galbany
\inst{1,2}\orcid{YY}
\and
Santiago Gonz\'alez-Gait\'an
\inst{3,4}\orcid{0000-0001-9541-0317}
}
          
\institute{
Institute of Space Sciences (ICE, CSIC), Campus UAB, Carrer de Can Magrans, s/n, E-08193 Barcelona, Spain. 
\and
Institut d'Estudis Espacials de Catalunya (IEEC), 08860 Castelldefels (Barcelona), Spain 
\and
Instituto de Astrof\'isica e Ci\^encias do Espaço, Faculdade de Ci\^encias, Universidade de Lisboa, Ed. C8, Campo Grande, 1749-016 Lisbon, Portugal
\and
European Southern Observatory, Alonso de C\'ordova 3107, Casilla 19, Santiago, Chile
}

\date{}
 
\abstract
{
We present a comprehensive analysis of the early spectra of type II and type IIb supernovae (SNe) to explore their diversity and distinguishable characteristics. Using 866 publicly available spectra from 393 SNe, 407 from type IIb SNe (SNe~IIb) and 459 from type II SNe (SNe~II), we analysed \ha\ and He~I 5876 \AA at early phases ($<40$ days from the explosion) to identify possible differences between these two SN types. By comparing the pseudo-equivalent width (pEW) and full width at half maximum (FWHM), we find that the strength of the absorption component of \ha\ and He~I lines serves as a quantitative discriminator, with SNe~IIb exhibiting stronger lines at all times. The most significant differences emerge within the first 10–20 days. 
To assess the statistical significance of these differences, we apply statistical methods and machine-learning techniques. Population density evolution reveals a clear distinction in both pEW and FWHM. Quadratic Discriminant Analysis (QDA) confirms distinct evolutionary patterns, particularly in pEW, while FWHM variations are less pronounced. A combination of t-distributed Stochastic Neighbour Embedding (t-SNE) and Linear Discriminant Analysis (LDA) effectively separates the two SN types. Additionally, a Random Forest Classifier (RFC) demonstrates the robustness of pEW and FWHM as classification criteria, allowing for accurate classification of newly observed SNe~II and IIb based on computed classification probabilities.
Applying our method to low-resolution spectra obtained from the Zwicky Transient Facility Bright Transient, a magnitude-limited survey, we identified 34 misclassified SNe. This revision increases the estimated fraction of SNe~IIb from $4.0\%$ to $7.26\%$. This finding suggests that misclassification significantly impacts the estimated core-collapse SN rate. Our approach enhances classification accuracy and provides a valuable tool for future supernova studies.
}

\keywords{supernovae: general,}
\authorrunning{Gonz\'alez-Ba\~nuelos, Guti\'errez et al.}
\titlerunning{SNe~II vs SNe~IIb}
\maketitle
%

\section{Introduction}

Core collapse supernovae (CCSNe) are those explosive phenomena associated with the death of massive stars ($>8\textup{M}_\odot$). When they reach the end of the nuclear fusion in hydrostatic burning, with no heavier elements to be synthesised, the core collapses as it can no longer be stabilised, causing supernovae \citep[SNe;][]{janka2007theory}. Among CCSNe, many subtypes have been established based on the different features shown in their spectra, which are determined by the composition of the exploding progenitor star. The main characteristic of the CCSNe spectral classification is the amount of retained hydrogen or helium. Those that have retained a significant fraction of their H envelope, i.e. H-rich CCSNe, are classified as type II (SNe~II) and are characterised by strong Balmer lines in the spectra \citep{minkowski1979spectra}. Moreover, SNe that have suffered a loss of their envelope before the explosion and thus have little or no hydrogen are known as Stripped-Envelope SNe (SESNe). SESNe that are hydrogen-poor (but still present) are classified as type IIb SNe \citep[SNe~IIb;][]{Filippenko93}, having stripped part of its outermost hydrogen envelope in the star's last stages before the explosion. Those SESNe that are H-deficient but He-rich are classified as type Ib SNe (SNe~Ib), exhibiting spectra dominated by He features. Notably, SNe~IIb are considered to be transitional objects between SNe~II and SNe~Ib. In early phases, their spectra are dominated by Balmer P-Cygni lines, resembling those of SNe II. However, as time passes, the H signatures disappear, and the He lines gradually dominate the spectra, comparable to SNe~Ib.
Finally, the H- and He-deficient are type Ic SNe (SNe~Ic). Thus, the mass loss drives the following spectroscopic sequence as it increases: II $\rightarrow$ IIb $\rightarrow$ Ib $\rightarrow$ Ic \citep{gal2016observational, shivvers2017revisiting}.

Historically, SNe~II are sub-classified based on the light curve (LC) shape. Type IIP SNe (SNe~IIP) exhibit a plateau right after the maximum peak, while type IIL SNe (SNe~IIL) show a fast decline after the peak \citep{barbon1979photometric}. However, recent works \citep[e.g.][]{ anderson2014characterizing, sanders2015toward,Gonzalez-Gaitan15, galbany2016ubvriz} have revealed that the observational properties of SNe IIP and IIL form a continuum rather than distinct groups, suggesting they represent a single population.

In addition to the general classification of SNe~II, two further subclasses have been identified: SNe IIn (defined spectroscopically) and 1987A-like events (defined photometrically). SNe~IIn are characterised by long-lasting narrow H emission lines \citep{Schlegel90} attributed to interaction with the circumstellar medium (CSM). On the other hand, the 1987A-like events are distinguished by long-rising LCs, similar to those observed in SN~1987A \citep{Taddia12, pastorello2012sn, taddia2016long}. 

The LC of H-poor SNe~IIb is characterised by a bell shape similar to those observed in other SESNe. Their main peak is generally attributed to the powering by radioactive decay of $^{56}$Ni \citep{ shigeyama1994theoretical, woosley1994sn, bersten2012type}. Photometric studies of SNe IIb have revealed that some objects exhibit double-peaked LC, with an initial peak before the main one powered by $^{56}$Ni. This initial peak has been suggested to be related to the cooling phase, a consequence of the cooling of the progenitor's envelope after the shock breaks out from the surface. That cooling phase will last longer or shorter depending on the size of the progenitor, although it also depends on its density distribution \citep{2011ApJ...728...63R, bersten2012type, 2014ApJ...788..193N, morales2015sn, 2017ApJ...838..130S}.

Pre-explosion imaging at the sites of SN~II explosions has identified Red Supergiant (RSG) stars with masses in the range of 8–18 \Msun\ as their progenitors \citep{van2003progenitor, smartt2004detection, smartt2009death}. For SNe~IIb, the direct detection of the progenitor of SN~1993J \citep{maund2004massive} and SN~2011dh \citep{folatelli2014blue, maund2019origin} suggests that some of these objects arise from stars in binary systems. Moreover, the detection of the possible companion stars for SN~1993J \citep{maund2004massive} and SN~2001ig \citep{ryder2018ultraviolet} further supports the binary scenario. 

Given the striking similarities in the early-phase spectra of SNe II and IIb, it is plausible to suggest that their progenitors form a continuum of decreasing hydrogen envelope mass \citep[e.g.][]{Hiramatsu2021, dessart2024sequence}. However, recent studies have found a discontinuity in the observational properties of their LCs. For instance, \citet{arcavi2012caltech} analysed the LC morphology of SNe~II and identified three distinct classes: IIP, IIL and IIb, which were interpreted as different progenitor populations. \citet{Faran14} found that fast-declining SNe~II (SNe~IIL) appear to be photometrically related to SNe~IIb. In contrast, \citet{pessi2019comparison} found a gap in the observed properties of SNe~II and IIb, concluding that they represent two distinct families rather than a continuous sequence.

From a theoretical perspective, the radiative transfer calculations of massive stars in binary systems with different initial orbital periods presented by \citet{dessart2024sequence} show how varying degrees of hydrogen envelope stripping can produce the observed diversity in CCSNe explosions, forming a continuum in the $V-$band LC morphology based on the orbital separation (i.e. retained hydrogen mass). This study also reveals distinctions in the \ha\ and helium lines features associated with different progenitor models, offering critical insights into the physical processes that drive the observed diversity among SNe~II and IIb.

Previous studies on SNe~II and IIb have focused on their observed photometric properties, leaving their spectral distinctions less explored. In this work, we present the first comprehensive analysis comparing and characterising the early spectra of these two SN classes. Drawing on methodologies from prior studies, such as those by \cite{liu2016analyzing, Shivvers19} and \citet{holmbo2023carnegie} for SESNe or \cite{gutierrez2017type} and \cite{deJaeger19} for SNe~II, we systematically examine the spectral features for each SN type. Additionally, this study introduces a novel classification method that exploits subtle differences in the spectral properties to distinguish between SNe~II and IIb. Our results demonstrate that these differences are reliable diagnostic tools. They help reduce misclassifications and enable a more accurate categorisation of SNe~II and IIb.

The paper is organised as follows. Section \ref{sec:sample} describes the data sample and measurements. Sections \ref{sec:analy} and \ref{sec:res} provide the analysis and results. Section \ref{sec:disc} presents the discussion, while Section \ref{sec:conc} lists the conclusions.

\section{Data sample and spectral measurements}
\label{sec:sample}

\subsection{Sample}

\begin{figure}
\includegraphics[width=\columnwidth]{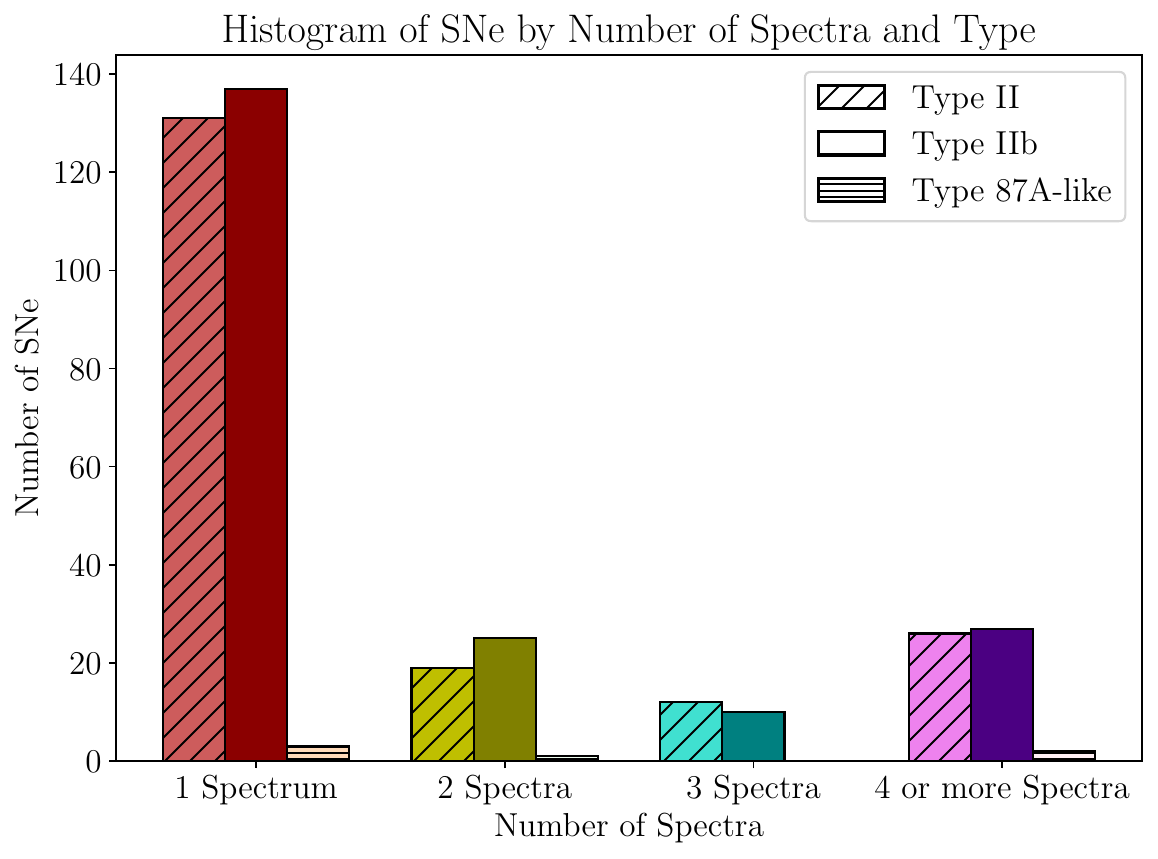}
\caption{Histogram of the number of spectra per SN. In total, our sample is composed of 866 spectra for 393 SNe. }
\label{fig:n_spectra}
\end{figure}

The sample analysed in this study includes all available SN~IIb spectra in the WISeREP\footnote{\url{https://wiserep.weizmann.ac.il/}} \citep{yaron2012wiserep} repository. We retrieved these data using the \textsc{wiserep\_api}\footnote{\url{https://github.com/temuller/wiserep_api}} tool. To ensure a comparable sample, and considering that SNe~IIb are less abundant than SNe~II ($\sim10$\% of the CCSNe vs $\sim70$\% of the CCSNe; \citealt{shivvers2017revisiting}), we selected a similar number of H-rich events. 
Our goal is not to recover the intrinsic population rates of these two SN types, but rather to investigate whether they can be reliably distinguished based on their spectral features, despite their similarities at early phases. To this end, we constructed balanced spectral samples for both types, allowing our method to focus on identifying the main features without being biased by the larger occurrence of SNe~II. Therefore, our dataset consists of 449 low-redshift ($z<0.12$) SNe observed between 1986 and 2024, comprising 222 SNe~II (including 5 of the 1987A-like objects) and 227 SNe~IIb. Given that we want to study the possible differences between SNe~II and IIb at early phases, some additional cuts were applied. First, we only included spectra obtained within the first 40 days post-explosion (see Section \ref{sec:explosion_date}). Second, since our line measurements were performed manually using a semi-interactive code (see Section \ref{sec:measurements}), we assessed the quality of the spectra, retaining only those with a sufficient resolution ($R$) to perform reliable analysis. These are spectra with $R\geq300$, where the spectral resolution is defined as $R=\lambda/\Delta \lambda$. Thus, we excluded the Spectral Energy Distribution Machine (SEDM) spectra ($R \approx 100$; \citealt{SEDM}). Given the larger number of available SNe II, we prioritised selecting those with well-constrained explosion dates. However, to test the robustness of our method under less optimal conditions, we also included a subset of approximately 15 SNe with lower-quality spectral and photometric data.

After applying these cuts, our final dataset contained 393 SNe, represented by 866 spectra: 407 of SN~IIb, 438 of SNe~II, and 21 of SNe~1987A-like. Among the SNe in our sample, 55 have four or more spectra, 22 have three, and 42 have two, while 271 have only one spectrum. Figure~\ref{fig:n_spectra} presents the histogram of the number of spectra by SN type. Table~\ref{tab:SNe} presents the whole dataset (Appendix \ref{app:tables}).

\begin{figure}
\includegraphics[width=\columnwidth]{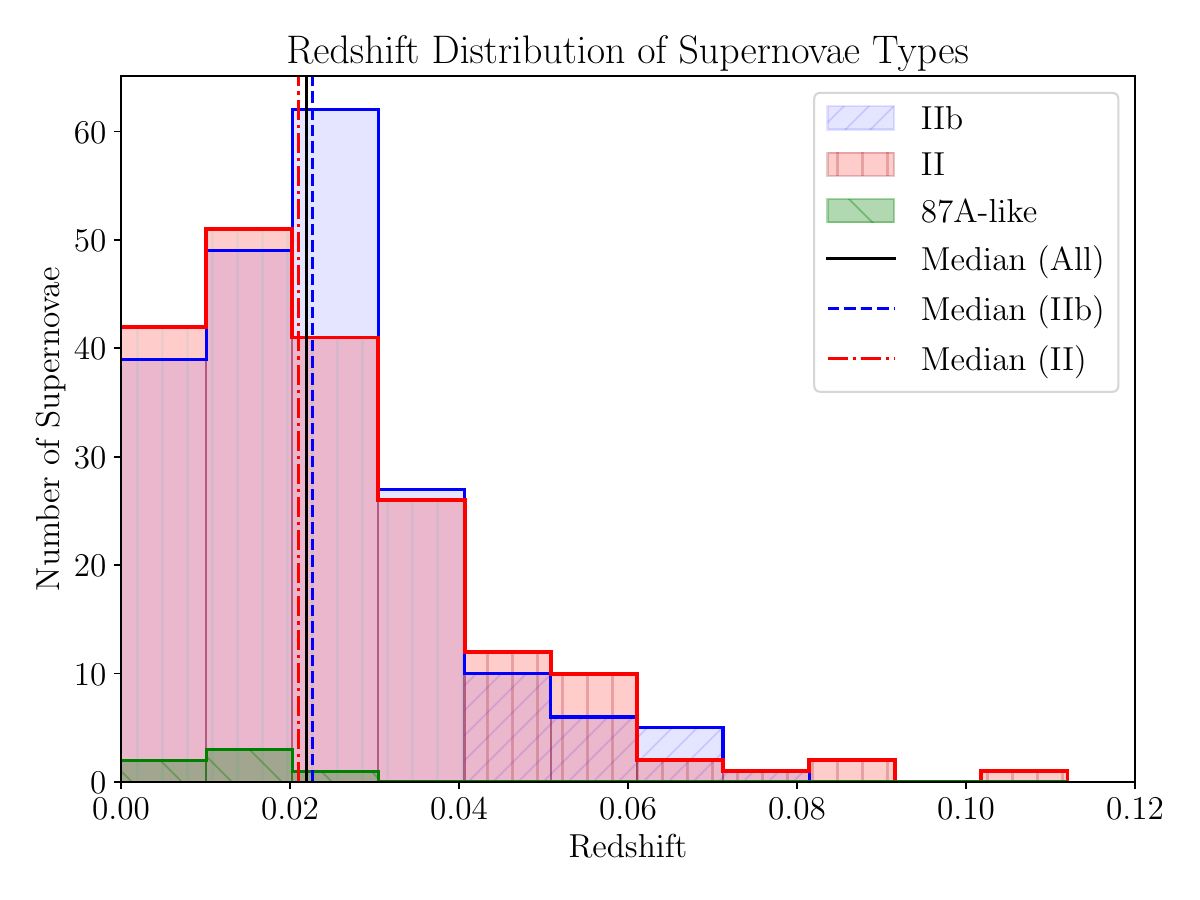}
\caption{Redshift distribution for the 393 SNe in our sample. The vertical lines indicated the median of the whole sample (solid black line) at 0.022, SNe~II (dashed-dotted red line) at 0.021 and IIb (dashed blue line) at 0.023.}
\label{fig:redshift}
\end{figure}

The redshift distribution of our sample is shown in Figure~\ref{fig:redshift}. The Figure shows that most of the sample has a redshift below 0.06. The median of the redshift distribution is 0.021 and 0.023 for SNe~II and IIb, respectively. These median values indicate that the central tendency of both distributions is very similar, suggesting that our sample is well-balanced in terms of redshift. This makes it suitable for the analysis. 

\subsection{Explosion date}
\label{sec:explosion_date}

We compute the explosion epoch for each SN in our sample following the methodology outlined in \cite{gutierrez2017type}. This approach uses the last non-detection\footnote{Non-detections correspond to fluxes below the detection threshold. In these cases, ATLAS and ZTF report upper-limit magnitudes based on the image depth, typically 3$\sigma$ for forced photometry in ATLAS, and 5$\sigma$ in ZTF.} and the first detection for each object, estimating the explosion date as the middle point between these two epochs. The detection limits were determined using publicly available photometry from the Asteroid Terrestrial-impact Last Alert System (ATLAS; \citealt{Tonry18, Smith20}) forced photometry server \citep{atlas_server} and the Zwicky Transient Facility \citep[ZTF;][]{Bellm19, Graham19} public stream through the Lasair\footnote{\url{https://lasair.roe.ac.uk/}} broker \citep{Smith19}. The average uncertainty in the estimated explosion dates is $\sim2.5$ days. When this method was not applicable either due to the absence of deep non-detections or the lack of any non-detections, we estimated the explosion epoch using spectral matching, as described in \cite{gutierrez2017type}. Thus, we compared our SN spectrum with templates of other SNe with well-established explosion dates. For this, we employed the Supernova Identification\footnote{Following \cite{blondin2007determining}, we only considered matches with rlap $\geq 5$ and lap $\geq 0.4$.} \citep[SNID;][]{blondin2007determining} and the NGSF\footnote{The best-fitting SN template is identified via a $\chi^{2}$ minimisation procedure.} (Next Generation SuperFit; \citealt{goldwasser2022next, howell2005gemini}), a Python-based implementation of the \textsc{superfit} template-matching classification tool. Finally, for the sample of SNe II analysed in \cite{gutierrez2017type}, we adopted their explosion dates.

\subsection{Measurements}
\label{sec:measurements}

We analyse the spectra of our SNe by characterising the strength/intensity of the absorption component of speciﬁc spectral lines. To achieve this, we compute the pseudo-Equivalent Width (pEW) and Full Width at Half Maximum (FWHM). Our methodology follows similar approaches presented in the literature, including works by \cite{liu2016analyzing, gutierrez2017type, holmbo2023carnegie}.

The pEW is defined as:

\begin{equation}
\text{pEW} = \sum_{i=1}^{N} \left( 1 - \frac{f(\lambda_i)}{f_0(\lambda_i)} \right) \Delta \lambda_i\;,
\end{equation}

where $\lambda_1$ and $\lambda_N$ are the wavelength boundaries of the spectral line to be measured, and $\Delta\lambda$ is the difference between two consecutive points, $f(\lambda_i)$ corresponds to the flux observed at $\lambda_i$ and $f_0(\lambda_i)$ the continuum flux. The pEW quantifies the overall strength of a spectral line relative to the pseudo-continuum, with higher pEW values indicating a stronger contribution of that feature in the SN spectrum. Variations in pEW can signal changes in the physical conditions of the ejecta, such as temperature, density, or ionisation state of the emitting or absorbing material. Therefore, its analysis provides valuable insights into the composition of the SN ejecta.

\begin{figure}
\centering
\vspace{-0.4cm}
\includegraphics[width=\columnwidth]{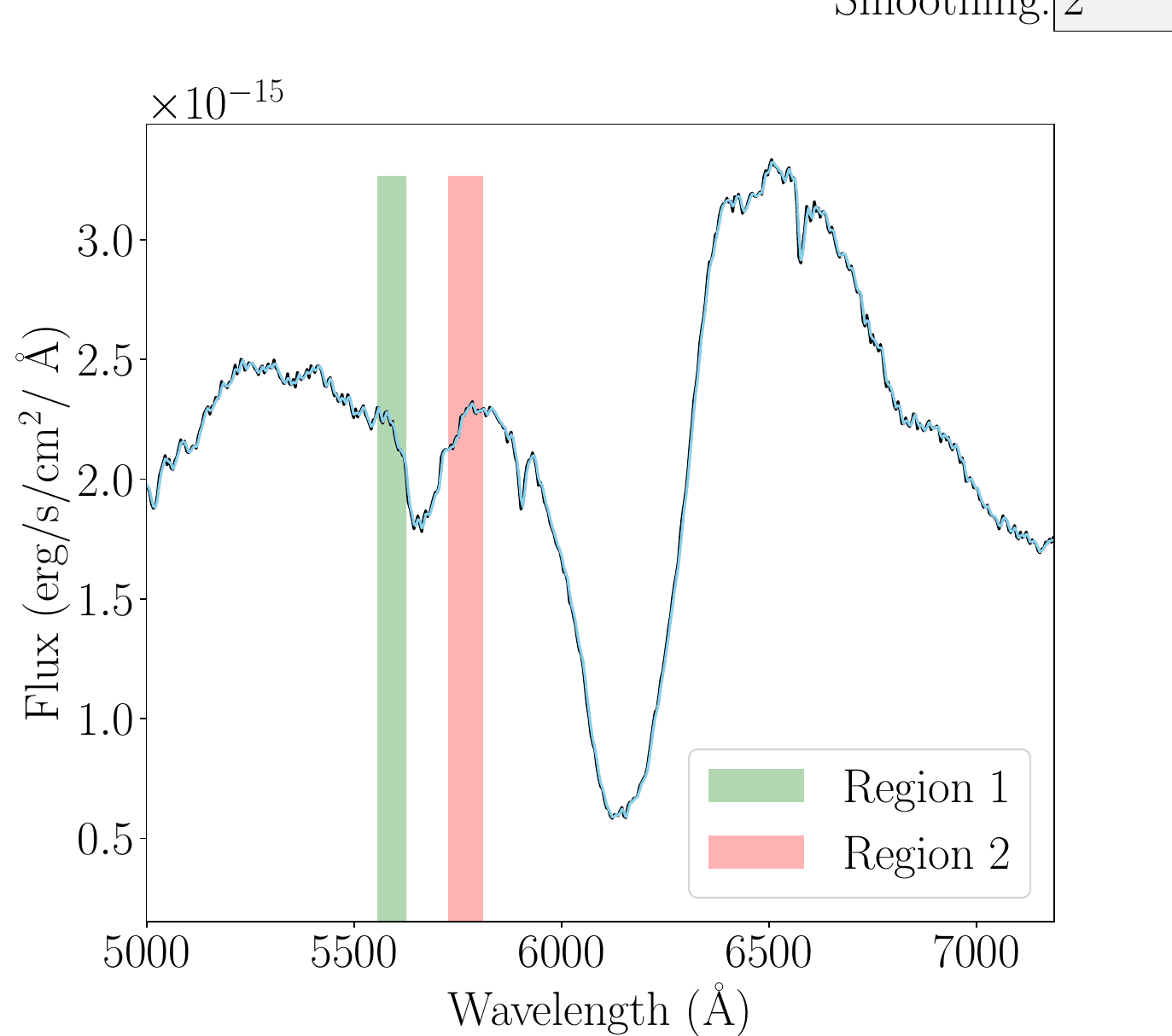}
\includegraphics[width=\columnwidth]{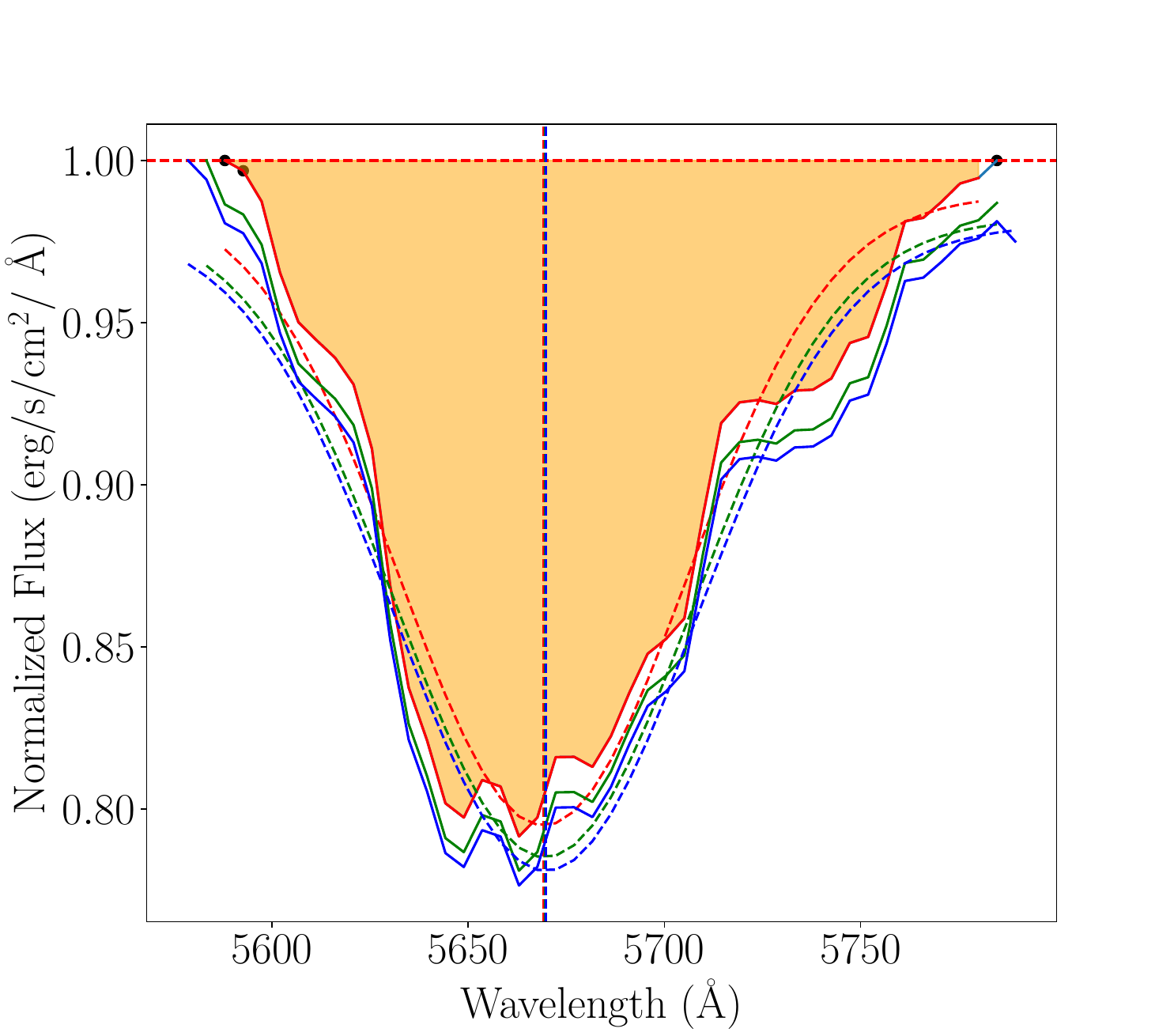}
\caption{An example of the pEW and FWHM measurement for the He I line. On the top panel, we show the selection of the two regions to define the line to be measured. The defined line and the Gaussian fitting for the pEW and FWHM computation are shown in the bottom panel. The different line colours represent the shifted measurements used to estimate the uncertainties.}
\label{fig:codigo_ejemplo}
\end{figure}

To define this continuum, we developed an interactive \textsc{Python} code, based on \textsc{Fit\_continuum}\footnote{\url{https://github.com/tdejaeger/Astronomy/blob/master/Codes/fit_continuum.py}}. We modified it so that it allows the user to select a specific feature by choosing two regions around it: a lower and an upper wavelength limit. Within those regions, the wavelength boundaries of the line ($\lambda_1$ and $\lambda_N$) are determined by identifying the points with the maximum slope change on each side. Using these boundary points, a pseudo-continuum is established, and the spectrum is normalised relative to it. Next, the difference between the normalised flux and the continuum is integrated across the absorption region by iterating over the range between $\lambda_1$ and $\lambda_N$, accumulating the pEW values using the previously defined formula (see Figure~\ref{fig:codigo_ejemplo}).

The FWHM is determined by fitting a Gaussian function to the spectral feature. The Gaussian model is widely used in spectral analysis because it can accurately represent symmetric, bell-shaped profiles. The relationship between the FWHM and the parameters of the fitted Gaussian function is given by:

\begin{equation}
\text{FWHM} = 2\sqrt{2\ln{2}}~\sigma.
\end{equation}

Here, $\sigma$ is the standard deviation of the fitted Gaussian function, which defines the width of the Gaussian curve. To calculate the FWHM, we use the function \textsc{specutils.gaussian\_fwhm}, which derives the FWHM directly from $\sigma$ based on the Gaussian profile. This value represents the width of the spectral feature at half of its maximum depth. The FWHM represents the velocity dispersion of the material in the line-forming region, directly reflecting how fast the ejecta is expanding along the line of sight. This velocity dispersion is related to the explosion energy of the SN, as more energetic explosions produce higher velocity spreads, resulting in broader spectral features.

The errors associated with the FWHM and pEW measurements were estimated by systematically varying the boundaries of the selected wavelength region and performing the measurements three times. The region boundaries were adjusted for each predefined shift, the continuum measured, and the FWHM and pEW recalculated, generating a set of shifted measurements. These measurements quantify the sensitivity of the FWHM and pEW to variations in the selection of the wavelength region. The deviations between the shifted and original measurements (obtained from the unshifted region) were analysed statistically. To calculate the errors, statistical metrics such as the standard deviation or root mean square error (RMSE) were used to characterise the spread of the shifted measurements. These values provide a robust estimate of uncertainties in the FWHM and pEW measurements.

To address noisy spectra, we incorporated a smoothing option into the code. This allows users to apply a smoothing degree before selecting the regions, ensuring accurate spectral line identification and measurements, without significantly affecting the pEW or FWHM values. After smoothing, the modified spectrum overlaps the original data. An example of how our interactive code works is shown in Figure~\ref{fig:codigo_ejemplo}. The top panel shows the two regions selected on each side of a specific line. The smoothed spectrum is shown in cyan. The bottom panel of Figure~\ref{fig:codigo_ejemplo} displays the final measured line after processing.

To validate the accuracy of our developed semi-interactive code, we compared its derived measurements with those obtained using manually \textsc{IRAF} \citep{tody1986iraf, tody1993iraf}. For this comparison, we selected a subsample of spectra and manually measured the pEW and FWHM for the same spectral lines with IRAF. The results of this comparison, presented in the Appendix~\ref{app:fig} (Figure \ref{fig:NIREA_vs_IRAF}), demonstrate the consistency between the two methods. 

Following the explained methodology and using our code, we estimated the pEW and FWHM for the spectral lines across all 866 spectra in our sample. Since the two SN types analysed in this work typically exhibit a blue featureless continuum at very early phases, with spectral features gradually emerging over time, we adopted the methodology of \citet{gutierrez2017type} by assigning a value of zero to a line measurement when the line is absent. Accordingly, a value of $pEW=0$ indicates that a specific line is not visible at that phase.

\section{Analysis}
\label{sec:analy}
 
\subsection{Re-classification}

\begin{table}
\begin{center}
\caption{Reclassified SNe based on their spectral evolution and/or photometry properties.}
\label{table:Re-classifications}
\setlength{\tabcolsep}{4pt}
\begin{tabular}{lccc}            
SN name  & Classification & Reference & Re-classification \\
\hline
\hline
SN~2015Y    & Ib   & [1]  & IIb      \\
SN~2016bam  & II   & [2]  & IIb      \\
SN~2018bti  & II   & [3]  & IIb      \\
SN~2020adnx & II   & [4]  & IIb      \\
SN~2019aax  & II   & [5]  & IIb      \\
SN~2019abu  & II   & [6]  & IIb      \\
SN~2019lqo  & II   & [7]  & IIb      \\
SN~2019lsk  & II   & [8]  & IIb      \\
SN~2019luz  & II   & [8]  & IIb      \\
SN~2019mrz  & II   & [9]  & IIb      \\
SN~2019rwd  & II   & [10] & IIb      \\
SN~2019sox  & II   & [11] & IIb      \\
SN~2019tua  & II   & [12] & IIb      \\
SN~2020abah & II   & [13] & 87A-like \\
SN~2020aqe  & II   & [14] & IIb      \\
SN~2021afdx & II   & [15] & IIb      \\
\hline
\hline
\end{tabular}
\begin{list}{}{}
\item Reclassifications done before performing the analysis. 
Objects considered in our sample with the new classification.\\
\textbf{References.} [1] \cite{zheng2015kait}, [2] \cite{Hosseinzadeh2016_TNS}, [3] \cite{hiramatsu2018global}, [4] \cite{srivastav2020},
[5] \cite{piranomonte2019epessto}, [6] \cite{Piranomonte2019}, [7] \cite{Bose2019}, [8] \cite{angus2019epessto+}, [9] \cite{wiseman2019epessto+}, [10] \cite{Arcavi2019}, [11] \cite{cartier2019transient}, [12] \cite{Burke2019}, [13] \cite{Sit2019}, [14] \cite{BurkeHira2019}, [15] \cite{Ragosta2019epessto+}. 
\end{list}
\end{center}
\end{table}

Before starting our analysis, we inspected all SN spectra in our sample to verify the accuracy of the assigned classifications. We visually checked the main features in the spectra and the LC morphology. For objects with multiple spectroscopic follow-up observations, the appearance of some spectral lines allows us to easily identify possible misclassifications. Through this inspection, we found 16 misclassified SNe. Table~\ref{table:Re-classifications} summarises these objects and their re-classifications. As seen, 14 objects were initially classified as SNe~II, and we re-classified them as SNe~IIb. SN~2015Y was classified as a SN~Ib; however, its earliest spectra show signatures of hydrogen features, leading to its re-classification as IIb \citep{childress2016anu}. Additionally, SN~2020abah was reclassified from SN~II to SN~87A-like.

\subsection{Spectral line identification}

\begin{figure}
\centering
\includegraphics[width=0.9\columnwidth]{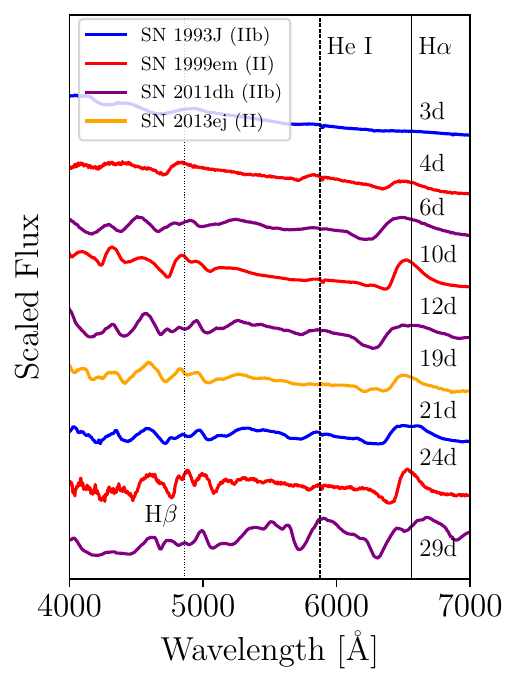}
\caption{Evolution of type II SNe~1999em \citep{leonard2001distance} (red) and 2013ej \citep{valenti2013first} (orange) and type IIb SNe~1993J \citep{barbon1995sn} (blue) and 2011dh \citep{valenti2013first, ergon2014optical} (purple). The epoch of each of the spectra is computed from the explosion date. The dotted vertical line marks the rest-wavelength corresponding to H$\beta$; the dashed line is the He I $\lambda5876$ feature, and the continued line is \ha.} 
\label{fig:visualanalysis}
\end{figure}

Given that the spectra of SNe~II and IIb appear similar at early phases due to the presence of H and He~I $\lambda5876$, we investigate whether they exhibit distinguishable properties. To do so, we examine how prototypical objects of each SN type evolve over time. Figure~\ref{fig:visualanalysis} shows the spectral evolution from the explosion to 30 days post-explosion for the type II SNe~1999em and 2013ej and the type IIb SNe~1993J and 2011dh. 

By examining the spectral evolution in each SN, we identified specific features that exhibit distinct tendencies based on their classification. In the first days after the explosion, both SNe~II and IIb often display a featureless blue continuum  \citep{matheson2000optical, arcavi2017hydrogen, hillier2019photometric}, which is the result of the high temperatures of the ejected material during that phase. As the ejecta cools down, broader lines, such as the hydrogen Balmer series and He I $\lambda5876$, emerge. Given that the H and He~I lines are present in both SN types, we considered them as primary indicators. However, despite being shared in these two classes, notable differences in the evolution and shape of \ha\ and He I $\lambda5876$ are apparent (Figure \ref{fig:visualanalysis}). In the first one to two weeks, both types exhibit strong H lines, but their evolutionary patterns begin to diverge. For instance, for SNe~II, \ha\ develops a well-defined P-Cygni profile. In contrast, in SNe~IIb, \ha\ transitions towards a flat-topped profile, which eventually begins to split into two components due to the emergence of He I $\lambda6678$ \citep{holmbo2023carnegie}. The presence and strength of this He I line can vary significantly between individual SNe, which can complicate the analysis of the \ha\ emission profile.

Subtle differences in the spectra of SNe~II and IIb are also visible in the \ha\ absorption and He I line $\lambda5876$. The \ha\ absorption profile is typically much broader throughout the evolution of SNe~IIb compared to SNe~II. Additionally, the He I line $\lambda5876$ emerges earlier and is generally stronger in SNe~IIb. This behaviour is likely a result of the partial stripping of the H envelope in SNe~IIb.

We also examined the H$\beta$ ($\lambda4861$) line, which, like \ha, shows differences between SNe~II and IIb. However, given that H$\beta$ is affected by the Iron lines around it, we excluded it from our analysis due to the complexities in its characterisation. The feature is marked with a dotted vertical line in Figure~\ref{fig:visualanalysis}. In consequence, our analysis is focused only on the \ha\ absorption feature and the He~I $\lambda5876$ line.

\subsection{Median spectra}

To visualise the main spectroscopic differences between SNe~II, IIb and 87A-like, we compute the median spectrum of each type by including all the spectra in our final sample, following a similar procedure as \cite{liu2016analyzing}. This includes 188 SNe~II, 199 SNe~IIb and 6 SNe~87A-like between 0-40 days after explosion. Due to the limited number of SN 87A-like objects, the median spectrum for this group may not fully represent its characteristics. Nevertheless, we decided to include it because some SNe~87A-like are often misclassified (see Section \ref{sec:generales}). The resulting median spectra for the three SN types are plotted in Figure~\ref{fig:medians_II_IIb_87like}. The spectra were normalised by scaling them to the median value of a common reference wavelength range, ensuring their fluxes match at a specific wavelength. This process allows a direct comparison of spectral features across different objects while preserving their relative shapes. The resolution of the available data constrains the quality of the median spectra. While spectra with very poor resolution were excluded from our sample, some noisy spectra worsened the quality of the median spectrum, which is evident in Figure~\ref{fig:medians_II_IIb_87like}.

A comparison between the three median spectra reveals common features with notable differences in the strength and width of the line depending on their type. In SNe~IIb, both \ha\ and He I $\lambda5876$ are broader than in SNe~II and 87A-like. In contrast, the spectrum of SNe~II appears to be still dominated by the continuum, with \ha\ and He I $\lambda5876$ barely seen. For SNe~87A-like, spectral differences are also observed, although they are classified based on their photometry. The SN~87A-like median spectrum displays a well-defined \ha\ P-Cygni profile with a broad emission line. The He~I $\lambda5876$ is also prominent. These findings align with our visual inspection.  

\begin{figure}
\centering
\includegraphics[width=0.9\columnwidth]{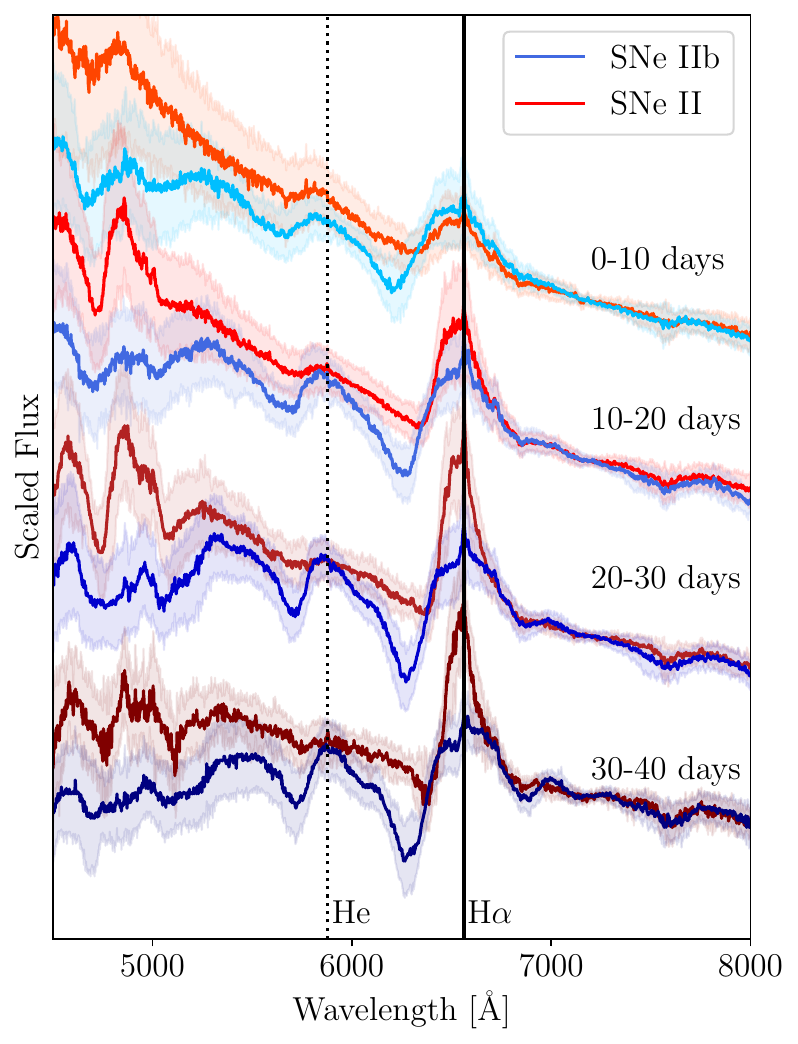}
\caption{Comparison between the median spectra of SNe~II and IIb in different time ranges: 0 -- 10 days post-explosion, 10 -- 20 days, 20 -- 30 days and 30 -- 40 days. Blue spectra correspond to SNe~IIb, and red spectra to SNe~II. The dashed vertical line marks the rest wavelength of He I ($\lambda5876$), while the continuous line marks the rest wavelength of \ha. The shaded regions represent the Median Absolute Deviation (MAD), which represents dispersion around the median.}
\label{fig:medians}
\end{figure}

Given that the spectral features change based on the SN phase, we perform a more detailed analysis of the median spectra over time. We computed it for SNe~II and IIb within the four-time intervals: 0 -- 10 days post-explosion, 10 -- 20 days, 20 -- 30 days and 30 -- 40 days. The resulting median spectra are presented in Figure~\ref{fig:medians}, where SNe~IIb are represented in blue and SNe~II in red. Due to the limited data for SNe~87A-like, we could not compute their median spectra for specific time intervals.

The median spectra constructed within time intervals show more explicit evolutionary tendencies for each SN type. Notably, the \ha\ line dominates the spectrum of both classes. In the early phases, the absorption of the H lines (\ha\ and H$\beta$) is stronger in SNe~IIb and remains so throughout their evolution up to 40 days. Furthermore, the He I line emerges within the first 10 days in SNe~IIb, whereas in SN~II, it appears at $\sim10$ days post-explosion. In SNe~II, both \ha\ and He I also increase their strength over time, which can complicate classification depending on the specific characteristics of each SN. However, a key distinction lies in the shape of the lines, particularly \ha. In SNe~II, the emission component remains strong at all phases, and the P-Cygni profile is well-defined.  

Furthermore, a comparison between the median spectra of our SNe~IIb sample and those presented by \cite{liu2016analyzing} reveals strong similarities. In both cases, the \ha\ line shows broad absorption with a flattened-topped emission component, accompanied by a strengthening He~I line that becomes more prominent as the SN evolves. Although \cite{liu2016analyzing} sample does not cover phases as early as those analysed in this work, the median spectra between 10 and 40 days post-explosion are consistent across both works. 

Examining the early spectra of SNe~II, IIb, and 87A-like reveals significant observational differences in the strength and shape of the H and He lines among the different classes. These subtle differences enable early characterisation and more reliable classifications. Therefore, the following analysis is focused on the \ha\ absorption feature and the He I $\lambda5876$ lines.

\section{Results}
\label{sec:res}

\subsection{Spectral measurements}
\label{sec:generales}

\begin{figure*}
\centering
\includegraphics[width=0.48\textwidth]{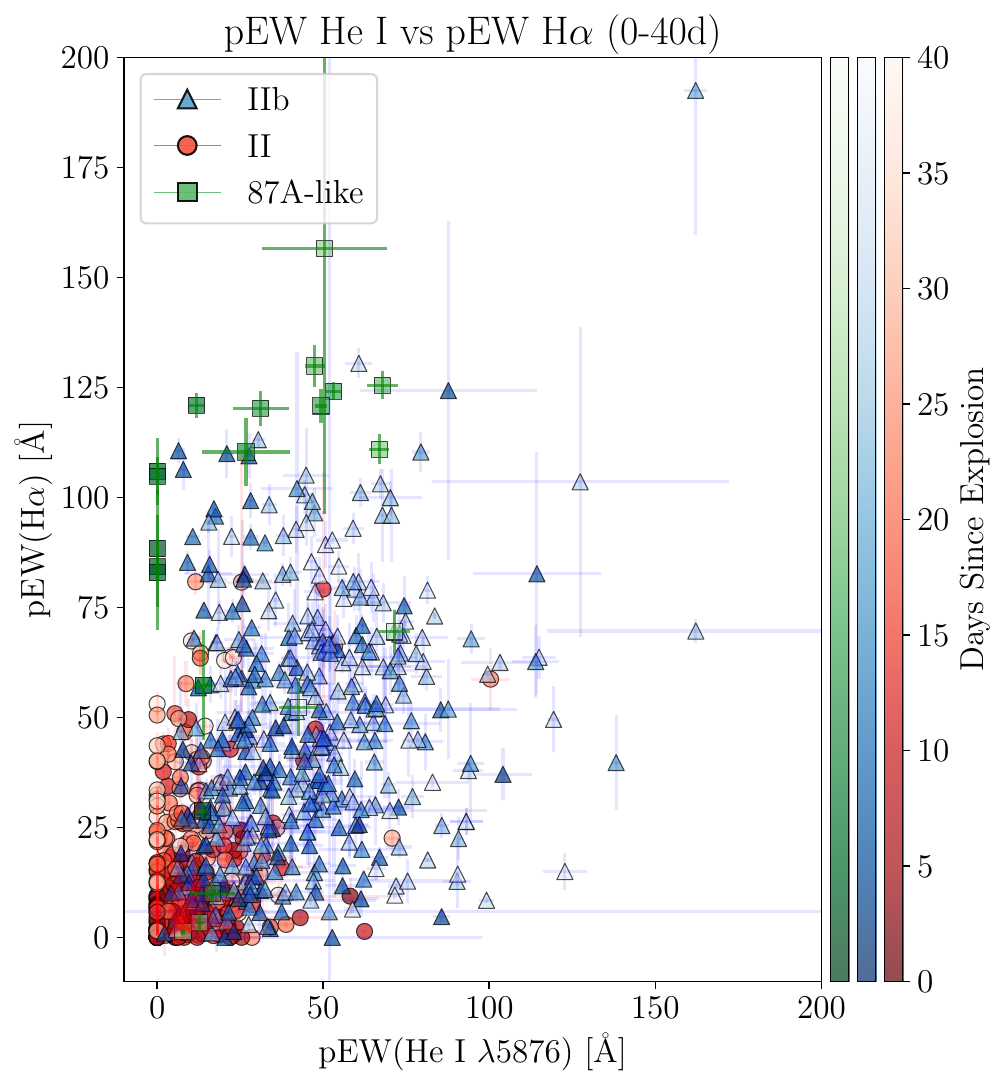}
\includegraphics[width=0.48\textwidth]{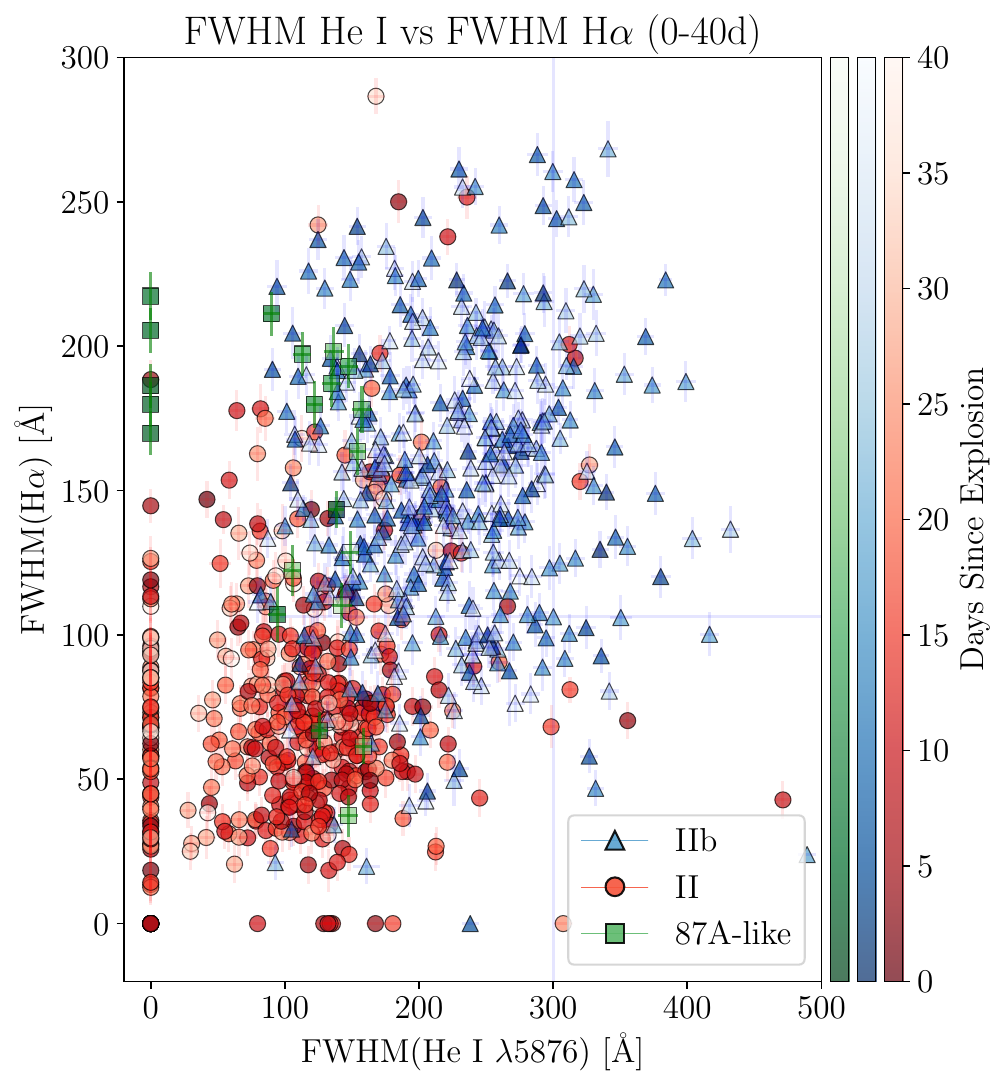}
\caption{\textit{Left panel:} pEW of the \ha\ absorption profile versus the pEW of the He I $\lambda5876$ line measured across the full-time interval (0 -- 40 days). The markers are established by SN type: SNe~IIb are shown as blue triangles, SNe II as red circles, and SNe 87A-like as green squares. The colour bars on the right side indicate the SN phase. \textit{Right panel:} Same as the left panel, but for the FWHM measurements.}
\label{fig:pEW_FWHM_0-40}
\end{figure*}

In this section, we present the results of our measurements. The values for the pEW and FWHM covering the first 40 days after the explosion are presented in Figure~\ref{fig:pEW_FWHM_0-40}. The colour bars on the right side indicate the SN phase. The analysis across the full-time interval (0 -- 40 days) does not reveal a distinct separation into two clusters; instead, it shows a continuum where SNe~IIb tend to dominate at higher values.   

\begin{figure*}
\centering
\includegraphics[width=0.4\textwidth]{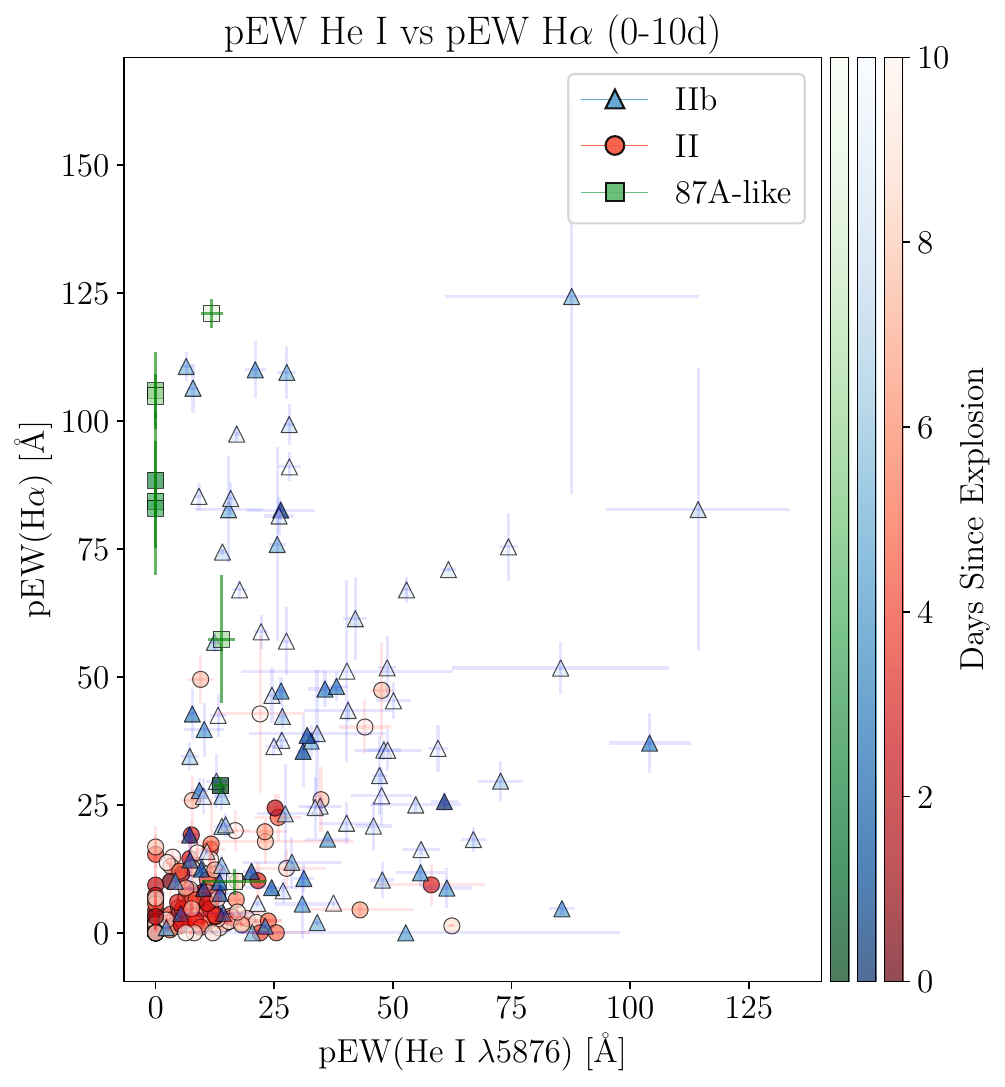}
\includegraphics[width=0.4\textwidth]{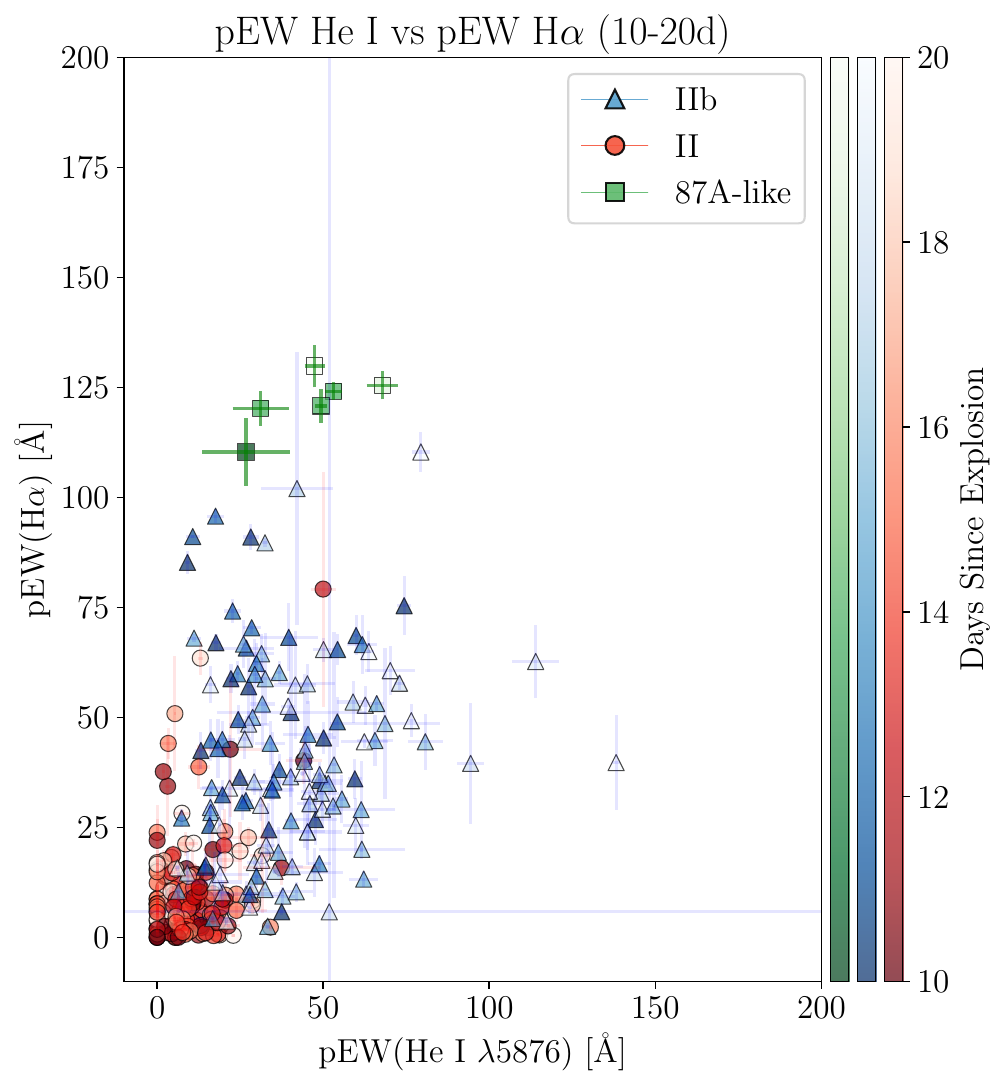}\\
\includegraphics[width=0.4\textwidth]{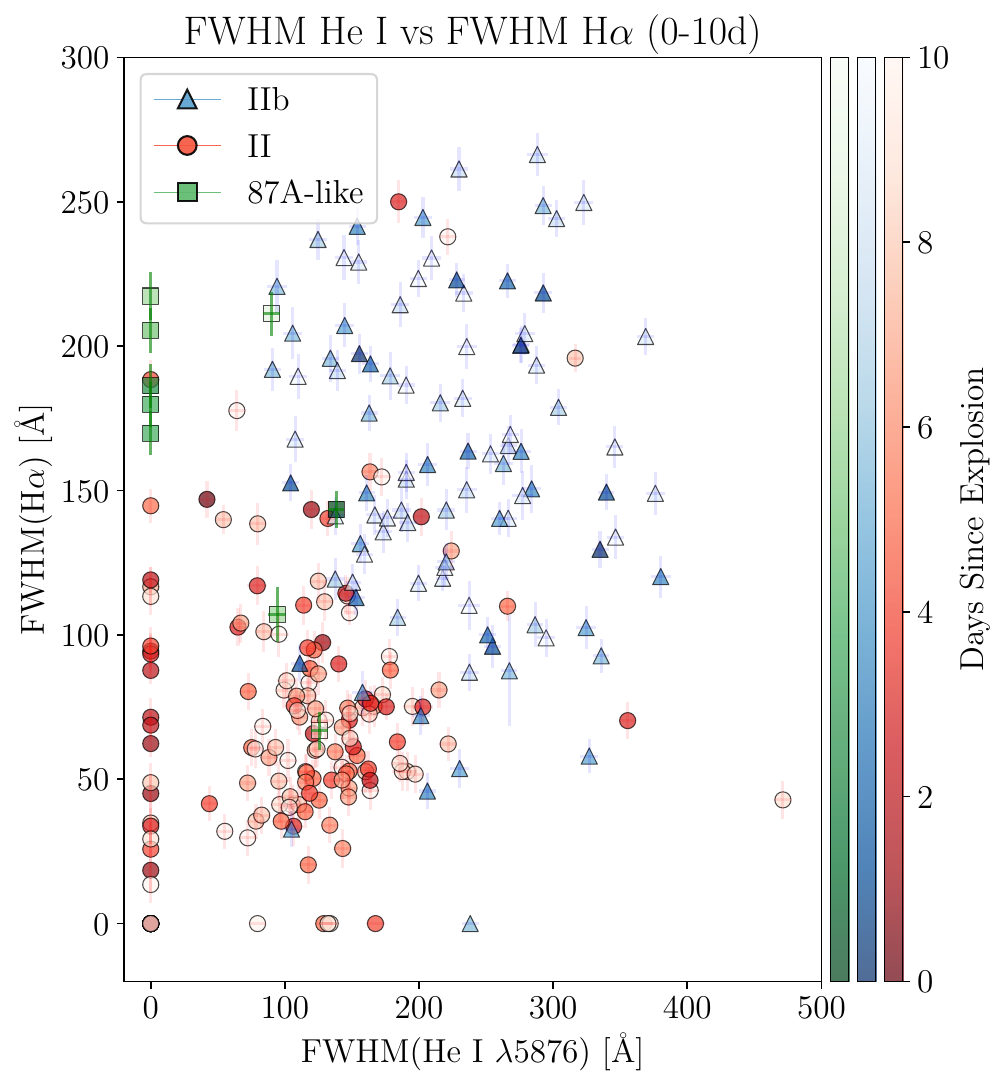}
\includegraphics[width=0.4\textwidth]{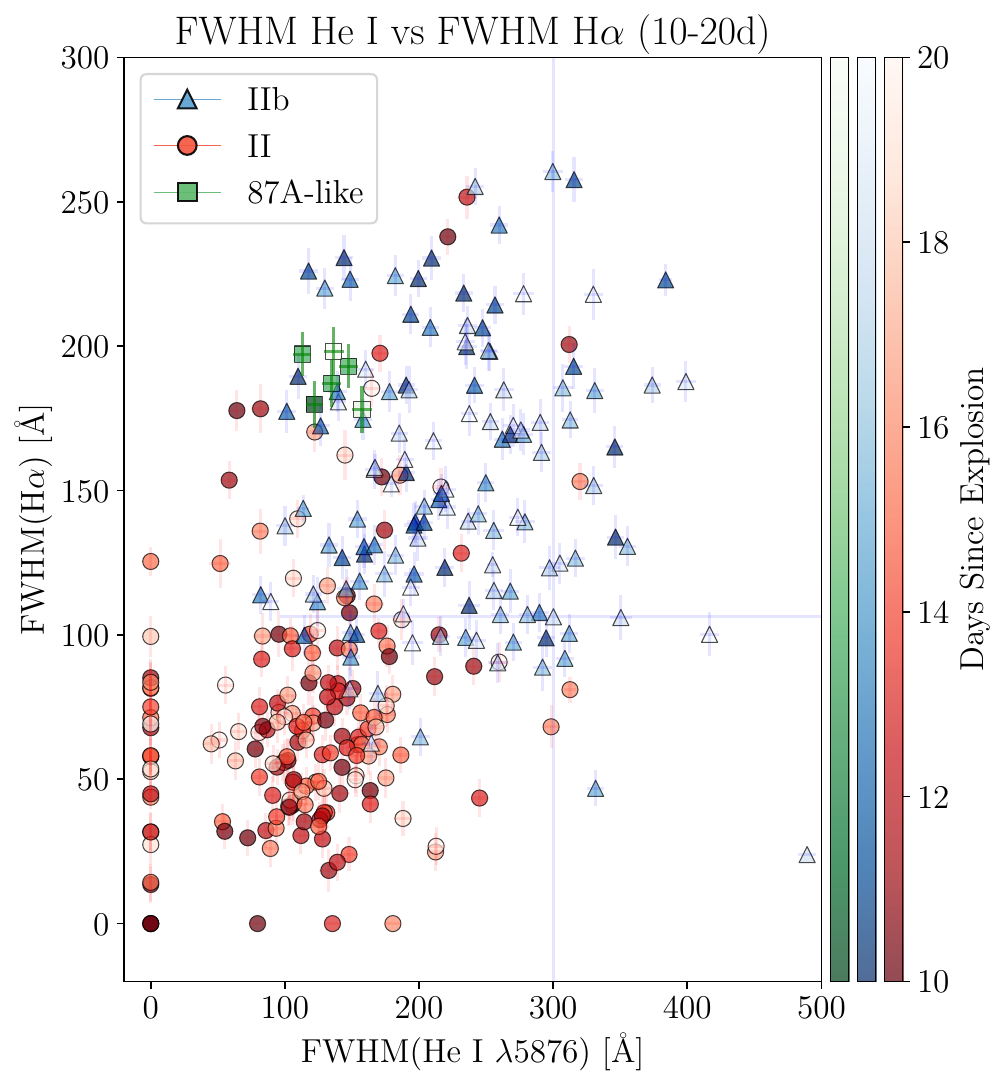}
\caption{pEW and FWHM values of \ha\ line and He I $\lambda5876$ for 0 -- 10 days and 10 -- 20 days. The markers are established by SN type: SNe~IIb are shown as blue triangles, SNe II as red circles, and SNe 87A-like as green squares. The colour bars on the right side indicate the SN phase. The pEW measurements are in the top panels, and the FWHM measurements are in the bottom panels. In the left panels, we show the measurements within 0 -- 10 days and in the right panels, between 10 -- 20 days.}
\label{fig:pEW-FWHM_0-10_10-20}
\end{figure*}

We examined the spectral features across smaller time intervals to investigate whether the observed continuum is intrinsic or a consequence of the broad time range considered (0 -- 40 days). The results for the earliest two intervals (0 -- 10 and 10 -- 20 days) are presented in Figure~\ref{fig:pEW-FWHM_0-10_10-20}. Within these intervals, the pEW and FWHM parameters reveal clear trends: SNe~IIb consistently exhibit stronger \ha\ and He I features throughout the entire early time range.  Notably, an intermediate region is observed around 25 \AA\ in the pEW of both \ha\ and He I, as well as around 125 \AA\ for \ha\ and 200 \AA\ for He I in the FWHM. This overlap between the two populations further supports the continuum observed in the full-time interval (0 -- 40 days). However, we note that the overlap between the groups becomes more significant as time passes. Specifically, the later the analysed time range, the greater the overlap, as the SN~II values grow stronger as time passes (see Figures~\ref{fig:pEW_timeranges} and \ref{fig:FWHM_timeranges} in the Appendix~\ref{app:fig}). This highlights the important role that the epoch of the SN will have in our analysis and the development of our method. A summary of the pEW and FWHM values for each SN type at each time interval is presented in Appendix~\ref{app:tables} (see Table  \ref{tab:main_values}).

We also find that, although SNe~87A-like are spectroscopically similar at all phases to SNe~II, they exhibit notably higher values, especially in the \ha\ absorption feature (Figures~\ref{fig:pEW_FWHM_0-40} and ~\ref{fig:pEW-FWHM_0-10_10-20}). 
Despite this, we excluded them from further analyses due to their low numbers.

To evaluate the statistical significance of our results, we applied the Kolmogorov-Smirnov test (KS test). This non-parametric method computes the maximum difference between the Cumulative Distribution Function (CDF) of two data sets under the null hypothesis that both groups are drawn from populations with identical distributions. The KS test results for the comparison of the pEW and FWHM of \ha\ and He I between SNe~II and IIb are summarised in Table~\ref{table:KS-test}. The Table includes the KS statistic (greatest difference between the two CDFs) and the corresponding p-value (significance of the difference obtained) for each time interval. A p-value$<0.05$ indicates the rejection of the null hypothesis, suggesting a statistically significant difference between the two groups. 

\begin{table*}
\setlength{\tabcolsep}{4pt}
\caption{The KS-statistic run for our measurements. Separately for the pEW and FWHM results for the lines of \ha\ and He I.}
\label{table:KS-test}
\begin{tabular}{cl|llll|llll}
        &                                           & \multicolumn{4}{c|}{\textbf{pEW}}                                                       & \multicolumn{4}{c}{\textbf{FWHM}}                                                           \\ 
\cline{3-10}
\multicolumn{1}{c}{\textbf{Time range}} &  \multicolumn{1}{|c|}{\# Sp.}                          & \multicolumn{2}{c}{\textbf{He I}}        & \multicolumn{2}{c|}{\textbf{\ha}}        & \multicolumn{2}{c}{\textbf{He I}}        & \multicolumn{2}{c}{\textbf{\ha}}                 \\
        &             \multicolumn{1}{|c|}{}               & \textit{KS-statistic} & \textit{p-value} & \textit{KS-statistic} & \textit{p-value} & \textit{KS-statistic} & \textit{p-value} & \textit{KS-statistic} & \textit{p-value}         \\ 
\cline{1-10}
\multicolumn{1}{c}{\textbf{0-10d}}& \multicolumn{1}{|c|}{220} & 0.349          & 0.0002                   & 0.281             & 0.005                & 0.303           & 0.002             &       \textit{0.205} & \textit{0.085}                  \\
\multicolumn{1}{c}{\textbf{10-20d}}& \multicolumn{1}{|c|}{278} & 0.482          & 2.012e-08               & 0.439             & 5.374e-07            & 0.386           & 1.609e-05               & 0.445          & 3.048e-07                       \\
\multicolumn{1}{c}{\textbf{20-30d}}& \multicolumn{1}{|c|}{233} & 0.348          & 1.494e-04               & 0.264             & 8.490e-03                & 0.340           & 2.366e-04               & 0.275          & 5.434e-03                           \\
\multicolumn{1}{c}{\textbf{30-40d}}& \multicolumn{1}{|c|}{135} & \textit{0.205} & \textit{0.227}          & \textit{0.141}    & \textit{0.667}       & \textit{0.195}  & \textit{0.273}          & \textit{0.259} & \textit{0.063}                  \\
\multicolumn{1}{c}{\textbf{0-40d}}& \multicolumn{1}{|c|}{866} & 0.732          & 7.914e-51               & 0.675             & 1.530e-42            & 0.635           & 3.131e-37               & 0.624          & 7.170e-36                       \\
\end{tabular}
\begin{list}{}{}
\item The table shows the two values returned by the KS-test: the KS-statistic and the p-value. Values that are not statistically significant are shown in italics.
\end{list}
\end{table*}

From Table~\ref{table:KS-test}, we observe that most distributions have p-values significantly below the 0.05 threshold, indicating a meaningful separation between the SN~II and IIb. The two samples exhibit statistically significant differences in the earliest 30 days post-explosion, with the most clear separation occurring within a 10 to 20-day interval. These findings support our hypothesis that SNe~II and IIb display differences in their early spectral evolution, reinforcing the possibility of identifying them based on pEW and FWHM values.

The obtained KS test values suggest that while early-time observations can effectively differentiate between the two types based on pEW values, this distinction becomes less clear as SNe evolve. Interestingly, the greatest difference between SN~II and IIb is observed during the 10 -- 20-day interval rather than the earlier range (0 -- 10 days). This is likely because both types are characterised by a blue featureless spectrum without prominent features soon after the explosion. Even as H lines emerge, they may still be weak, lacking sufficient signatures to differentiate between SNe~II and IIb. Consistent with this, we find no significant difference in the \ha\ FWHM distribution during the first 10 days post-explosion.

From 30 days post-explosion, classifying SNe~II and IIb based on the analysed spectral features becomes increasingly challenging due to the significant overlap between the two populations. For the 30 -- 40-day interval, the p-values exceed the significance threshold across all analysed distributions, indicating that the differences between the pEW and FWHM of the two SN types decrease as their spectra evolve. Although the reduced number of spectra may influence the results, the physical convergence of spectral features also contributes. Around 30 days after explosion, the \ha\ line in SNe~II typically strengthens, reaching intensities comparable to those observed in SNe~IIb. At the same time, the He~I line is often blended with the emerging Na~I feature, particularly for SNe~II \citep{gutierrez2017type}. Nevertheless, the distinction between the two types remains clear at this stage when considering additional spectral features beyond \ha\ and He~I. Indeed, by $\sim30$ days post-explosion, the \ha\ P-Cygni profile in SNe~IIb begins to exhibit distinct characteristics of this subtype. Notably, the He I $\lambda6678$ feature, located near the peak of \ha\, becomes more prominent, indicating an interference in the emission of the H feature. Additionally, He I lines emerge at this phase, providing robust criteria for distinguishing between SNe~II and IIb.

\subsection{Machine learning methods}

Machine learning methods offer powerful tools for identifying and analysing differences between the studied classes. These techniques are particularly useful for uncovering patterns and relationships that may remain hidden by traditional approaches. By handling complex, non-linear patterns, machine learning methods can effectively identify the features that distinguish the classes. This subsection explores our previous results by employing various machine-learning techniques. These include density contours, Quadratic Discriminant Analysis (QDA), t-SNE (T-distributed Stochastic Neighbour Embedding), Linear Discriminant Analysis (LDA), and Random Forest Classifier (RFC).

\subsubsection{Density Contours}
\label{sec:densitycontourns}

To further investigate the effect of overlapping values and the potential separation between the two SN types, we analysed their density evolution using a kernel density estimate (KDE). This method, which applies a Gaussian kernel with bandwidth$=1.0$ by default, smooths the observations and produces a continuous estimate of the bivariate probability density. As a result, KDE offers a clearer representation of the data through continuous probability density curves.

\begin{figure*}
\centering
\includegraphics[width=0.33\textwidth]{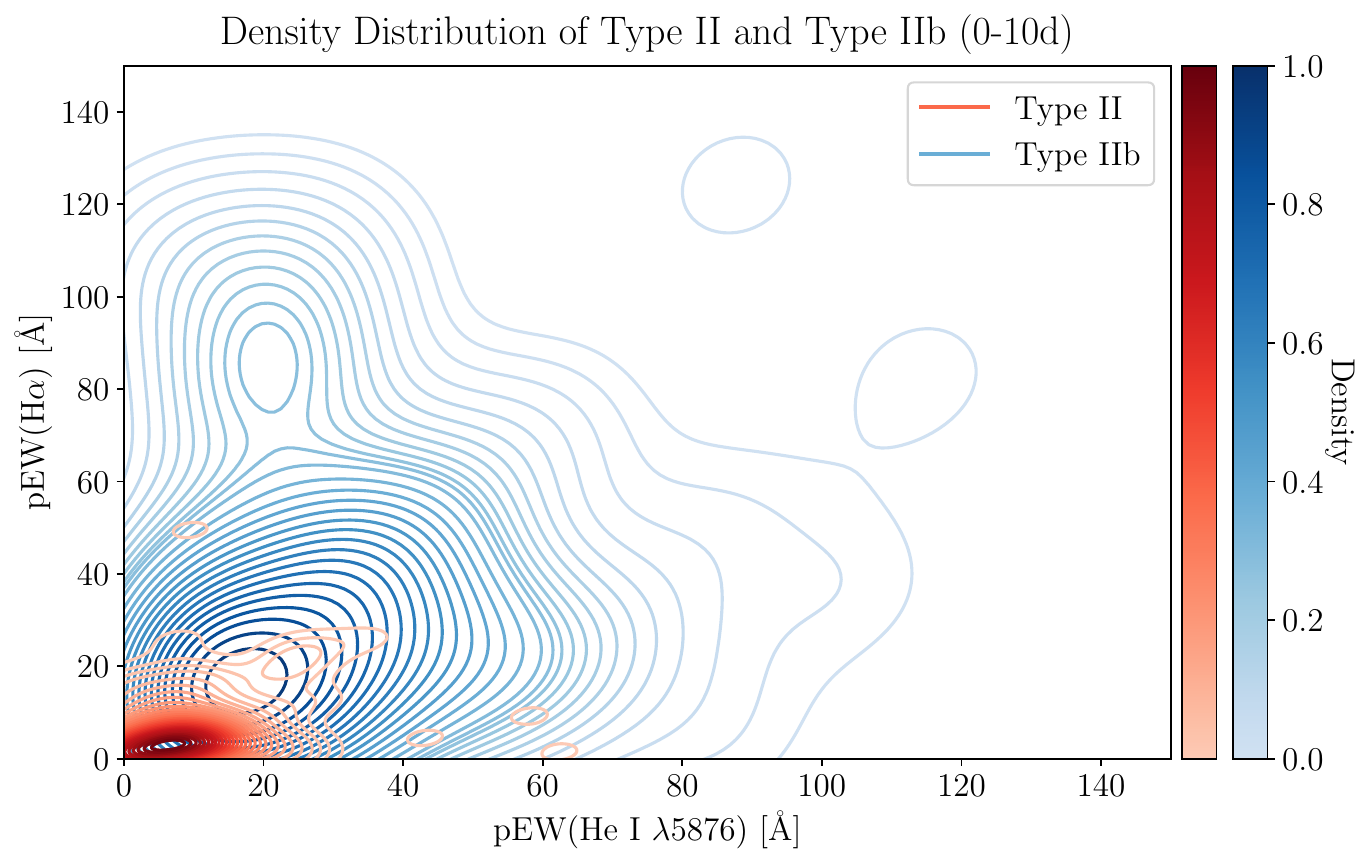}
\includegraphics[width=0.33\textwidth]{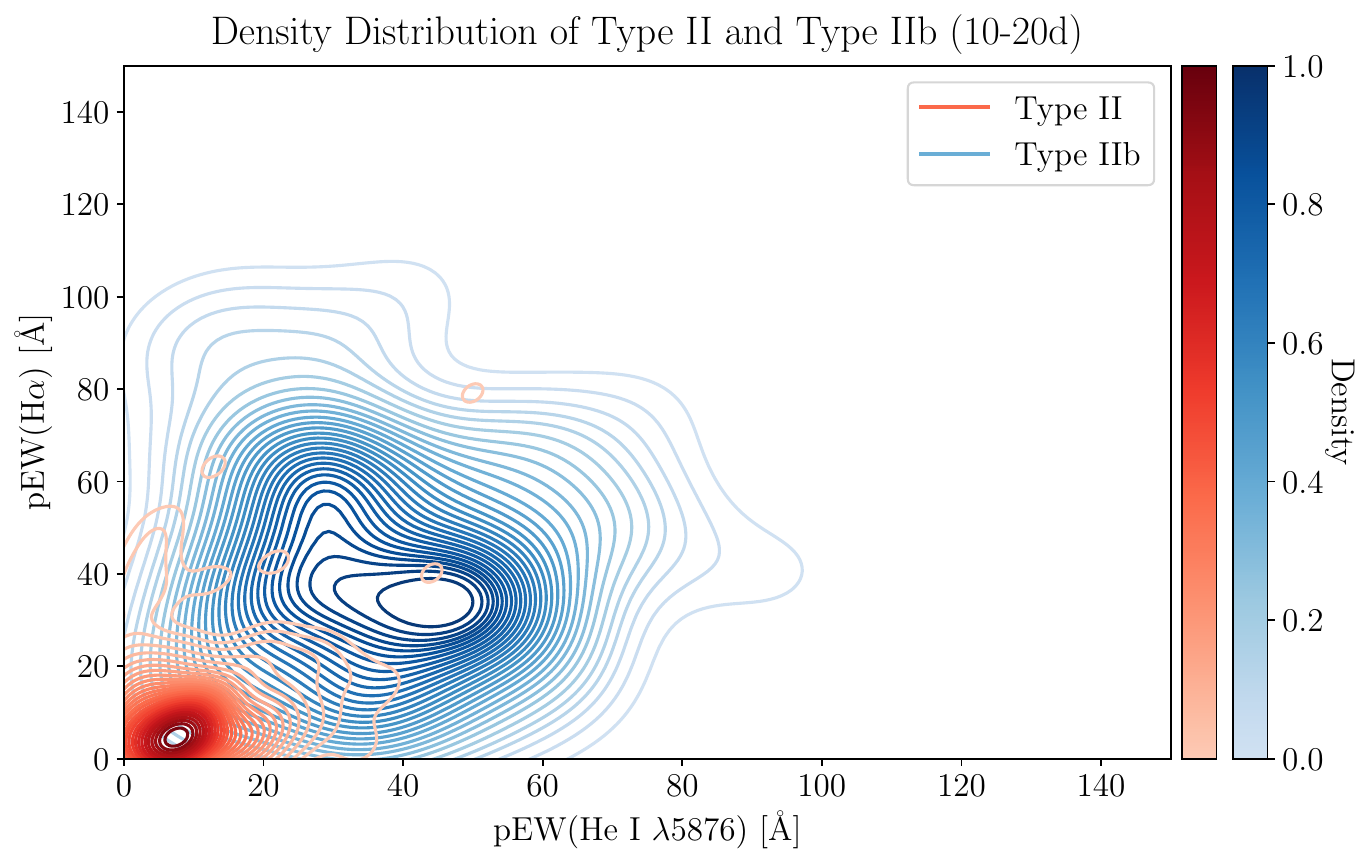}
\includegraphics[width=0.33\textwidth]{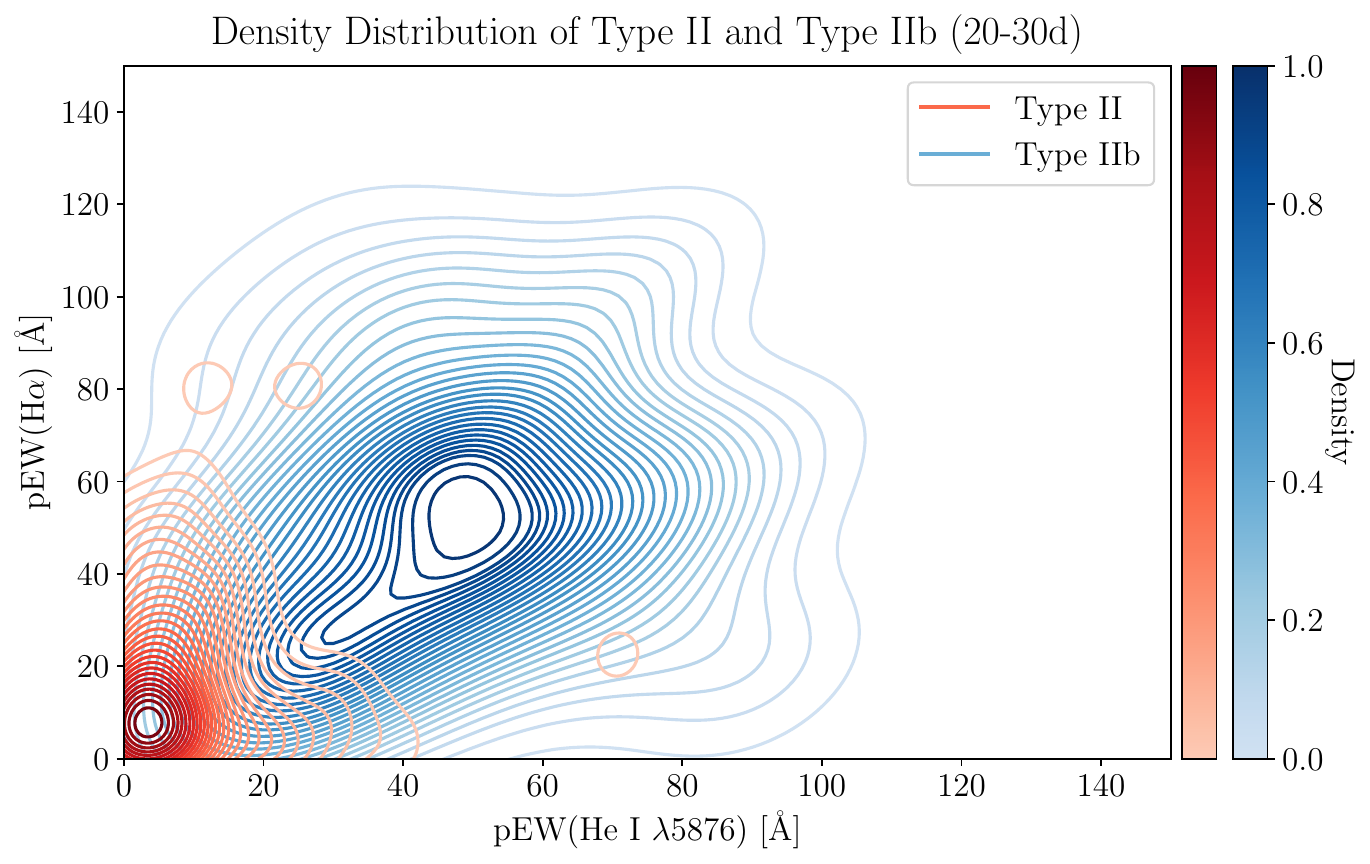}
\includegraphics[width=0.33\textwidth]{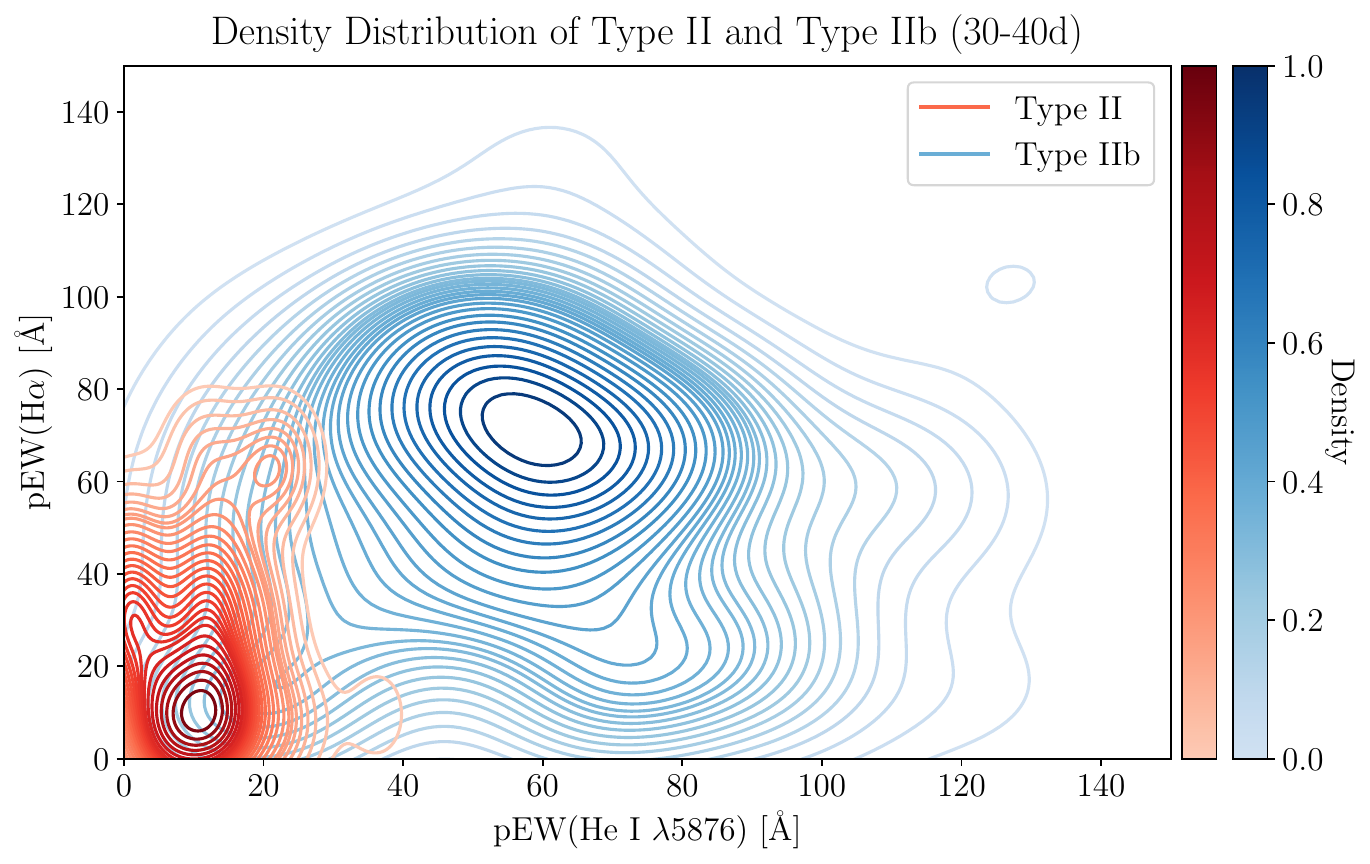}
\includegraphics[width=0.33\textwidth]{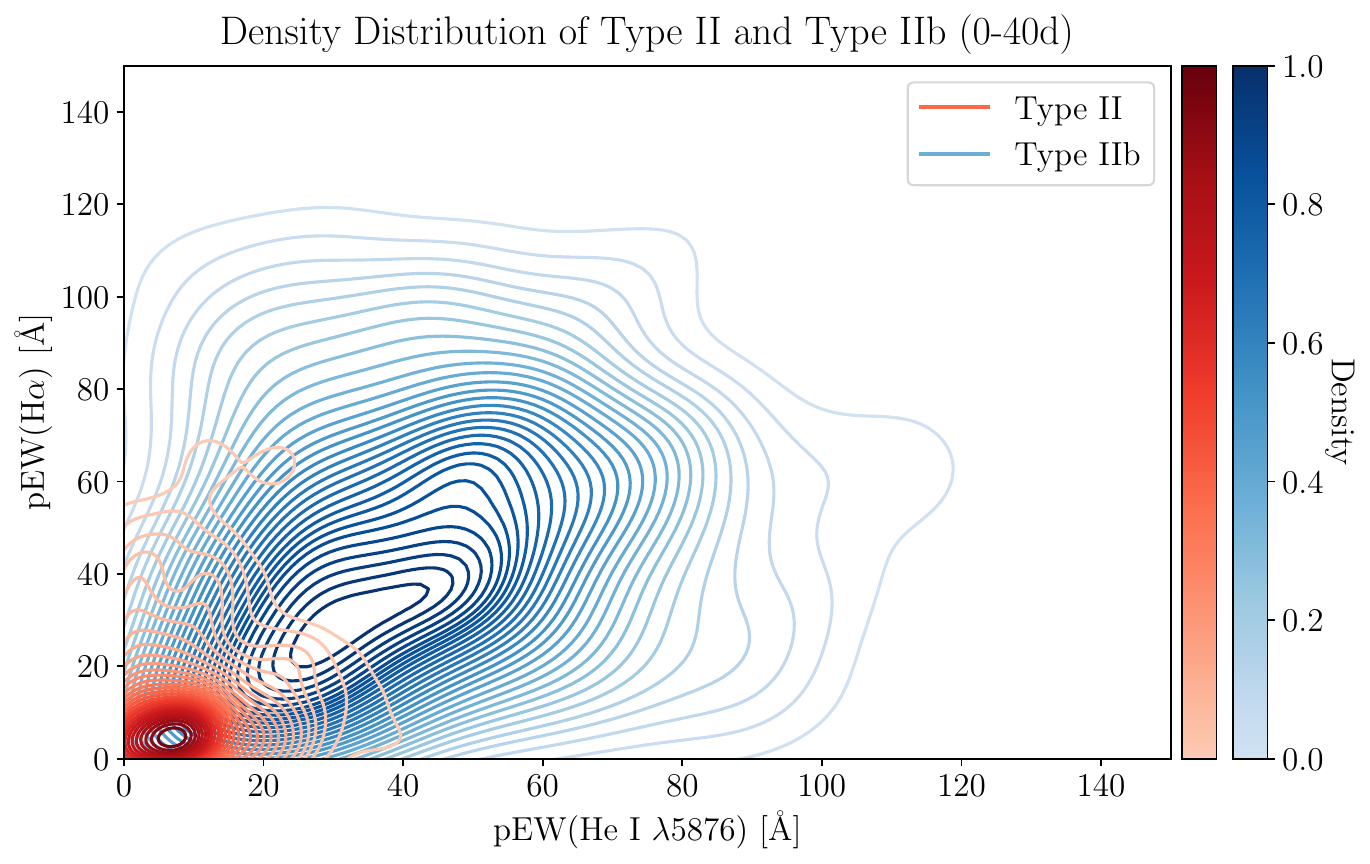}
\caption{Density contours of SNe~II (in red) and IIb (in blue) based on the pEW measurements. The colour bars on the right side indicate the normalised density values, with darker colours representing higher densities. Each plot represents the density contours at different time intervals, as indicated at the top of the figure.}
\label{fig:pEW_densitycontours}
\end{figure*}

Figure \ref{fig:pEW_densitycontours} illustrates the density contours of the pEW measurements. The comparison between the pEW of the \ha\ and He I demonstrates that SN~II measurements are intensely concentrated within a narrow region below (20, 20) \AA\ throughout the 0 -- 40-day period. However, a noticeable increase in the density contours for SNe~II appears after approximately 20 days, with the contours continuing to expand over time. Along the y-axis, a significant strengthening of the \ha\ absorption is observed, reaching values of approximately 80 \AA\ by 20 days post-explosion. In contrast, the He I feature gradually increases, extending its outer density contours to around 40 \AA.

In comparison, the SN~IIb population exhibit a broader range of values and less concentration compared to SNe~II. Although some overlap exists between SNe~IIb and II, their density contours separate with distinct central values. For SNe~IIb, the central values shift over time, increasing from approximately (30, 30) \AA\ in the early phases to around (70, 70) \AA\ at later epochs. Notably, during the later epochs, the values for SNe~IIb start to disperse, mimicking the behaviour observed in SNe~II. This dispersion is particularly evident in the 30 -- 40-day interval, where a secondary high-density contour emerges closer to the SN~II contour centre, resulting in a broader overlap between the two populations. The increasing pEW values for both SN types contribute to this overlap, as highlighted in the comparison plots presented in the previous section.  

A similar density contour analysis was also performed for the FWHM measurements (Figure~\ref{fig:FWHM_densitydistribution} in Appendix~\ref{app:fig}). The results also indicate a separation between the two SN populations. However, the overlap between the datasets is significantly more pronounced, and the concentration of points within specific regions for each population is less distinct.

\subsubsection{Quadratic Discriminant Analysis}

\begin{figure*}
\centering
\includegraphics[width=0.335\textwidth]{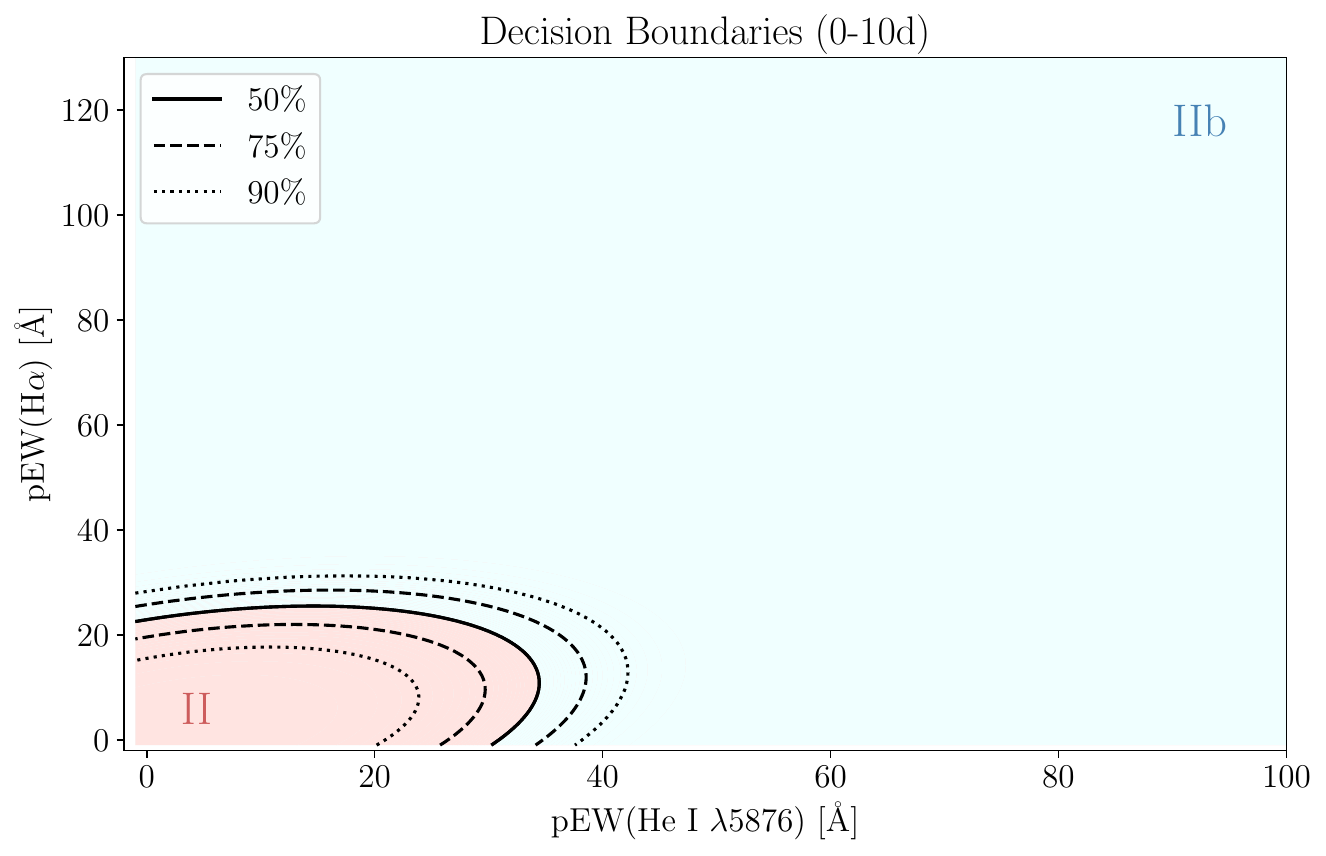}
\hspace{-0.3cm}
\includegraphics[width=0.335\textwidth]{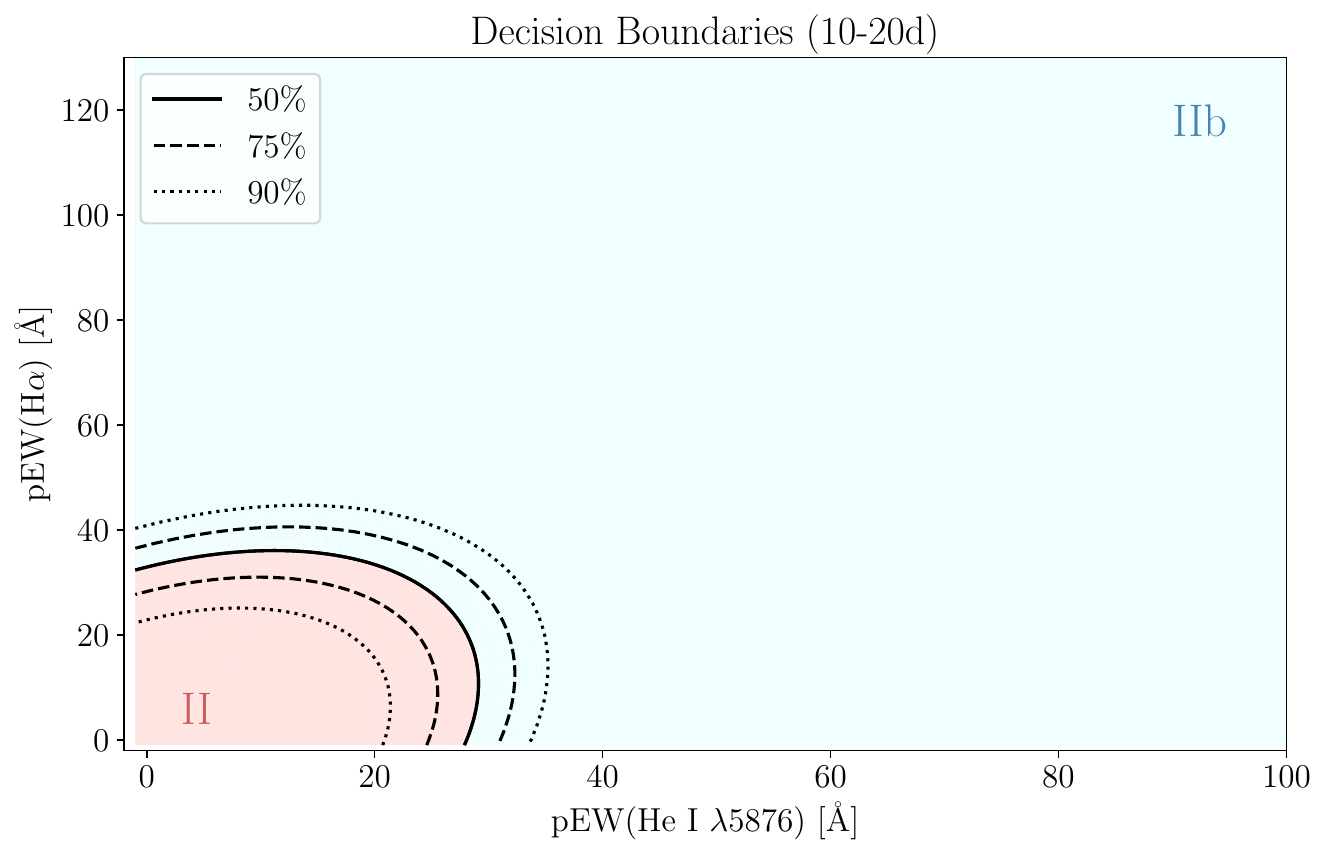}
\hspace{-0.3cm}
\includegraphics[width=0.335\textwidth]{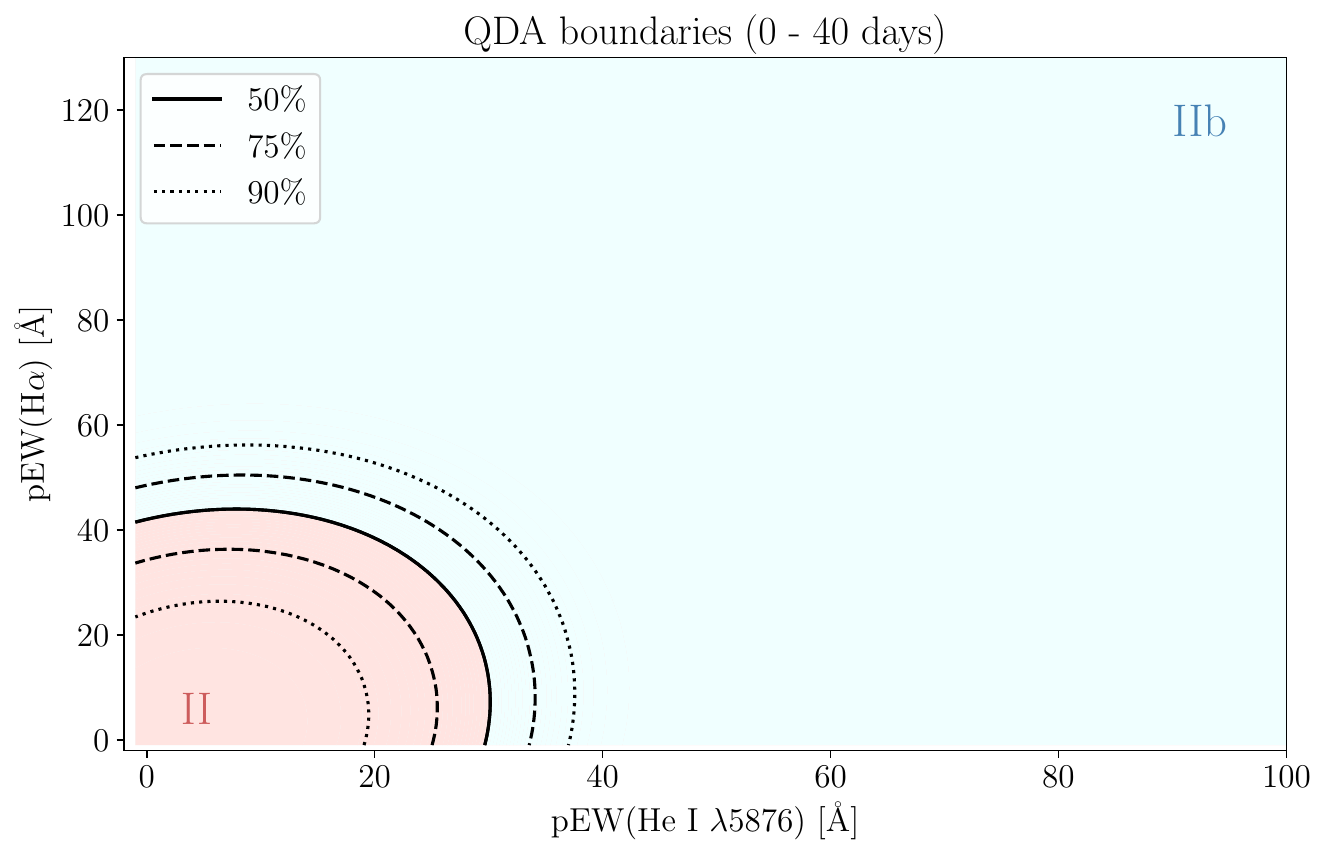}
\caption{Decision boundaries defined by the QDA classifier based on the pEW measurements for SN~II and IIb pEW are shown across three time intervals: 0 -- 10 days (left), 10 -- 20 days (middle) and 0 -- 40 days after the explosion (right). The red region corresponds to predictions for SNe~II, while the blue region represents SNe~IIb. The decision regions with a filled contour indicate where the model assigns a given SN type. Black contour lines represent classification probability thresholds, with a solid line marking a 50\% probability, dashed lines for 75\%, and a dotted line for 90\% of belonging to a specific class.}
\label{fig:regiones}
\end{figure*}

Considering the significance of the separation between the two SN types confirmed by the KS test for both pEW and FWHM across time intervals, we define the separation boundaries using a machine-learning approach. Specifically, we apply the Quadratic Discriminant Analysis (QDA) to establish the optimal decision boundary for classifying our data points. QDA uses quadratic decision surfaces to discriminate between the measurements of the two samples. It can be considered a generalisation of the Linear Discriminant Analysis (LDA), as it is not limited to dividing the data by a straight line.

Defining these regions not only demonstrates that the two populations are separable but also visually captures the distinct evolution of the pEW and FWHM. In Figure~\ref{fig:regiones}, we present the separation defined by the QDA classifier for our sample, using the pEW measurements of He I $\lambda$5876 and \ha. The decision boundaries are shown for the full dataset over the first 40 days post-explosion and for the 0 -- 10 and 10 -- 20-day intervals. Notably, the overlap between the two SN types is significantly reduced in the earliest intervals compared to the full 40-day range, as the 90\% probability region is more compact. A similar analysis was performed for other time intervals and for the FWHM measurements (details in Appendix~\ref{app:fig}, Figures~\ref{fig:QDA_pEWtimeranges} and \ref{fig:QDA_FWHMtimeranges}). However, as demonstrated by the density contours and the KS test results (Table \ref{table:KS-test}), the separation based on FWHM values is less distinct than that based on the pEWs.

\subsubsection{tSNE + LDA:}

As an alternative method to demonstrate and visualise the behaviour of the two SN populations, we applied the t-SNE (t-distributed Stochastic Neighbour Embedding; \citealt{van2008visualizing}). This statistical technique is designed for visualising high-dimensional data by reducing it to lower-dimensional spaces, offering intuitive insights into how data is structured in higher dimensions. Unlike linear methods, t-SNE can separate data that cannot be divided by a straight line, overcoming the limitations posed by the shape of our measurements. This approach is particularly useful for uncovering underlying patterns and relationships within the data.

We use t-SNE to visualise our data, incorporating all the measured values for each SNe. This process reduced our 5D dataset—including the pEW and FWHM of H$\alpha$ and He~I, along with the SNe epoch — into a 2D representation. The two resulting t-SNE components represent the reduced dimensions, capturing the spatial arrangement of the data points while preserving the relationships from the high-dimensional space. Additionally, to further analyse the 2D t-SNE distribution, we classified the data using LDA.

\begin{figure}
\centering
\includegraphics[width=0.48\textwidth]{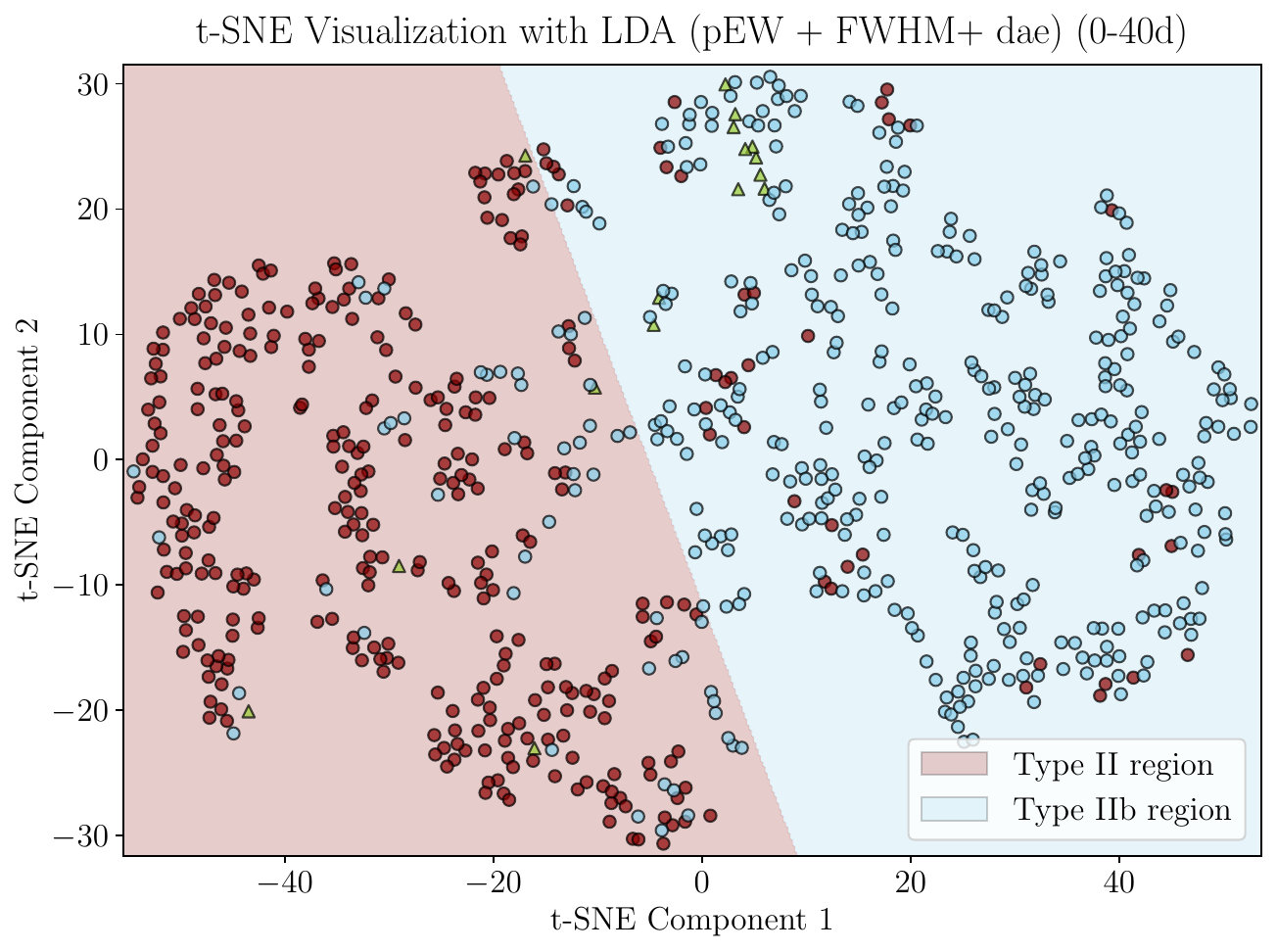}
\caption{Resulting t-SNE 2D reduced plot, using a perplexity value of 20. The red and blue dots represent SNe~II and IIb, respectively, while green triangles indicate 87A-like events. The figure is separated into two regions defined by the LDA, with the solid line representing the decision boundary found for the t-SNE projection.}

\label{fig:tSNE+LDA}
\end{figure}

The resulting 2D distribution for the 0 -- 40-day interval is presented in Figure~\ref{fig:tSNE+LDA}. The plot reveals a separation between SNe~II and IIb, with each type exhibiting a higher distribution in distinct regions of the plot. 
This finding supports the hypothesis that the two SN types can be distinguished based on the observed features (pEW, FWHM, and epoch). However, a complete separation is not evident. This continuity aligns with the trends observed in the individual pEW and FWHM comparisons of \ha\ and He~I. Visualising the data with the t-SNE method, as opposed to the QDA analysis, highlights a noticeable density decrease between the two populations, delineating two primary clusters, even though a mixed region remains visible. This 2D visualisation highlights, as discussed in Section~\ref{sec:res}, the behaviour of the 87A-like SNe, which appear in the region predominantly occupied by SNe~IIb. This may be a consequence of the strong \ha\ line exhibited by the 87A-like events (see Figure~\ref{fig:pEW_FWHM_0-40}).
Results for the 2D distributions in other time intervals are shown in Appendix~\ref{app:fig} (Figure~\ref{fig:t-SNE+LDA_timeranges}).

\subsubsection{Random Forest Classifier}
\label{sec:RFC}

Finally, we employed a Random Forest Classifier (RFC) to analyse the differences in the behaviour of the two SN types. This machine learning approach utilises multiple decision trees to classify SNe based on their specific characteristics.
To prepare the data for the RFC, we constructed a feature matrix combining the spectral parameters (FWHM and pEW for both \ha\ and He~I) with the SN phase. The dataset was split into training and testing sets using a 60/40 ratio, ensuring a robust training sample while maintaining a significant portion for testing predictive accuracy. The classifier was trained on the training sample and evaluated on the test sample, predicting the SN types based on the input parameters. Importantly, we introduced only one value per parameter for each SN to ensure accurate outputs.

\begin{figure}
\centering
\vspace{-0.4cm}
\includegraphics[width=\columnwidth]{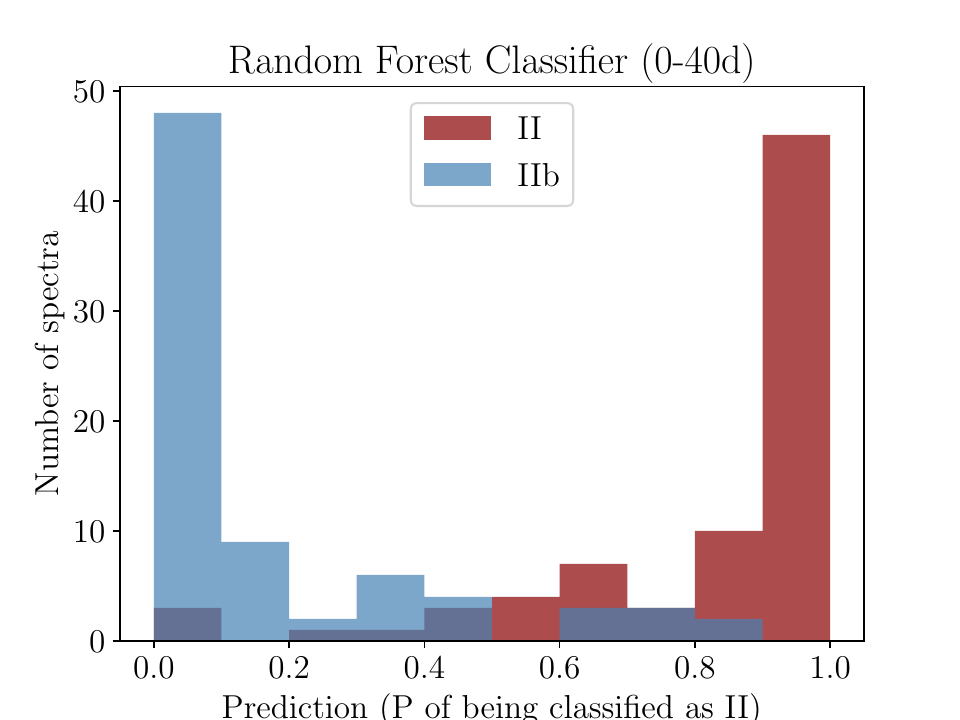}
\includegraphics[width=\columnwidth]{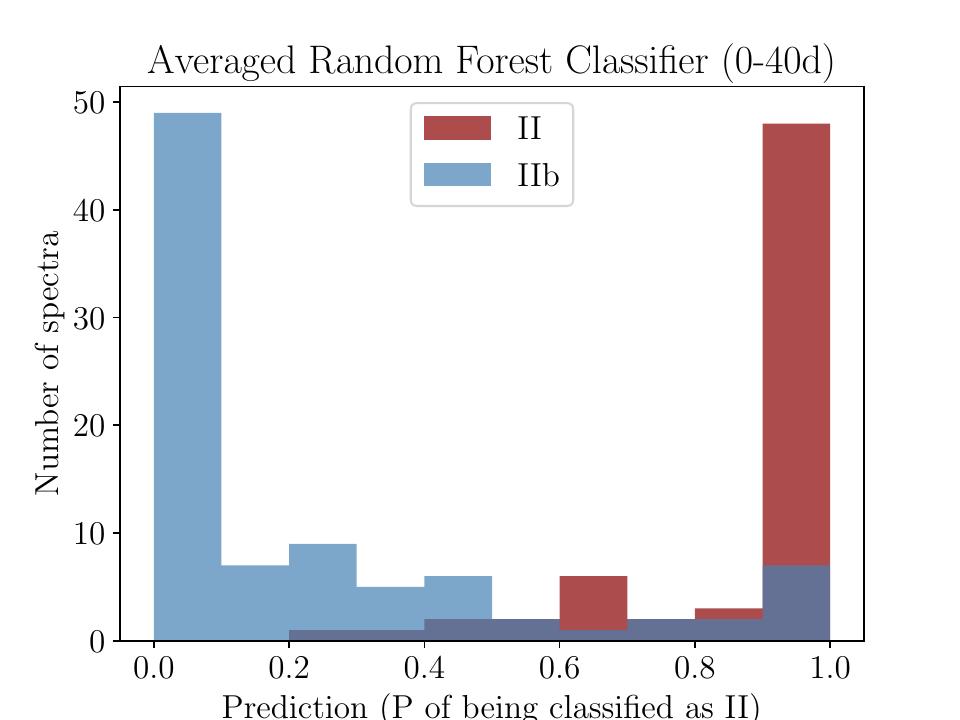}
\caption{Probabilities assigned by the RFC based on five parameters (pEW and FWHM of \ha\ and He~I, along with the epoch) within the first 40 days post-explosion. \textbf{Top panel:} Top panel: Classification probabilities obtained by randomly selecting one spectrum per SN from those with multiple observations. \textbf{Bottom panel:} Probabilities computed using the average spectrum for each SN, combining all available spectra.}
\label{fig:RF_0-40}
\end{figure}

To evaluate the RFC’s effectiveness in distinguishing between the two SN types, we applied two different sampling strategies. In the first approach, for SNe with multiple spectra, we randomly selected a single spectrum per object for inclusion in the analysis. In the second approach, we computed the average of all available spectra for each SN, including the phase, and used these averaged measurements as input to the RFC. Figure~\ref{fig:RF_0-40} presents the predicted probabilities for test sample SNe classified as Type II under both strategies. As shown in Table~\ref{tab:prercison_recall}, the classifier performed slightly better with the random selection method.

\begin{table}
\resizebox{\columnwidth}{!}{%
\begin{tabular}{llccccl}
                                                     &                                   & \shortstack {\textbf{RFC}\\(averaged)}         & \shortstack {\textbf{RFC}\\(random)}         & \textbf{t-SNE+LDA}   & \multicolumn{2}{c}{\textbf{QDA}} \\ \hline
\textbf{}                                            & \multicolumn{1}{l|}{}             & \multicolumn{1}{l}{} & \multicolumn{1}{l}{}  & \multicolumn{1}{l}{} & \multicolumn{1}{l}{pEW}  & FWHM  \\
\multicolumn{1}{c}{\textbf{Precision}}               & \multicolumn{1}{l|}{\textit{II}}  & 0.813                & 0.897           & 0.848                & 0.837                    & 0.871 \\ 
                                                     & \multicolumn{1}{l|}{\textit{IIb}} & 0.950                & 0.961           & 0.907                & 0.909                    & 0.879 \\ \hline
\multicolumn{1}{c}{\multirow{2}{*}{\textbf{Recall}}} & \multicolumn{1}{l|}{\textit{II}}  & 0.938                & 0.897           & 0.890                & 0.925                    & 0.890 \\
\multicolumn{1}{c}{}                                 & \multicolumn{1}{l|}{\textit{IIb}} & 0.844                & 0.961           & 0.871                & 0.806                    & 0.857\\
\end{tabular}%
}
\caption{Comparison of precision and recall scores for the different classification methods.}
\label{tab:prercison_recall}
\end{table}

 Based on the RFC probabilities, SNe classified as Type II (in red) consistently received a high probability of obtaining the same classification by the RFC. Conversely, SNe~IIb (in blue) were assigned a low probability of being classified as Type II. These results confirm that the measured features used in our analysis effectively distinguish the two SN types. We also tested alternative data combinations introduced in the RFC, which resulted in less effective classifications; For comparison, these results are presented in Appendix \ref{app:fig} (see Figures \ref{fig:RF_AZKENA} and \ref{fig:RF_timeranges2}).

The RFC provides a statistical method for identifying differences between these two SN types and is a practical tool for classifying or reclassifying new SNe. By adjusting the training sample percentage from 60\% to the entire dataset and using the values of interest for a specific SN as the test sample, it is possible to estimate the likelihood of that SN belonging to each type. For this purpose, and given the demonstrated utility of the method, we selected the RFC model based on averaged spectral values as our preferred classification tool. This approach ensures consistent, non-variable input for each SN by integrating all available spectral data from the main sample.

\section{Discussion}
\label{sec:disc}

\subsection{A continuum in the \ha\ pEW and FWHM?}
 
The diversity observed in CCSNe arises from the varying amounts of outermost layers retained by their progenitor stars before the explosion. SNe~IIb exhibit a transitional behaviour, evolving from hydrogen-dominated to helium-dominated spectra, suggesting that their progenitors retained less hydrogen before explosion than those of SNe~II.

Our results show a complex picture: \textbf{(i)} while the pEW and FWHM of \ha\ and He I are generally higher in SNe~IIb than in SNe~II, \textbf{(ii)} no definite separation exists between these parameters, indicating a continuum. The first finding is somewhat counterintuitive, as one might expect SNe~II, which retain a more significant amount of H, to exhibit stronger spectral features. However, this can be explained by the differences in progenitor structure. In progenitor stars with a substantial hydrogen envelope, a greater fraction of the shock-deposited energy is radiated away, leading to lower kinetic energy per unit mass ($E_{kin}/M_{ej}$) and, consequently, slower expansion velocities and narrower spectral lines. In contrast, SNe~IIb, with their low-mass hydrogen envelopes, experience faster ejecta expansion, resulting in broader spectral lines \citep{dessart2024sequence}.

As for the second finding, theoretical models with varying orbital separations predict a continuum of light-curve morphologies for Type Ib, IIb, and II SNe, reflecting a range of H-rich envelope masses \citep{dessart2024sequence}. However, the distinct differences in the LCs of SNe~II and IIb suggest their progenitors have different properties \citep[e.g.][]{arcavi2012caltech, pessi2019comparison}, challenging the idea of a continuous distribution in spectral parameters.
Our results indicate that, while hydrogen retention plays a key role in shaping the observed properties of SNe~II and IIb, other factors likely contribute to their observed diversity. Even if hydrogen retention follows a continuum, this does not necessarily imply that their progenitors belong to a single, continuous evolutionary sequence. Instead, they likely represent distinct endpoints influenced by multiple parameters.

\subsection{Effectiveness of our method}

After applying several statistical techniques to quantitatively analyse the behaviour and differences between \ha\ and He~I in SNe~II and IIb, we evaluated the potential utility and reliability of these findings as a foundation for a novel classification method.

Having already included all available higher-resolution spectra in our primary sample, we tested our results using SNe with lower-resolution spectra. To this end, we gathered all available SEDM spectra for SNe~II and IIb from the Transient Name Server (TNS\footnote{\url{https://www.wis-tns.org/}}), comprising 609 SNe~II and 25 SNe~IIb. Objects without accompanying light curve information were excluded, as the explosion date was necessary to constrain the SN phase. Thus, the final sample consisted of 106 SNe, as detailed in Table~\ref{table:SEDM_list_ONA} (Appendix~\ref{app:tables}). We note the significant disparity in the number of SEDM spectra classified as SNe~II versus SNe~IIb: among the 106 SNe, 92 were classified as SNe~II and 14 as SNe~IIb. Additionally, we incorporated newly classified SNe that were released after compiling our main sample. This includes 36 newly classified objects (28 SNe~II and 8 SNe~IIb) along with two SNe (SN~2013gg and SN~2015dj) that were classified as SNe~II on TNS but as SNe~IIb on WISeREP and SN~2019oxi, which is a type II but was showing broad features. As a result, the comparison sample comprised 123 SNe~II and 22 SNe~IIb.

Following the same procedure as before, we measured the pEW and FWHM of \ha\ and He I $\lambda5876$ for the newly collected spectra. The results were then plotted within the density contours (Figure~\ref{fig:CONTOURNS+SEDM_3}) and the QDA decision regions (Figure~\ref{fig:QDA+SEDM_timeranges}, in Appendix~\ref{app:fig}). Each SN was represented according to its official classification type as recorded in TNS. This approach enabled us to validate our method by re-evaluating the classification of spectra that exhibited potential signs of misclassification. 

\begin{figure*}
\centering
\includegraphics[width=0.495\textwidth]{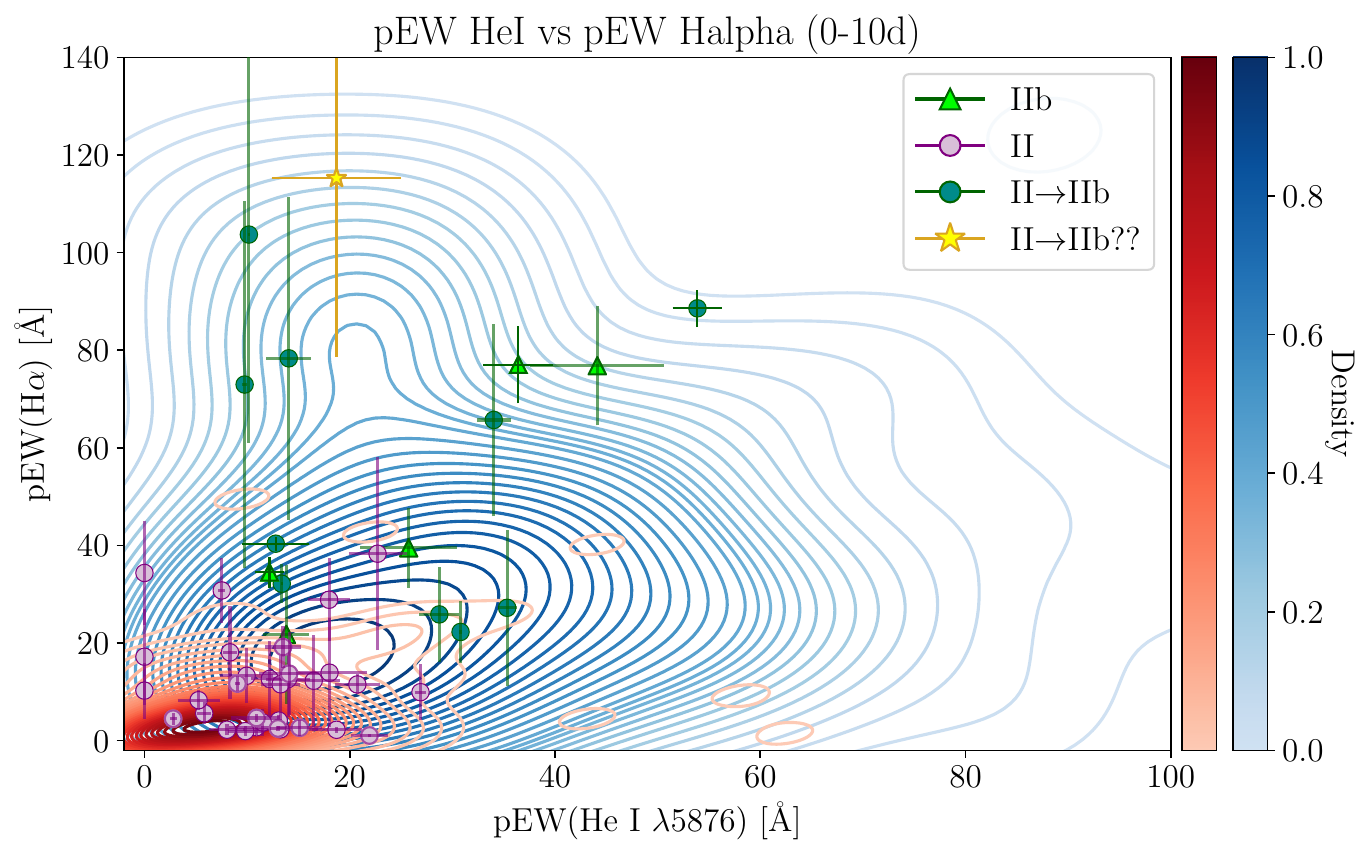}
\includegraphics[width=0.495\textwidth]{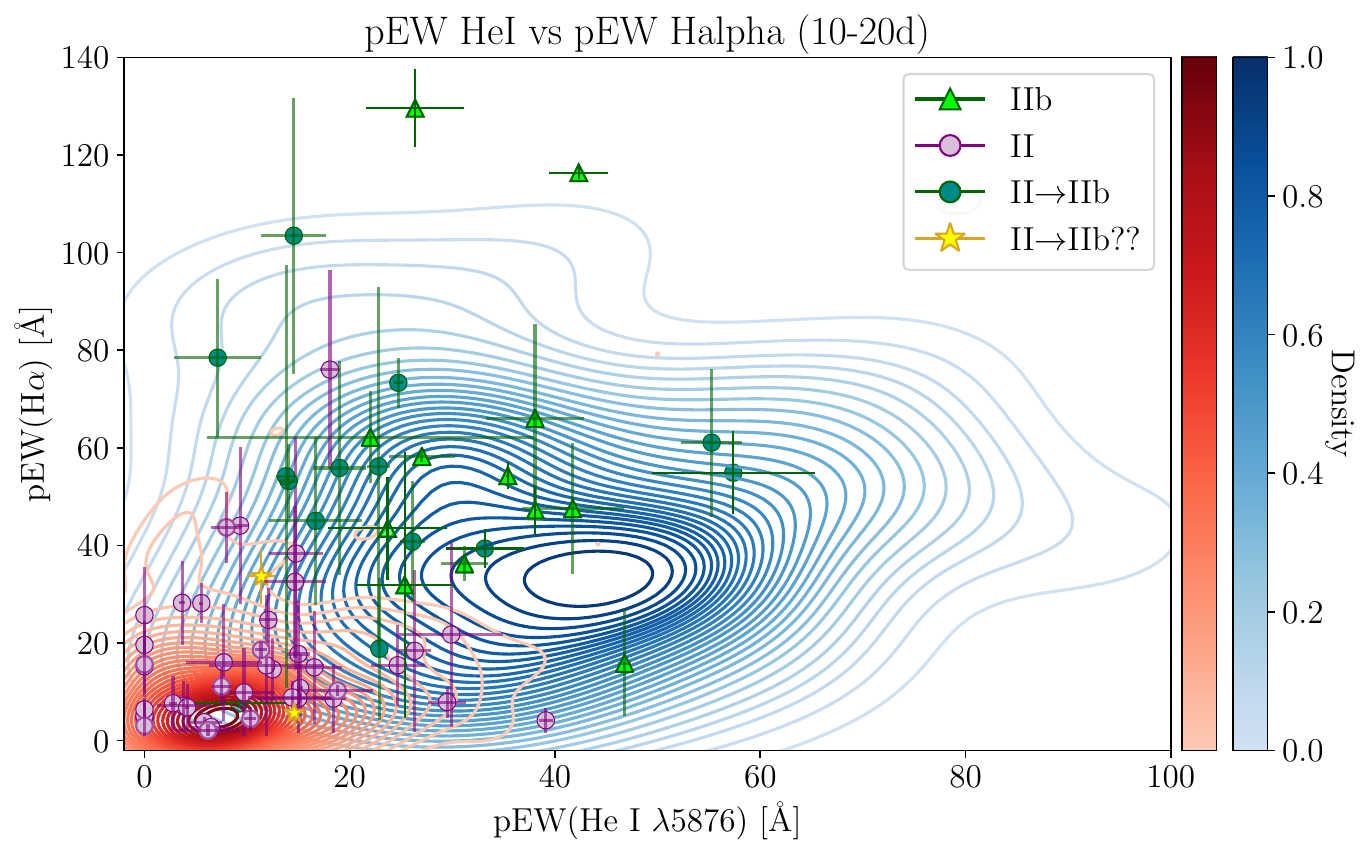}
\includegraphics[width=0.495\textwidth]{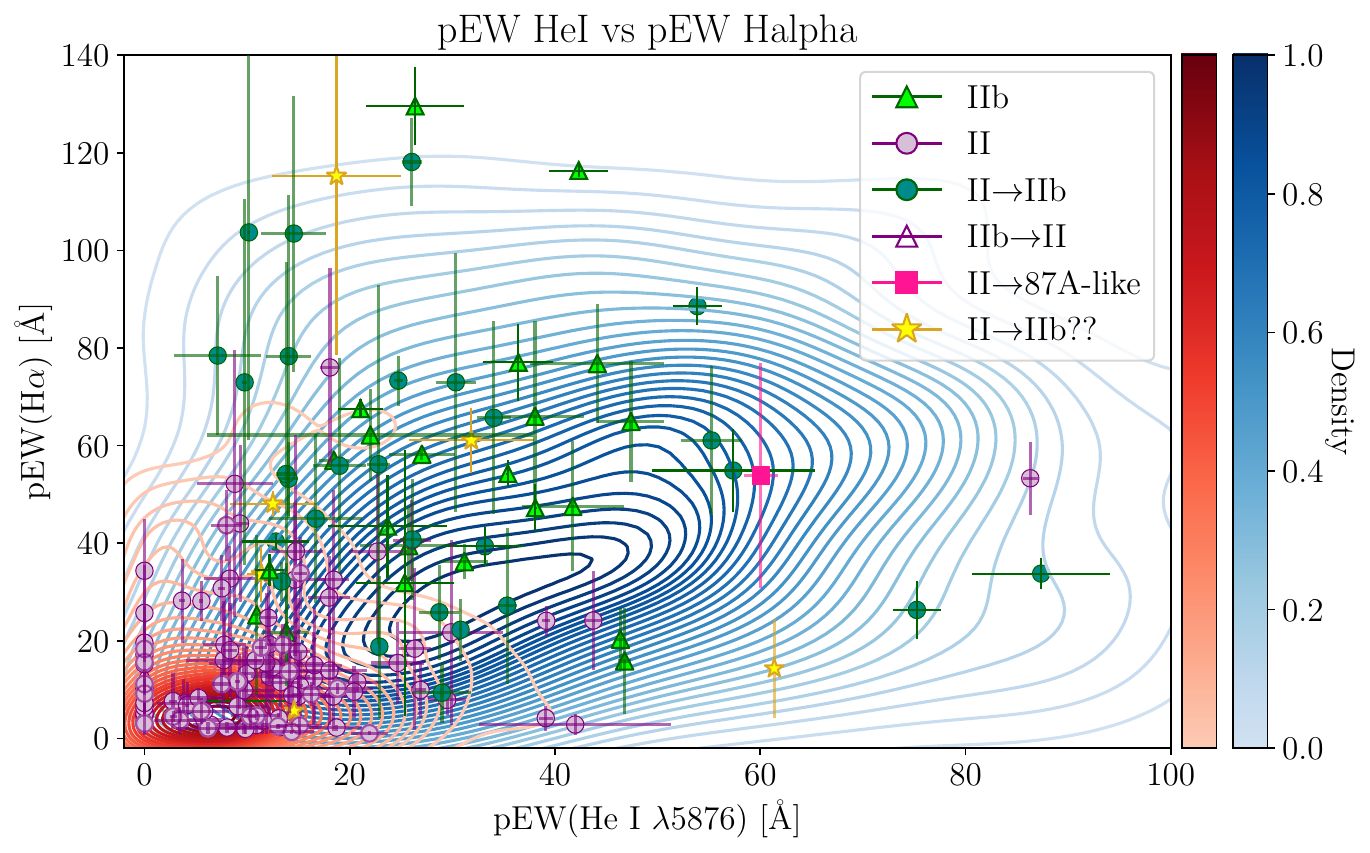}
\caption{pEW measurements of \ha\ and He~I for the comparison sample (SEDM plus newly classified SNe) overlaid on the density contours of SNe~II (red) and IIb (blue) obtained in Section~\ref{sec:densitycontourns} (same as Figure~\ref{fig:regiones}), for three time intervals: 0 -- 10 days (left), 10 -- 20 days (middle) and 0 -- 40 days (right). The comparison sample is represented by SNe~II as filled purple dots and SNe~IIb as filled green triangles. Dark green circles correspond to SNe initially classified as Type II but exhibiting SNe~IIb-like properties. Open purple triangles represent SNe, which were initially classified as Type IIb but spectroscopically belong to the SN~II class. Pink squares denote SNe~II showing 87A-like characteristics. Yellow stars highlight SNe~II that appear to exhibit SNe~IIb characteristics but lack sufficient data to confirm this classification.}
\label{fig:CONTOURNS+SEDM_3}
\end{figure*}

As shown in Figure~\ref{fig:CONTOURNS+SEDM_3}, most of the objects from the comparison sample were correctly classified; however, a few SNe exhibit spectroscopic properties that more closely align with the opposite class. For instance, some SNe initially classified as Type II display characteristics of SNe~IIb (i.e. higher pEW values), and vice versa. In these cases, we cross-verified the classification by examining their LC shapes and rise times. Specifically, this decision is based on the observation that SNe~II typically have shorter rise times than SNe~IIb  \citep{pessi2019comparison} and that most of the SNe~II exhibit a quasi-constant luminosity (a plateau) lasting several months. However, when the LC did not provide definitive confirmation, we conducted a more detailed spectral matching to refine the classification. 

Through this re-evaluation process, we proved a potential misclassification of 34 SNe. Among these, 26 objects initially classified as Type II were found to align more closely with the properties of SNe~IIb, while one SN, originally classified as SNe~IIb, was determined to be better matched with SNe~II. Notably, six SNe with insufficient photometric data (SN~2013gg, SNe~2018hqt, 2019noj, 2019oxi, 2024cgi and 2024lsi) show characteristics of both classes, providing inconclusive classifications. Lastly, SN~2020rmk was reclassified as a 87A-like event based on its spectral properties and exceptionally long-rise time ($>50$ days). The re-classified objects are shown in Figure~\ref{fig:CONTOURNS+SEDM_3} and summarised in Tables~\ref{table:reclassification_SEDM} and \ref{table:new_classifications} in the Appendix~\ref{app:tables}. 

To achieve a more accurate classification for those objects without a conclusive designation, we developed a Monte Carlo simulation with 500 iterations to compute their uncertainty. In each iteration, we resampled the base training set using bootstrapping, where some objects may be repeated and others omitted, and retrained the RFC model. This process yielded a distribution of classification probabilities for each object, enabling a more informed assessment of SN type in cases where photometric classification is not feasible. As a result, we classified SNe~2024lsi and 2018hqt as Type II and SNe~2024cgi and 2019noj as Type IIb. For the SN~2019oxi and SN~2013gg, the RFC assigned 83\% probability of being a SN~II and 17\% of being a SN~IIb, and 55.6\% and 44.4\%, respectively, for the latter.

To evaluate how well each model balances the trade-off between identifying SNe of each type, we computed the precision and recall scores for each classification method. These metrics provide an overview of each method’s ability to classify the sample accurately. The precision measures the classifier's ability to identify positive samples while minimising false positives correctly. This measures the proportion of correctly classified SNe among all predicted SNe, ensuring reliability. The recall reflects the classifier's ability to identify all actual positive samples, minimising false negatives. It quantifies the proportion of actual SNe that were correctly identified by the model, ensuring completeness. Then, a high precision score indicates fewer false positives, while a high recall score means fewer wrong classifications. The equations for each of these metrics are as follows:

\textbf{\begin{equation}
    \text{Precision:}~~\frac{TP}{TP+FP},
\end{equation}    
\begin{equation}
    \text{Recall:}~~\frac{TP}{TP+FN},    
\end{equation}}

where TP are the true positives, cases where the model correctly classifies a SN, FP are false positives, where the given classification is incorrect, and FN are false negatives, cases where the model fails to identify the SN type. The metric scores for the methods are presented in Table \ref{tab:prercison_recall}. They all have similar precision and recall scores above 0.8 for both types.

\subsection{Relative SN fractions}

\begin{figure}
\centering
\includegraphics[width=\columnwidth]{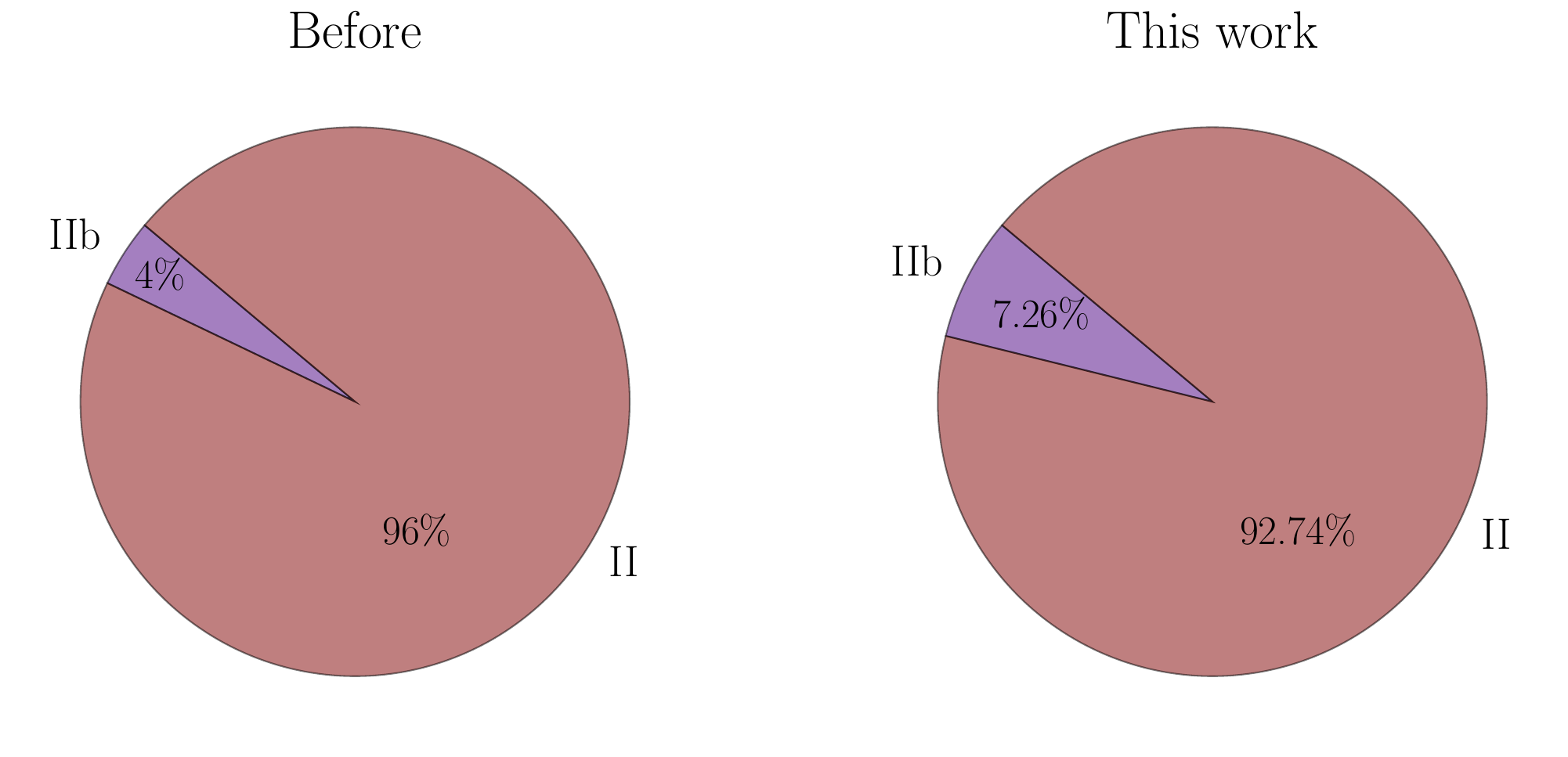}
\caption{Comparison of the fraction of SEDM spectra of each SN type before and after the performance of our re-classifications. }
\label{fig:SEDM_comparison_pie}
\end{figure}

Once all SNe in our test sample were assigned a SN type following the methodology previously described, we compared these results with their initial classifications. We specifically analysed the impact of re-examined SEDM spectra on the original SN-type fractions. This comparison, presented in Figure~\ref{fig:SEDM_comparison_pie}, reveals a noticeable increase in the fraction of SNe~IIb, with a $3.26\% \pm 0.13\%$ rise (almost twice the initial fraction) in SEDM spectra classified as SNe~IIb.

The rates of CCSNe using the Lick Observatory Supernova Search volume-limited sample have been estimated in several works \citep[e.g.][]{leaman2011nearby, li2011nearby, li2011nearbyb, maoz2011nearby} and recently revised by \citet{shivvers2017revisiting}. They found that SNe~II constitute approximately 70\% of all CCSNe, while SESNe account for about 30\%. Within SESNe, there are SNe~Ib, Ic, and IIb. Based on known classifications and reclassifications, they estimated the true CCSNe fractions to be 69.6\% for SNe~II, 10.94\% for SNe~IIb, and 19.46\% for other SESNe. More recently, \citet{Perley20}, using the ZTF Bright Transient Survey (BTS) magnitude-limited sample, found that hydrogen-rich SNe (SNe~II, IIb, IIn, and superluminous SNe~II) constitute 72.2\% of all CCSNe. Within the hydrogen-rich SN class, SNe~II are 69.05\% of all the CCSNe, while the fraction corresponding to IIb is 5.10\%. The remaining 25.85\% is formed by SNe~Ib, Ic, Ic-BL and Ibn SESNe. These percentages correspond to the fractions excluding the SLSNe that were included in the analysis of \citep{Perley20}.

\begin{figure}
\centering
\includegraphics[width=\columnwidth]{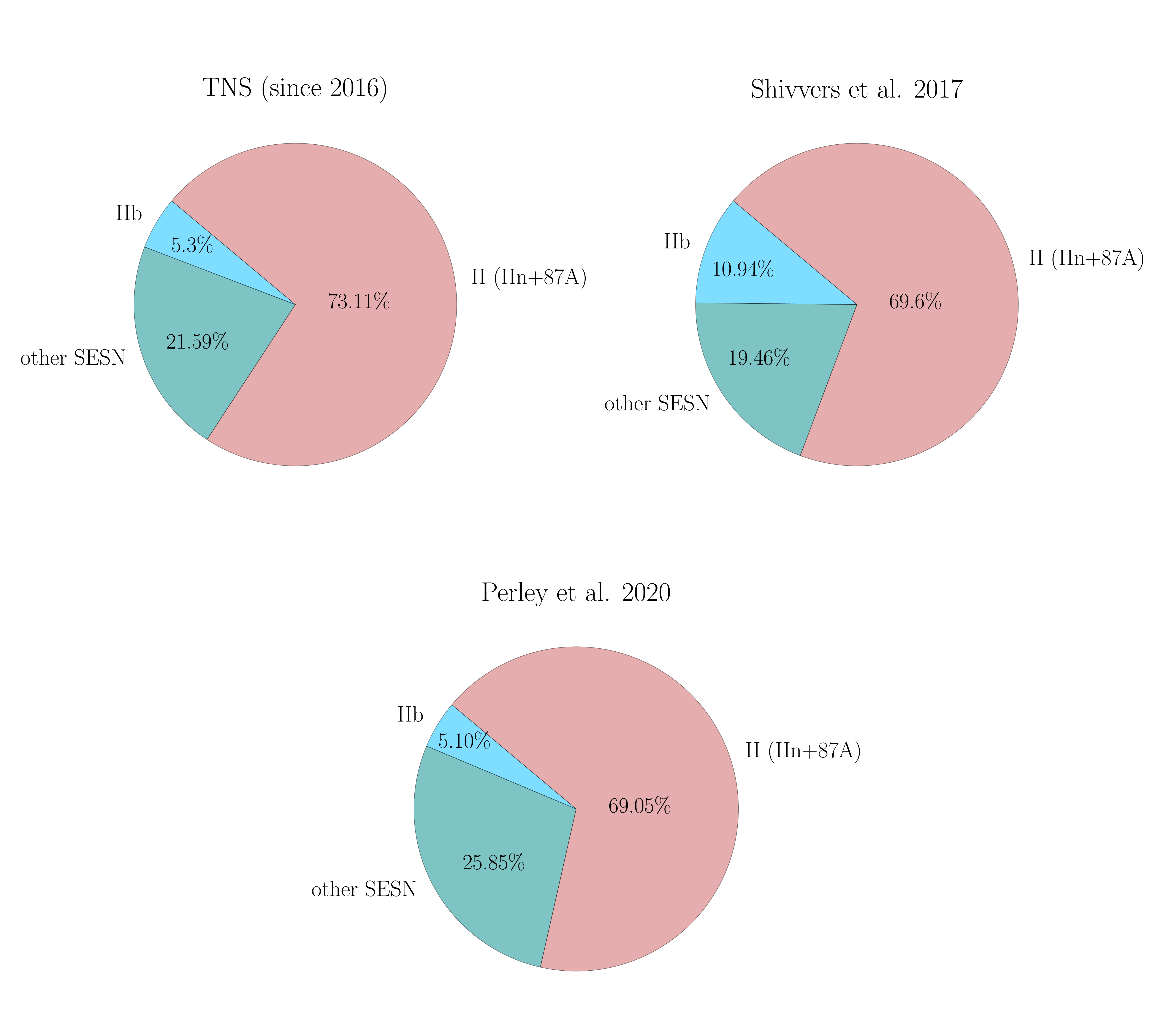}
\caption{CC-SNe fractions computed by \cite{shivvers2017revisiting} (right-top), \citet{Perley20} (bottom) and the fractions computed in this work by using the available data on TNS since 2016 (left-top). }
\label{fig:pienew}
\end{figure}

Based on the subtype ratio presented in \citet{shivvers2017revisiting}, we performed a similar analysis, including the same SN types, using publicly available TNS data collected from 2016 to 2024. Our results, presented in Figure~\ref{fig:pienew}, reveal discrepancies compared to \citet{shivvers2017revisiting}. The fraction of SNe~II differs by 3.5\%, while the SNe~IIb fraction shows a disagreement of 5.6\%. Compared to \citet{Perley20}, the SNe~IIb fraction is almost the same, but their SNe~II fraction is $4\%$ smaller (this percentage is added to the SESN). While we recognise that this comparison is neither direct nor entirely equitable,  since different selection effects than ours influence the referenced samples, we include it to provide a general perspective on how the relative fractions of SNe II and IIb vary when comparing our results with those of other studies.

Considering our findings, even just the reanalysis of SEDM spectra resulted in a measurable shift in the SNe~IIb fraction. This suggests that classification uncertainties, especially in distinguishing SNe~II and IIb at early times, could significantly impact the CCSNe fraction estimates. If SNe~IIb are not followed long enough, they may be misclassified as SNe~II. This effect could partly explain the observed difference in the SN~IIb fractions between our TNS analysis and previous works.

\subsection{Transitional events}

During our analysis, we identified several peculiar objects, which are discussed in this section. SNe showing ambiguous behaviour in their spectra or LCs are considered "transitional objects". These cases present conflicting classification indicators, as their photometry and spectroscopic characteristics suggest different SN types. The type II SNe~2019nyn, 2019uqp, 2019yyf, 2021agle, 2023rhj and 2024dgy were included in this discussion after their luminosity decay over 30 days post-peak was measured, revealing a magnitude difference comparable to that of the SNe~IIb. Similarly, SN~2018arx (classified as a SN~IIb) exhibited a LC shape similar to that of SNe~II. For each transitional event, we also computed the RFC probability of belonging to each classification category (II/IIb/87A-like). 

\subsubsection{SNe~II with Type IIb-like luminosity declines}

The SNe presented in Table \ref{tab:pekuliar} were classified as Type II based on their spectral features; however, their luminosity decline within 30 days after peak is more consistent with SNe~IIb. This discrepancy raises questions about their classification, suggesting possible transitional properties between the two types. 

The similar post-peak decline rates observed could be indicative of fast-declining SNe~II (SNe~IIL) or SNe~IIb, both of which exhibit comparable light-curve behaviour during this phase. However, their early-time evolution diverges, with SNe IIb typically showing a slower rise to maximum light \citep{pessi2019comparison}. Some studies have suggested a degree of overlap between SNe~IIL and IIb \citep[e.g.,][]{Faran14, Reguitti18ivc}, which can complicate photometric classification. Nevertheless, the spectroscopic features in our data strongly support a Type II classification, making a IIb interpretation unlikely. Indeed, the probabilities assigned by the RFC strongly support their official classification as SNe~II.

\begin{table}
\centering
\caption{RFC probabilities for the 'transitional objects' identified through luminosity decline measurements.}
\label{tab:pekuliar}
\begin{tabular}{l|ccc}
\multicolumn{1}{c}{SN}                   & \multicolumn{3}{c}{Probabilities} \\
                     & \textbf{II}       & \textbf{IIb}      & \textbf{87A-like}    \\
\hline
2019nyn  & 1.0      & 0.0      & 0.0         \\
2019uqp              & 0.76     & 0.21     & 0.03         \\
2019yyf              & 0.99     & 0.01      & 0.0        \\
2021agle             & 0.74     & 0.25     & 0.01        \\
2023rhj              & 1.0      & 0.0      & 0.0         \\
2024dgy (10d)        & 1.0      & 0.0      & 0.0         \\
2024dgy (17d)        & 0.99     & 0.01      & 0.0       
\end{tabular}

\end{table}

\subsubsection{SNe~II with mixed spectral and photometric properties}

Some SNe display a combination of spectral and photometric characteristics that place them in the overlapping region between SNe~II and IIb.

\begin{itemize}
\item \textbf{SN~2019oxi:} Officially classified as a SN~II, this object present notable deviations from the typical properties of SNe~II. Its spectral features closely resemble those of SNe~IIb, suggesting a possible reclassification. Although its photometric data benefits from a well-coverage follow-up, it does not provide a clear type identification, as it shows a slow rise over $\sim40$ days followed by a slow decline. By applying the RFC classification method combined with a Monte Carlo approach, we found that both techniques favour a Type IIb classification. RFC probabilities: II: 0.60; IIb: 0.40; 87-like: 0.0.

\item \textbf{SN~2018hoa:} Officially classified as SNe~II. The spectral line measurements are consistent with this classification, showing small values typical for SNe~II. However, the LC exhibits a clear double peak morphology, a feature more commonly associated with SNe~IIb. RFC probabilities: II: 0.50; IIb: 0.49; 87-like: 0.0.

\item \textbf{SN~2024ehs:} Officially classified as a SN~II, it shows a bell-like shape LC, with a rising time of about 20 days and a rapid decline after the peak. The spectral measurements of H and He lines are not conclusive either, as they fall into the overlapping region. RFC probabilities at 2 days post-explosion: II:0.32; IIb:0.67; 87A-like:0.01; RFC probabilities at 16 days post-explosion: II:0.60; IIb:0.39; 87A-like:0.01.
\end{itemize}

\subsubsection{SNe~IIb showing Type II-like photometric evolution}

These SNe were classified as Type IIb based on spectral features, yet their photometric behaviour resembles SNe~II.

\begin{itemize}
\item \textbf{SN~2018arx:} Although classified as a SN~IIb and exhibiting a broad, flat-topped \ha, the He I line is very weak, causing it to fall into the overlapping region between the two populations. Regarding the photometry, it shows a peculiar, relatively constant luminosity that resembles a plateau. The RFC probabilities for this SN are: II: 0.36; IIb: 0.64; 87A-like: 0.00.

\item \textbf{SN~2022emz:} Spectroscopically, it exhibits the characteristics of a slowly evolving SN~IIb, with a broad and deep He I line and a flat top \ha. However, upon analysing its LC, a constant luminosity is observed for nearly 100 days after the peak. This behaviour is usual of slow-declining SNe II (SNe~IIP). On the other hand, the rise to the peak aligns with the average time exhibited by SNe~IIb \citep{pessi2019comparison}. The RFC probabilities for this object are as follows: II:0.0; IIb:1.00; 87A-like:0.00.
\end{itemize}

\subsection{A new tool to classify SNe~II and IIb based on early spectra}

We also defined two alternative classification methods that could indeed be used to complement the density contours and QDA regions. First, we performed again a RFC (see section \ref{sec:RFC}), taking this time our entire sample as a training sample and a newly measured SN as a test sample. This way, we achieve the probabilities for that SN to belong to each type. The probabilities for the SEDM spectra and new classifications are presented in Table \ref{table:new_classifications} and \ref{table:reclassification_SEDM}. Second, we developed the possibility of plotting the newly measured SNe on our t-SNE plots. So far, we can see if it falls in the cluster formed by a majority of type II or IIb SNe. This would allow us to have different perspectives on interpreting the values obtained for the new SN.

\section{Conclusions}
\label{sec:conc}

In this paper, we presented a detailed analysis of the early spectra of SNe~II and IIb to uncover their diversity. The 866 spectra of 393 SNe, composing our main sample, were analysed using the pEW and FWHM of the \ha\ and He~I $\lambda5876$ features. Statistical methods and machine-learning techniques were applied to identify differences between SNe~II and IIb. Our main results are:  

\begin{itemize}
\item We find that pEW and FWHM of \ha\ and He~I $\lambda5876$ in SNe~IIb display larger values than those of SNe~II during the entire evolution, with most significant differences occurring in early phases, particularly in the 10 -- 20-day interval. 

\item Population density evolution reveals a clear separation in the pEW of SNe~II and IIb. SNe~II consistently exhibit high-density concentrations around 20 \AA, while SNe~IIb reach 60–80 Å values. Despite this distinction, the two populations seem to form a continuum in spectral properties. 

\item The QDA analysis shows that SNe~II and IIb exhibit distinct pEW evolution patterns, with the greatest separation occurring within the first 10 days. In contrast, differences based on FWHM measurements are less pronounced.

\item t-SNE+LDA effectively separated the two types, supporting their distinction based on observed features. However, continuity between the two clusters persists, suggesting a gradual transition rather than a strict boundary between the populations.

\item The RFC confirms the distinct characteristics of the two groups and demonstrates the effectiveness of these parameters (pEW and FWHM of \ha\ and He I) as classification criteria. Additionally, it enables the classification of newly observed SNe.

\item We demonstrate the method validation using low-resolution SEDM spectra and newly classified SNe. We identify 34 possible misclassified SNe within the re-analysed sample. Based on these revised classifications, we recalculated the SEDM fraction for each SNe type, leading to a notable increase in the percentage of SNe~IIb reported in WISERep from 4.0\% to 7.26\% (see Figure \ref{fig:SEDM_comparison_pie}), nearly doubling its original value.

\end{itemize}
    
We have developed a new classification method for early SNe~II and IIb based on quantified differences in spectral features during their early evolution.  Using the obtained results, we computed classification probabilities with the RF classifier and plotted the measurements within defined regions, enabling a clear classification assessment. This approach not only improves our understanding of the early phases of SN evolution but also serves as a valuable tool for enhancing classification accuracy in future observations.
Our findings indicate that misclassification significantly impacts the actual CCSNe rate. The reclassification of SEDM spectra has improved the accuracy of previously available classification data, highlighting the robustness of our method. \\
All classification tools developed in this study, along with the computed median spectra for each time interval, are publicly available at \url{https://doi.org/10.5281/zenodo.15756198} (\url{https://zenodo.org/communities/sndata/}) to facilitate broader use.

\begin{acknowledgements}

We thank the anonymous referee for the comments and suggestions that have helped us to improve the paper. We thank the anonymous referee for the comments and

We thank Maria Kopsacheili, Koraljka Muzic, Joseph Anderson, Luc Dessart and Basti\'an Ayala for valuable discussions.
M.G.B.’s work has been carried out within the framework of the doctoral program in Physics of the Universitat Autònoma de Barcelona.
The M.G.B., C.P.G., L.G. research group acknowledges financial support from the Spanish Ministerio de Ciencia e Innovaci\'on (MCIN) and the Agencia Estatal de Investigaci\'on (AEI) 10.13039/501100011033 under the PID2020-115253GA-I00 HOSTFLOWS and PID2023-151307NB-I00 SNNEXT project, from Centro Superior de Investigaciones Cient\'ificas (CSIC) under the projects PIE 20215AT016, ILINK23001, COOPB2304, and the program Unidad de Excelencia Mar\'ia de Maeztu CEX2020-001058-M, and from the Departament de Recerca i Universitats de la Generalitat de Catalunya through the 2021-SGR-01270 grant.
C.P.G. acknowledges financial support from the Secretary of Universities and Research (Government of Catalonia) and by the Horizon 2020 Research and Innovation Programme of the European Union under the Marie Sk\l{}odowska-Curie and the Beatriu de Pin\'os 2021 BP 00168 programme.

\end{acknowledgements}

%
\bibliographystyle{aa} 
\bibliography{Biblio} 
%



\appendix

\section{Figures}
\label{app:fig}

\begin{figure}
\includegraphics[width=\columnwidth]{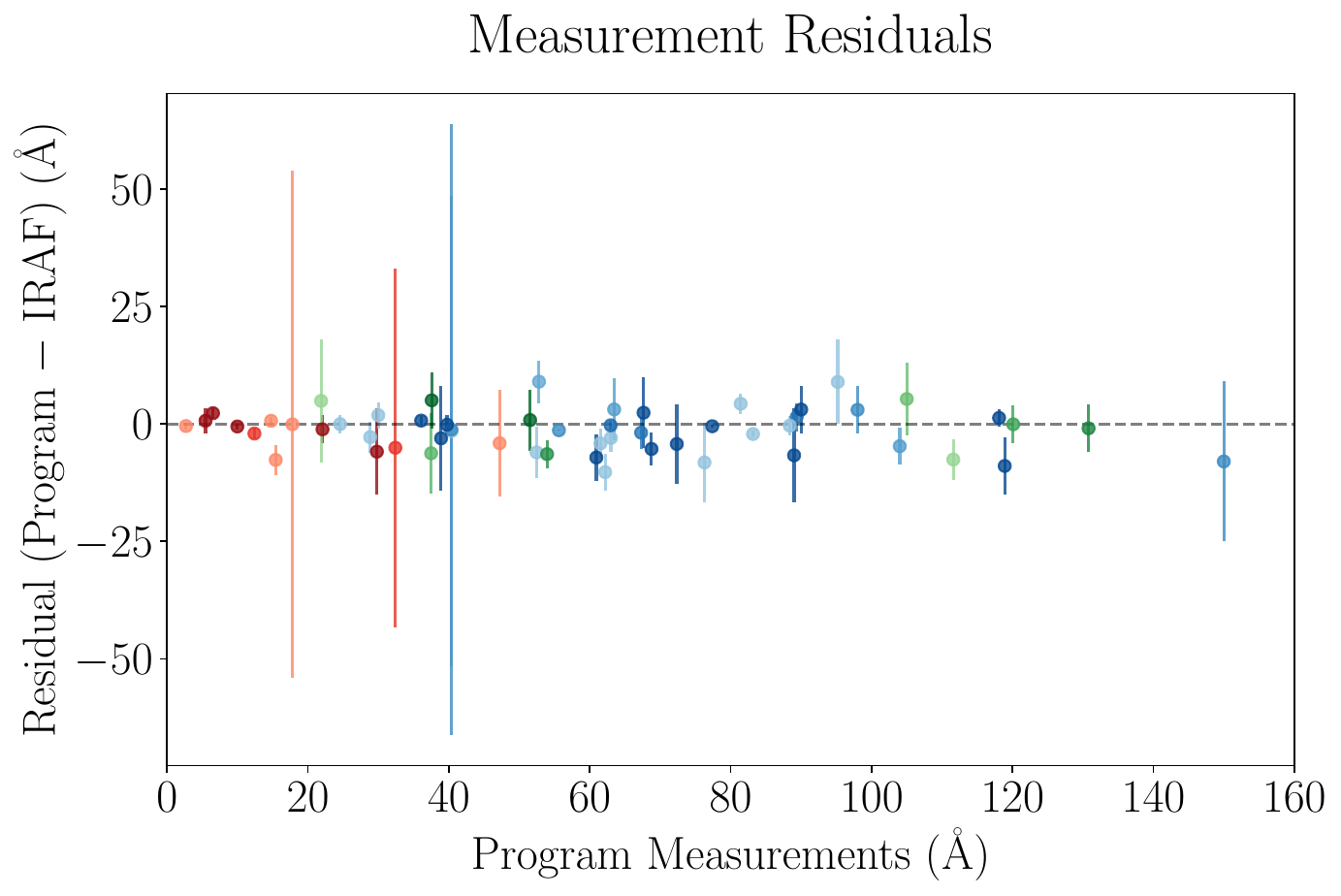}
\caption{Residuals between the pEW and FWHM values measured with our interactive program and those obtained with IRAF are shown as (program – IRAF) plotted against the programme-derived values. Data points are colour-coded by SN type: SNe~II are shown in warm colours, while SNe IIb are in cool tones. Different colours represent different SNe.}
\label{fig:NIREA_vs_IRAF}
\end{figure}

\begin{figure}
\includegraphics[width=\columnwidth]{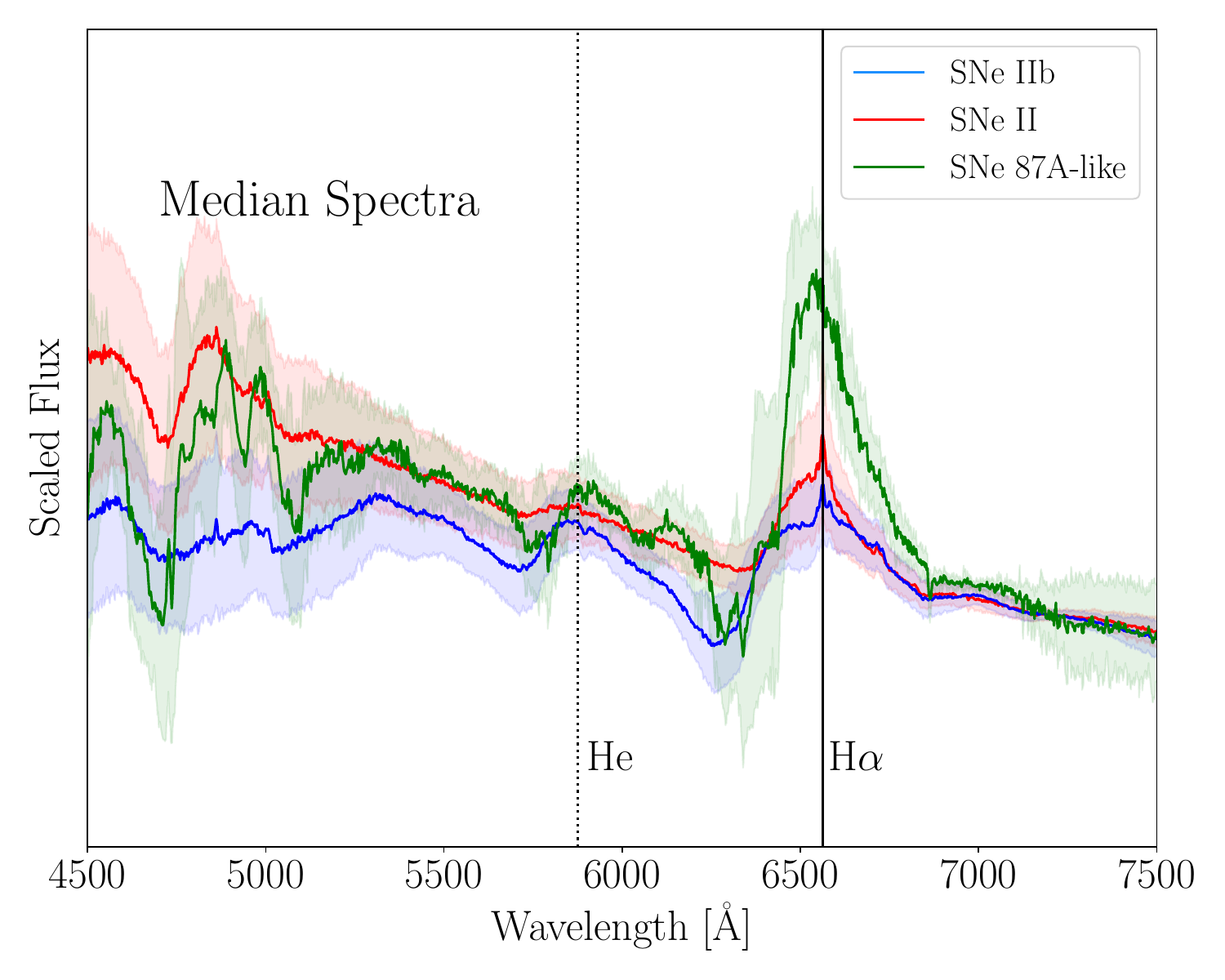}
\caption{Comparison between the median spectrum for SNe~II (red), SNe~IIb (blue) and SN~87A-like (green) constructed from all spectra in our sample (from explosion to 40 days post-explosion). The shaded regions represent the Median Absolute Deviation (MAD), providing a measure of dispersion around the median. }
\label{fig:medians_II_IIb_87like}
\end{figure}

\begin{figure*}
\centering
\includegraphics[width=0.5\columnwidth]{FIGURES/pEW_0-10d_abs_BATERA_AZKENA_LASTSAMPLE.pdf}
\centering
\includegraphics[width=0.5\columnwidth]{FIGURES/pEW_10-20d_abs_BATERA_AZKENA_LASTSAMPLE.pdf}
\centering
\includegraphics[width=0.5\columnwidth]{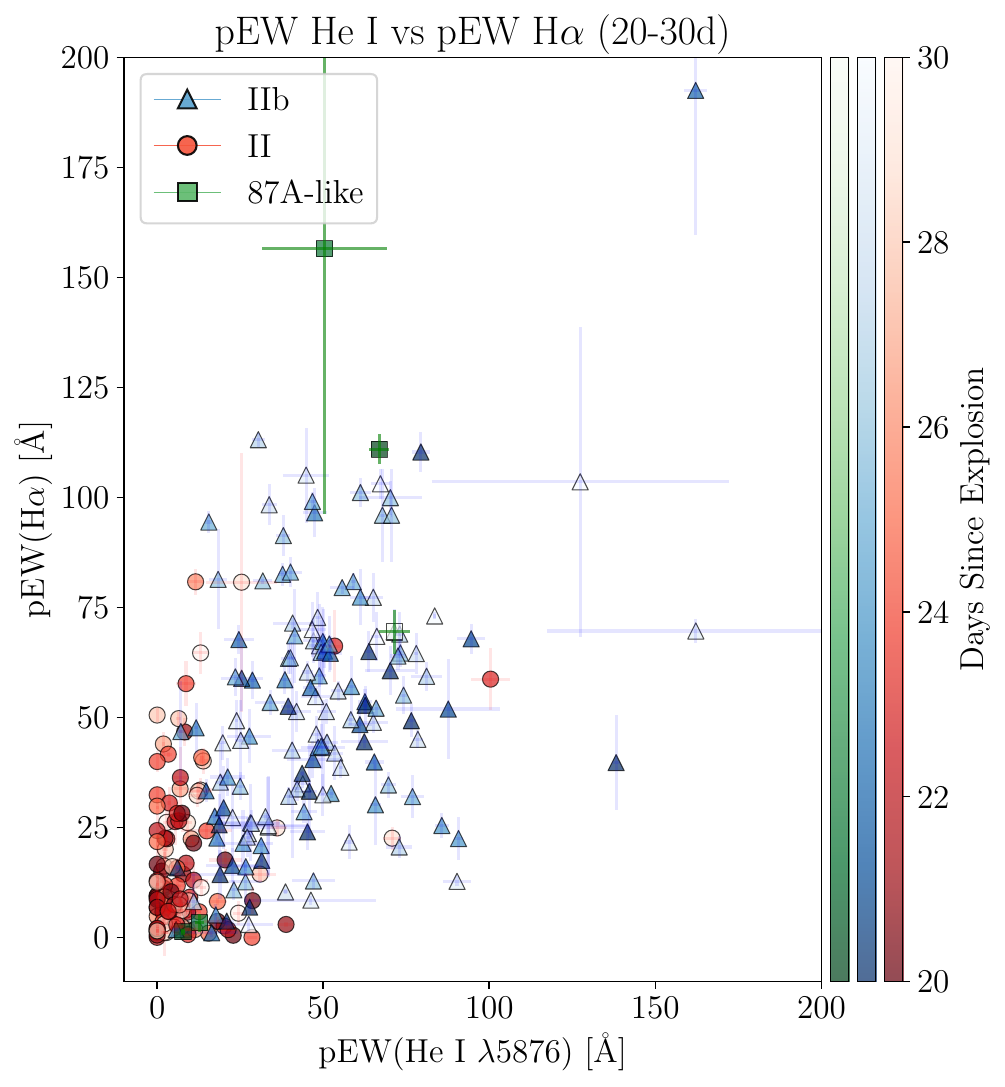}
\centering
\includegraphics[width=0.5\columnwidth]{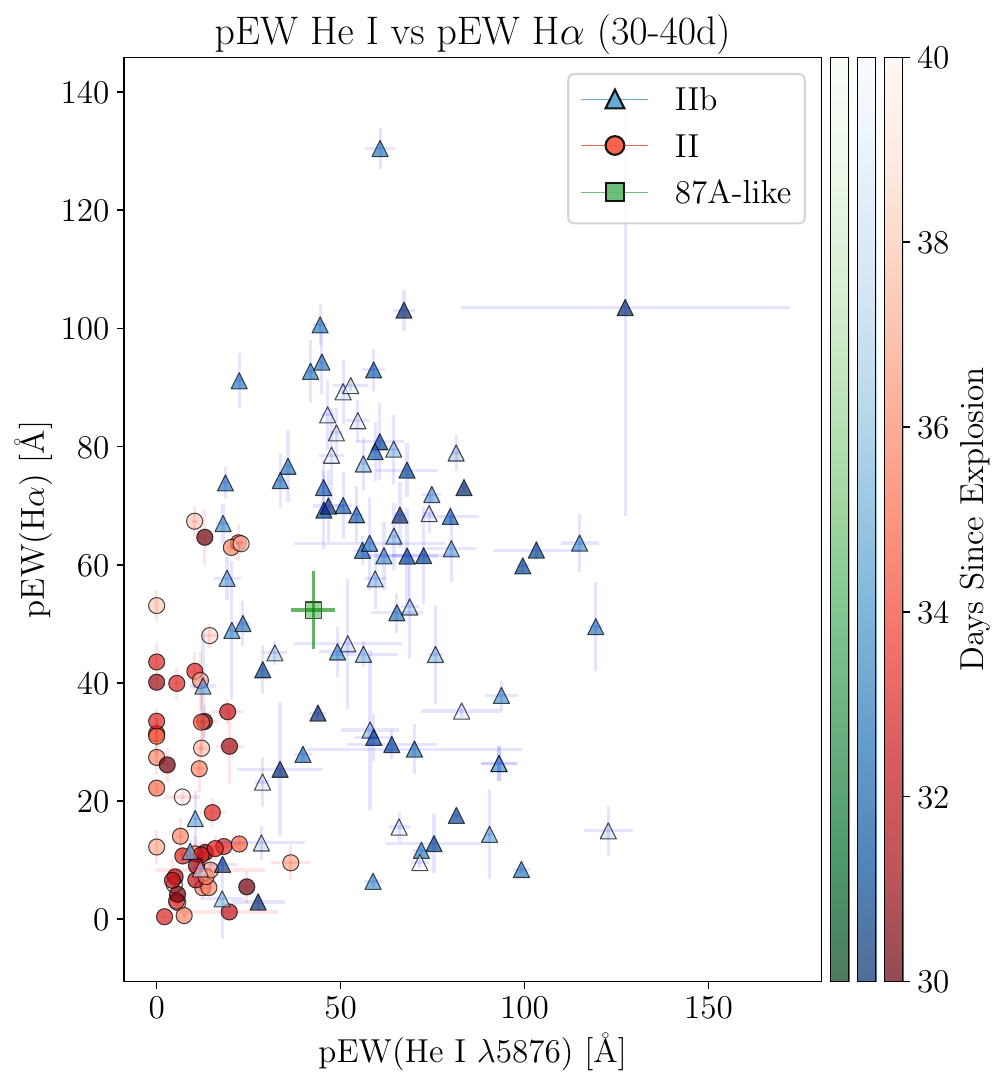}
\caption{pEW values of the \ha\ vs the same measurements for He I $\lambda5876$ for separated time ranges since explosion: 0-10 days, 10-20 days, 20-30 days and 30-40 days. The markers are established by SN type, corresponding to the type SNe IIb, the blue triangles, SNe II, the red circles and SNe 87A-like, the green squares. The colour bars show how the colour intensity changes depending on the number of days since the explosion.}
\label{fig:pEW_timeranges}
\end{figure*}

\begin{figure*}
\includegraphics[width=0.5\columnwidth]{FIGURES/FWHM_0-10d_BATERA_AZKENA_LASTSAMPLE.pdf}
\centering
\includegraphics[width=0.5\columnwidth]{FIGURES/FWHM_10-20d_BATERA_AZKENA_LASTSAMPLE.pdf}
\centering
\includegraphics[width=0.5\columnwidth]{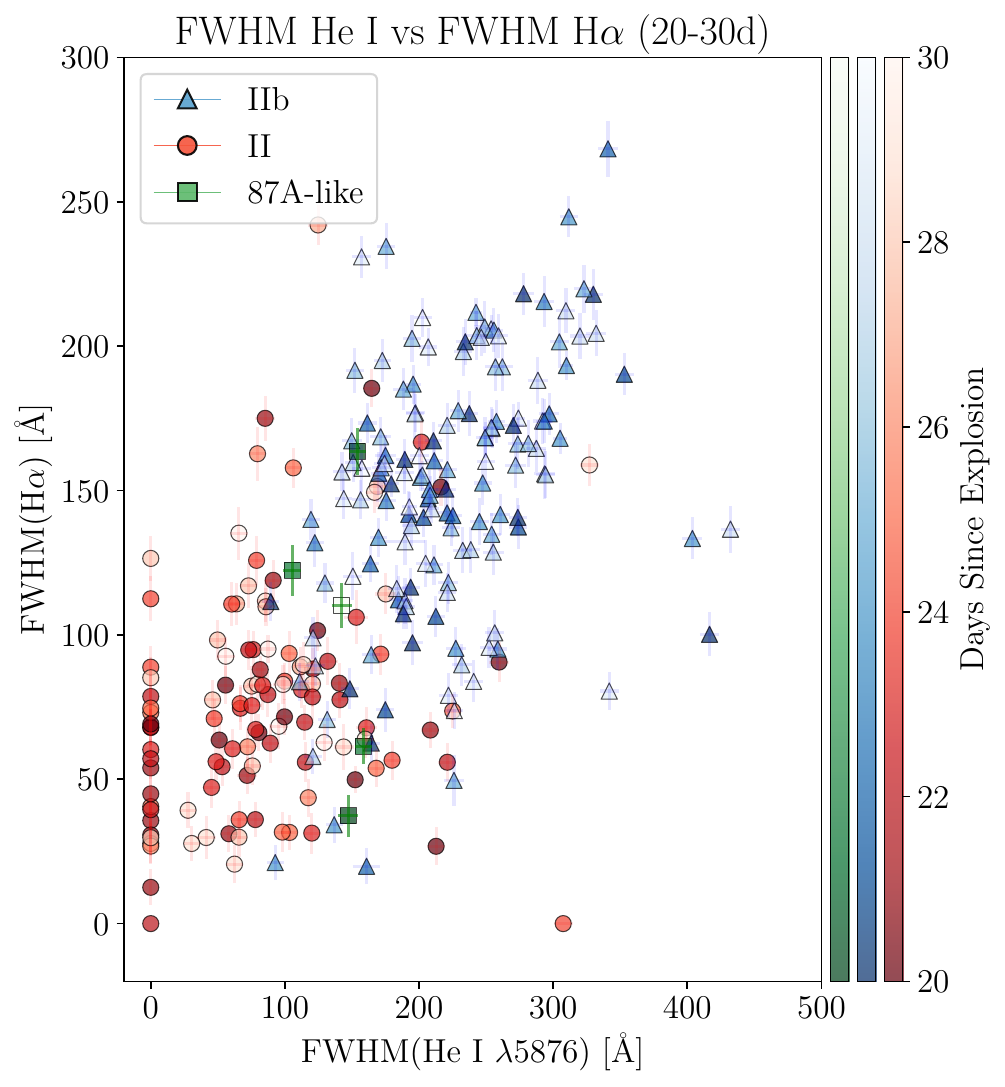}
\centering
\includegraphics[width=0.5\columnwidth]{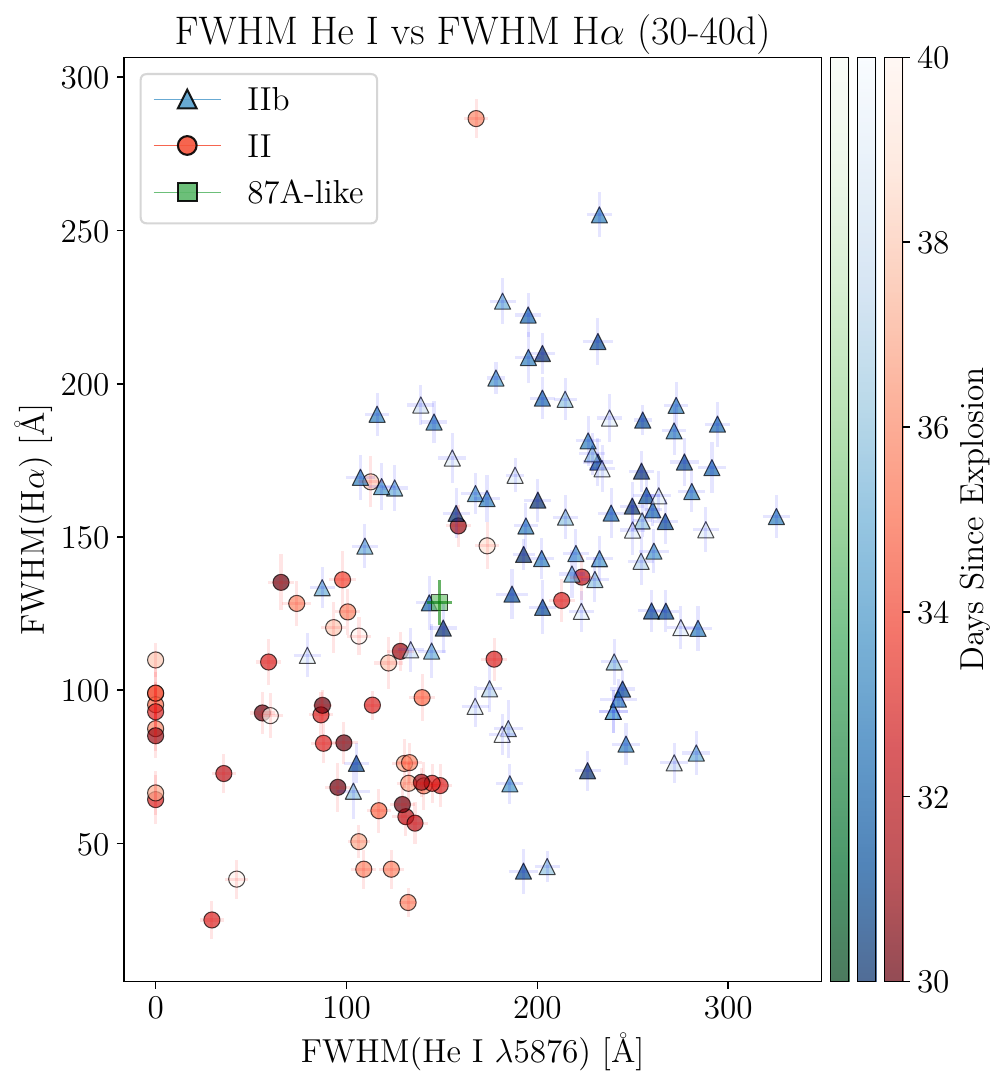}
\caption{FWHM of the \ha\ ($\lambda6563$) line vs the same measurements for the He I ($\lambda5876$) line for separated time ranges since explosion: 0-10 days, 10-20 days, 20-30 days and 30-40 days. The markers are established by SN type, corresponding to the type SNe IIb, the blue triangles, SNe II, the red circles and SNe 87A-like, the green squares. The colour bars show how the colour intensity changes depending on the number of days since the explosion. }
\label{fig:FWHM_timeranges}
\end{figure*}

\begin{figure*}
\centering
\includegraphics[width=0.32\textwidth]{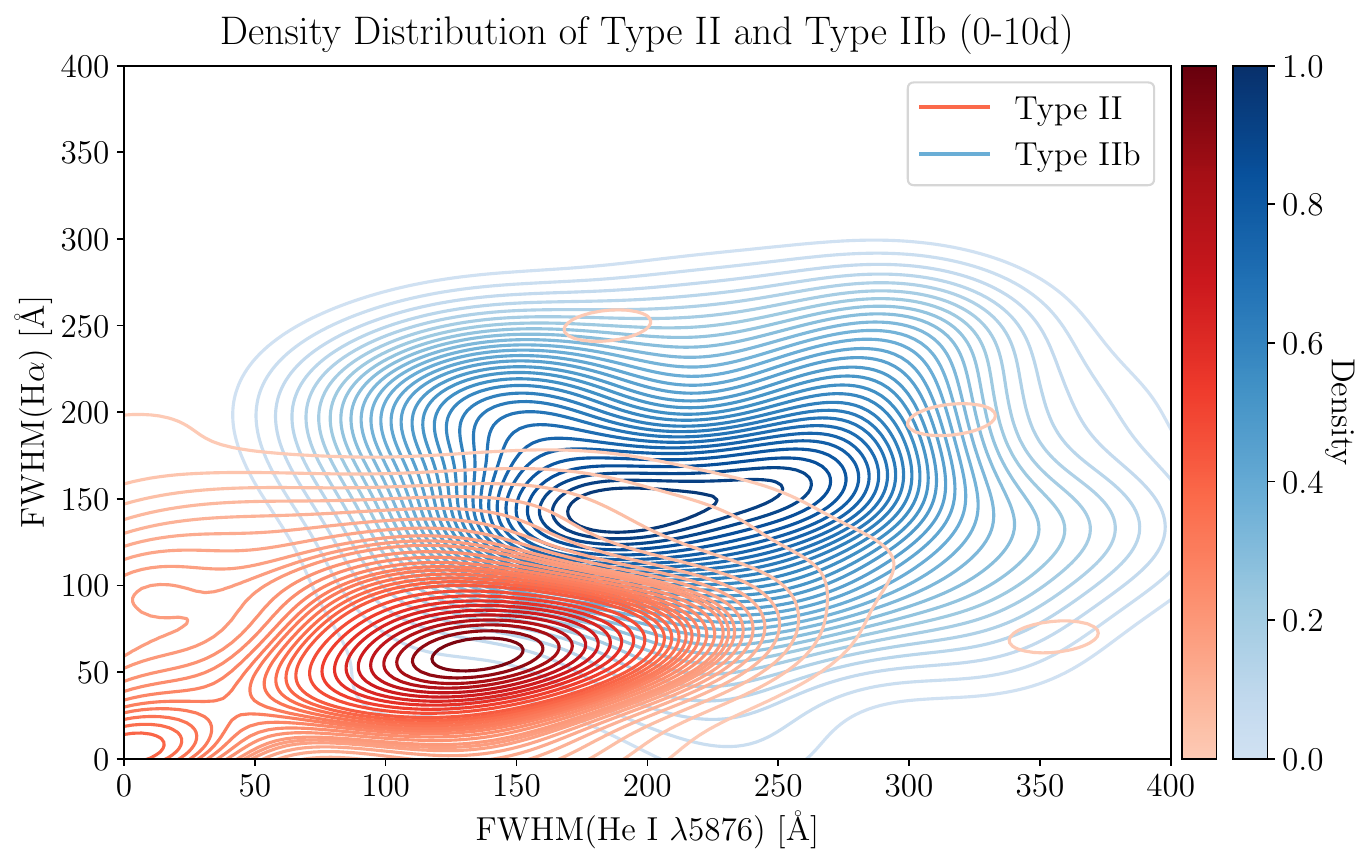}
\includegraphics[width=0.32\textwidth]{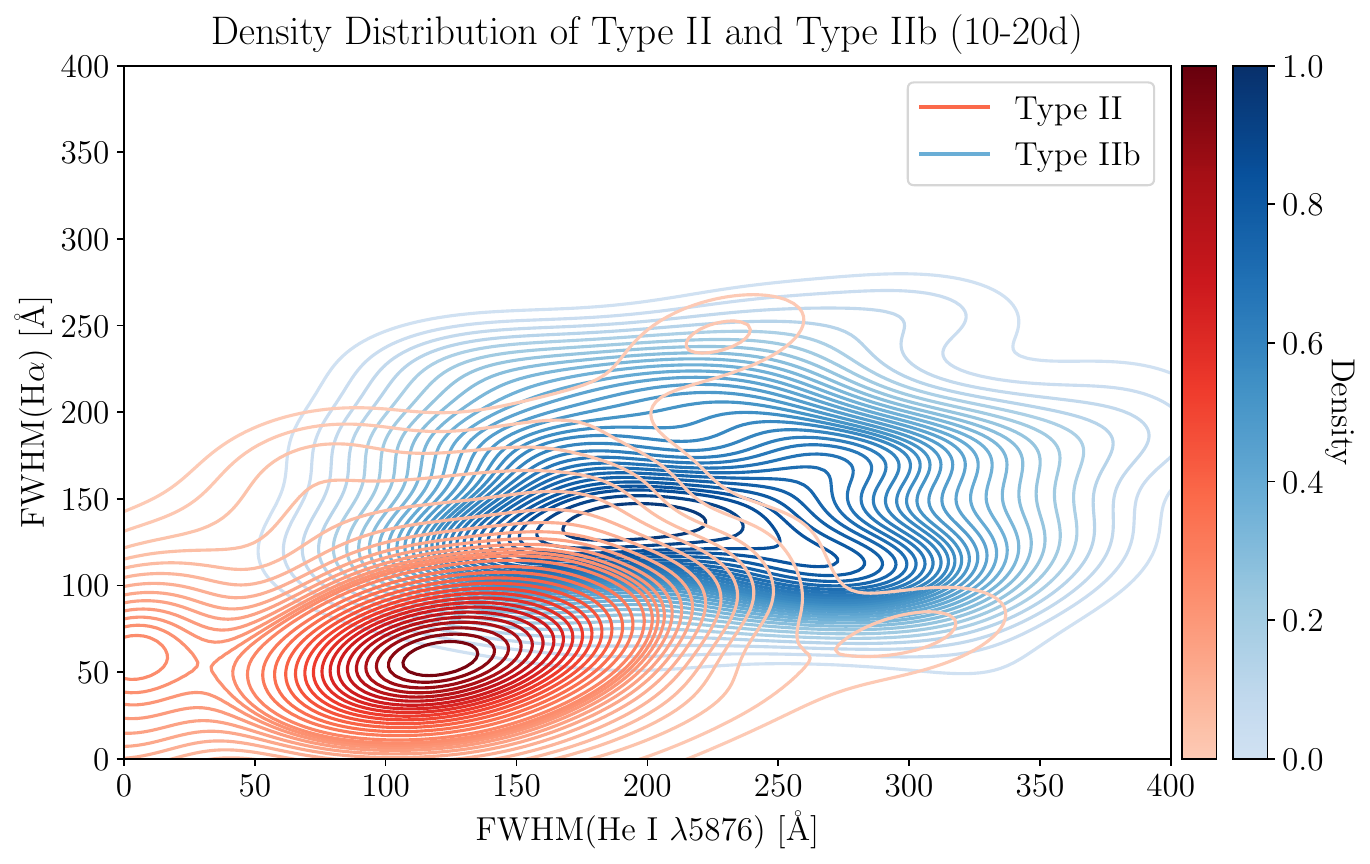}
\includegraphics[width=0.32\textwidth]{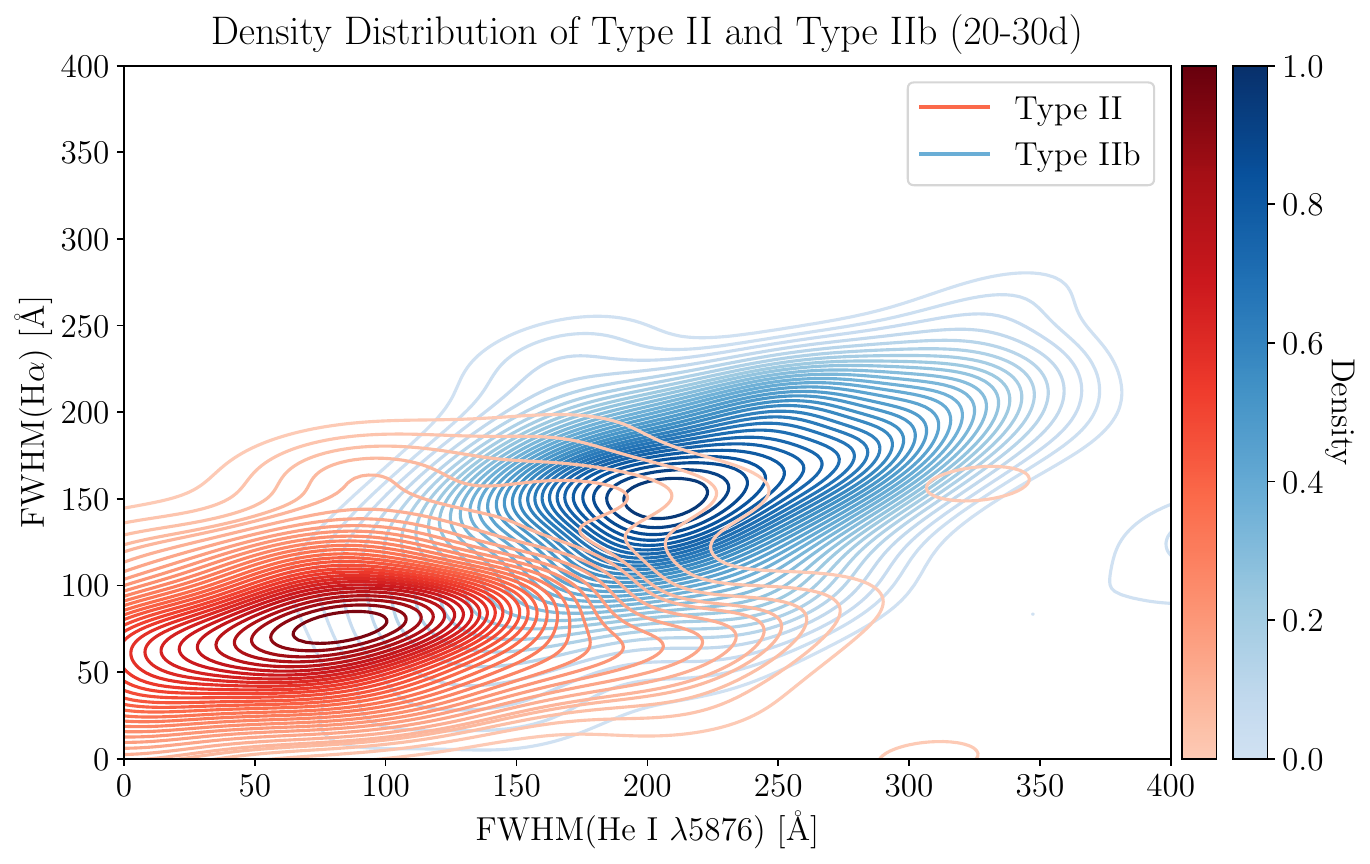}
\includegraphics[width=0.32\textwidth]{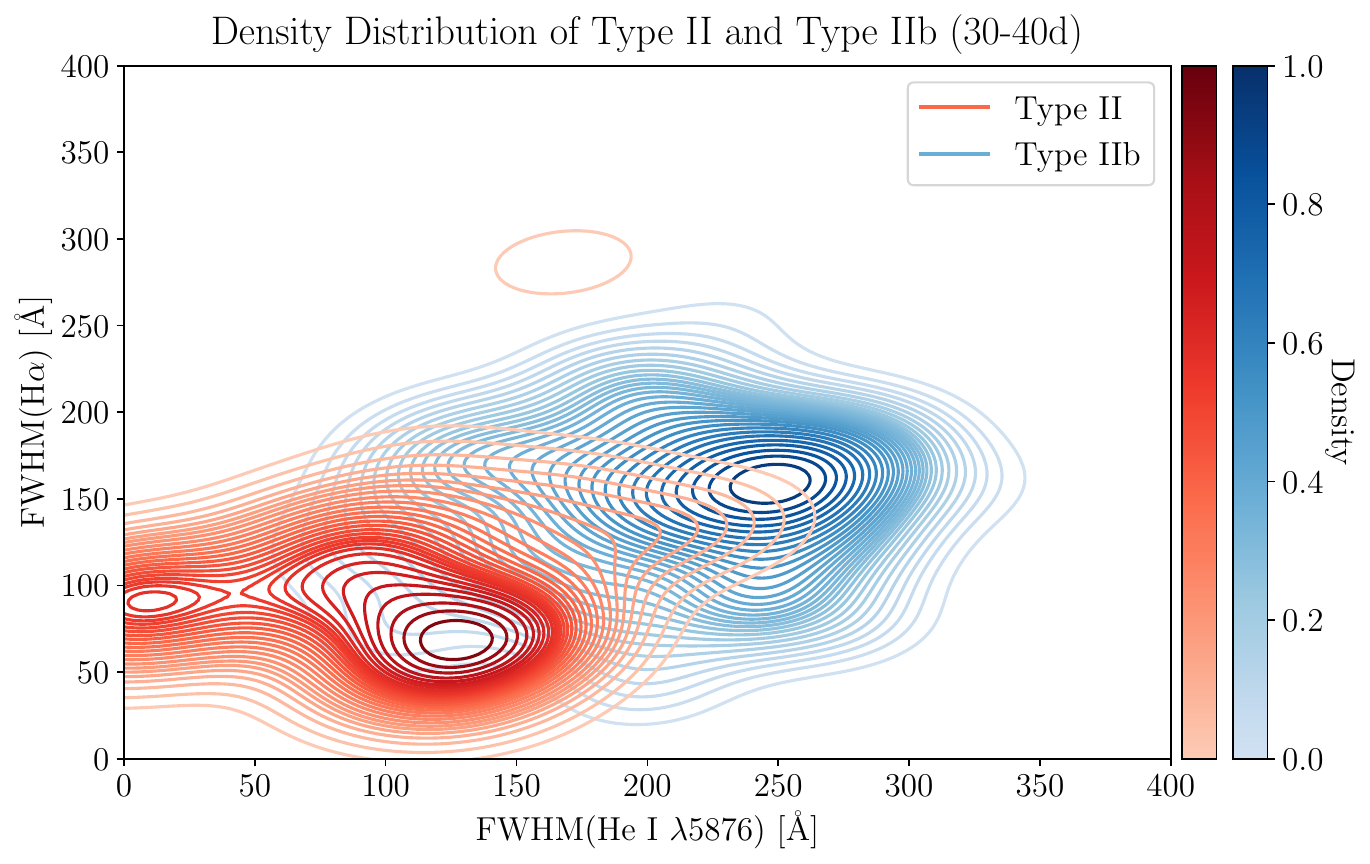}
\includegraphics[width=0.32\textwidth]{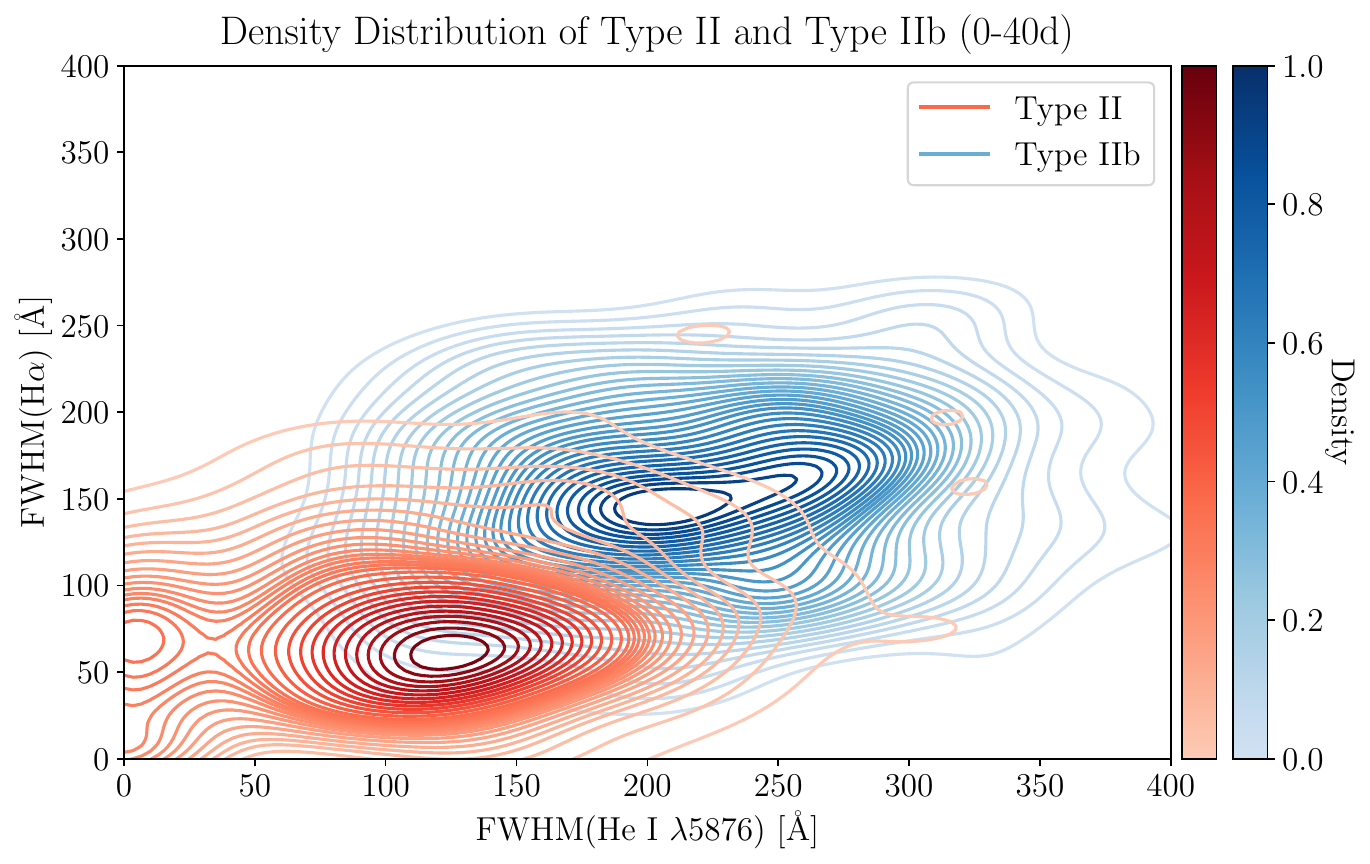}
\caption{The density distribution of type II (in red) and IIb (in blue) population of our sample based on the FWHM measurements. We present the distribution for the different time ranges that we have analysed. }
\label{fig:FWHM_densitydistribution}
\end{figure*}

\begin{figure*}
\centering
\includegraphics[width=0.48\textwidth]{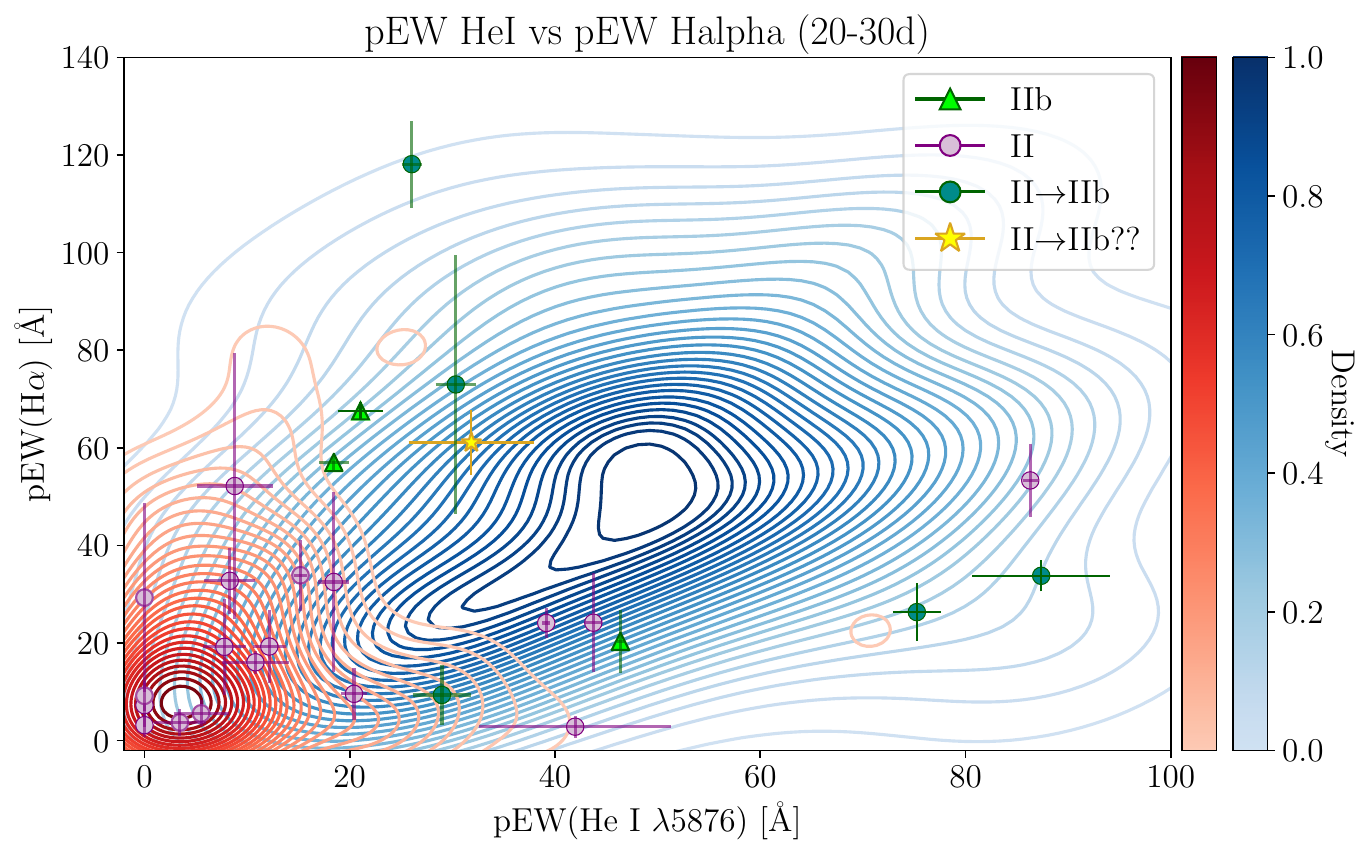}
\includegraphics[width=0.48\textwidth]{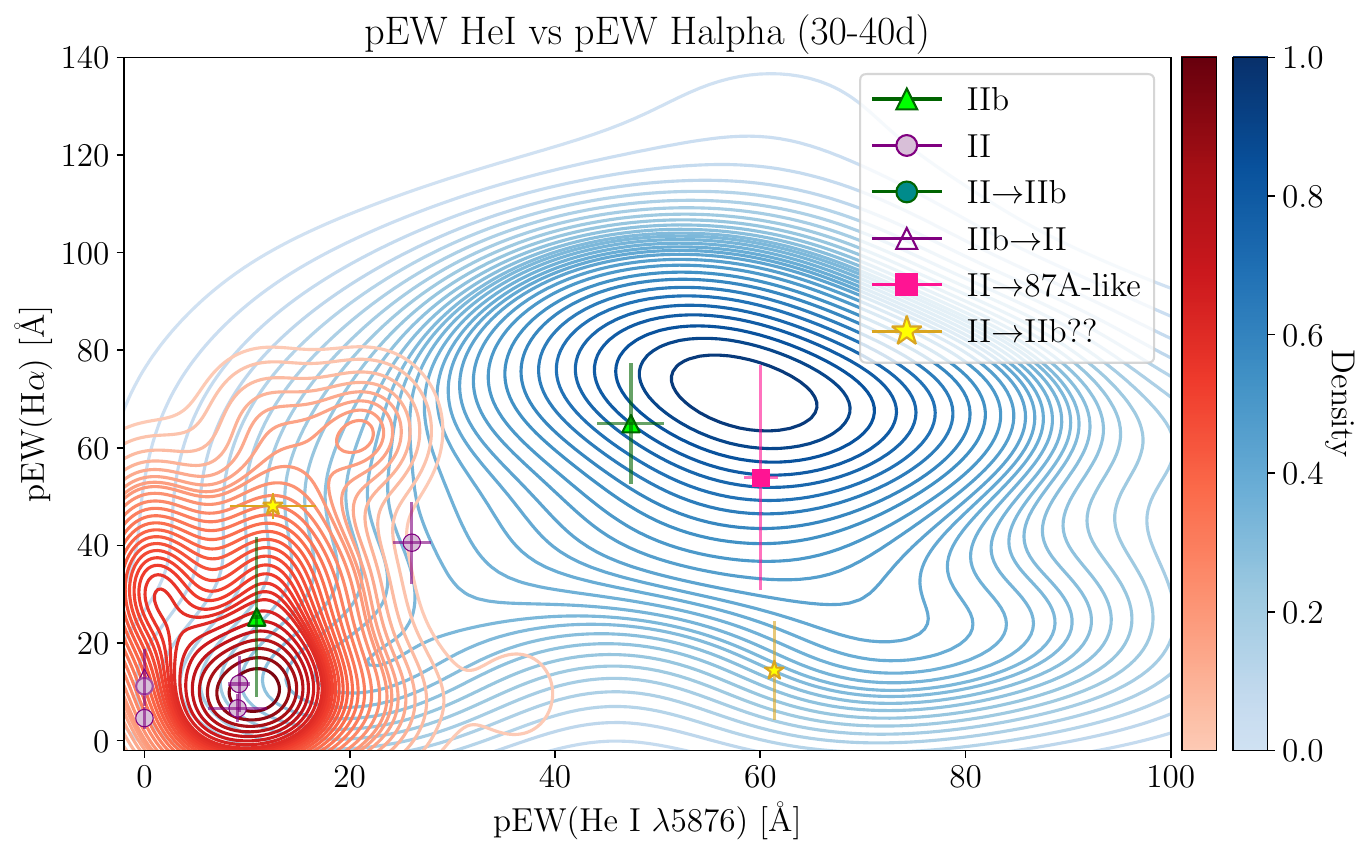}
\caption{The density contours of the pEW along with the SEDM and the new classified SNe used as test sample for 20-30 and 30-40 days after explosion. The comparison sample is represented by SNe~II as filled purple dots and SNe~IIb as filled green triangles. Dark green circles correspond to SNe initially classified as Type II but exhibiting SNe~IIb-like properties. Open purple triangles represent SNe initially classified as Type IIb but spectroscopically belonging to the SN~II class. Pink squares denote Type II SNe showing 87A-like characteristics. Yellow stars highlight Type II SNe that appear to exhibit SNe~IIb characteristics but lack sufficient data to confirm this classification. }
    \label{fig:CONTOURNS+SEDM}

\end{figure*}

\begin{figure*}
\centering
\includegraphics[width=0.32\textwidth]{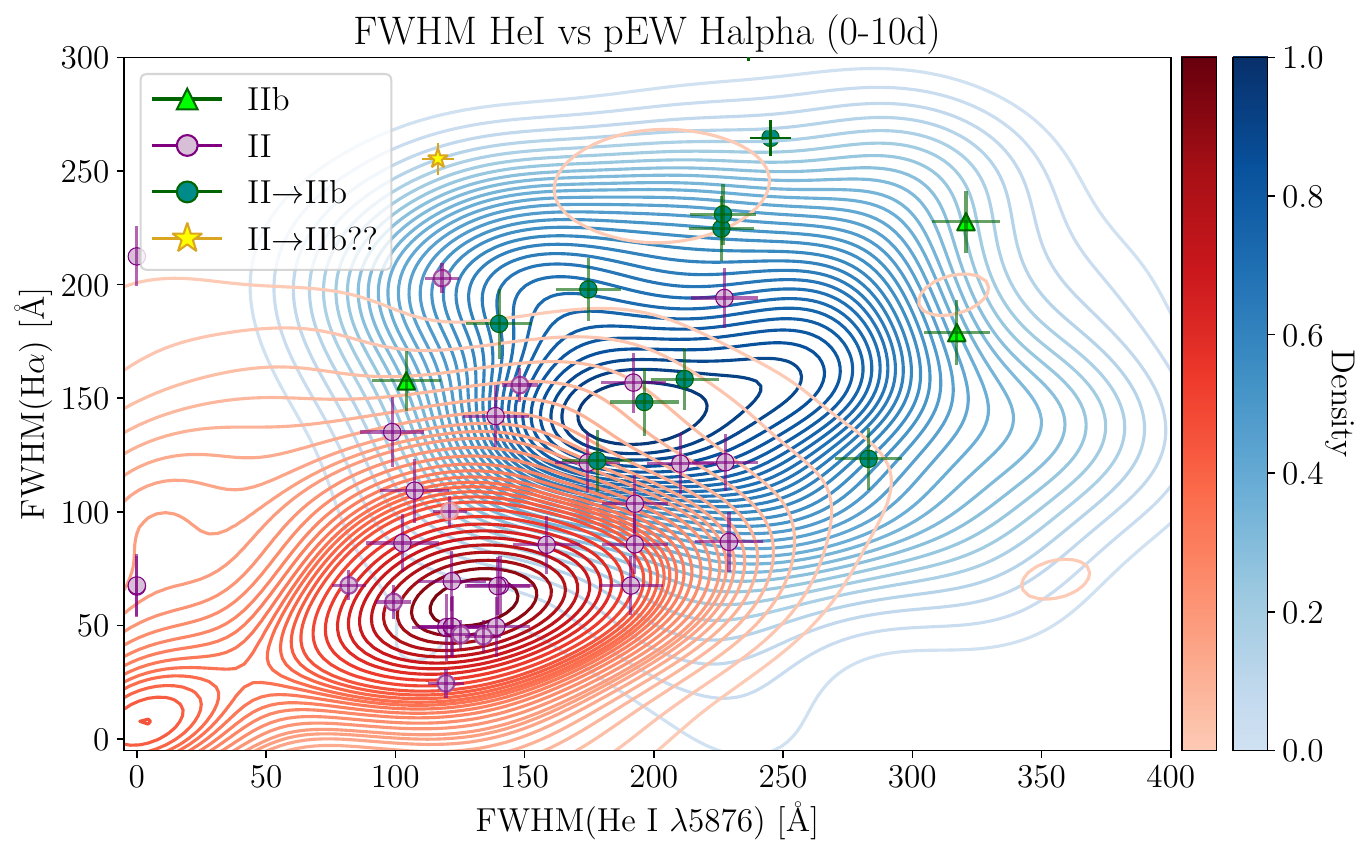}
\includegraphics[width=0.32\textwidth]{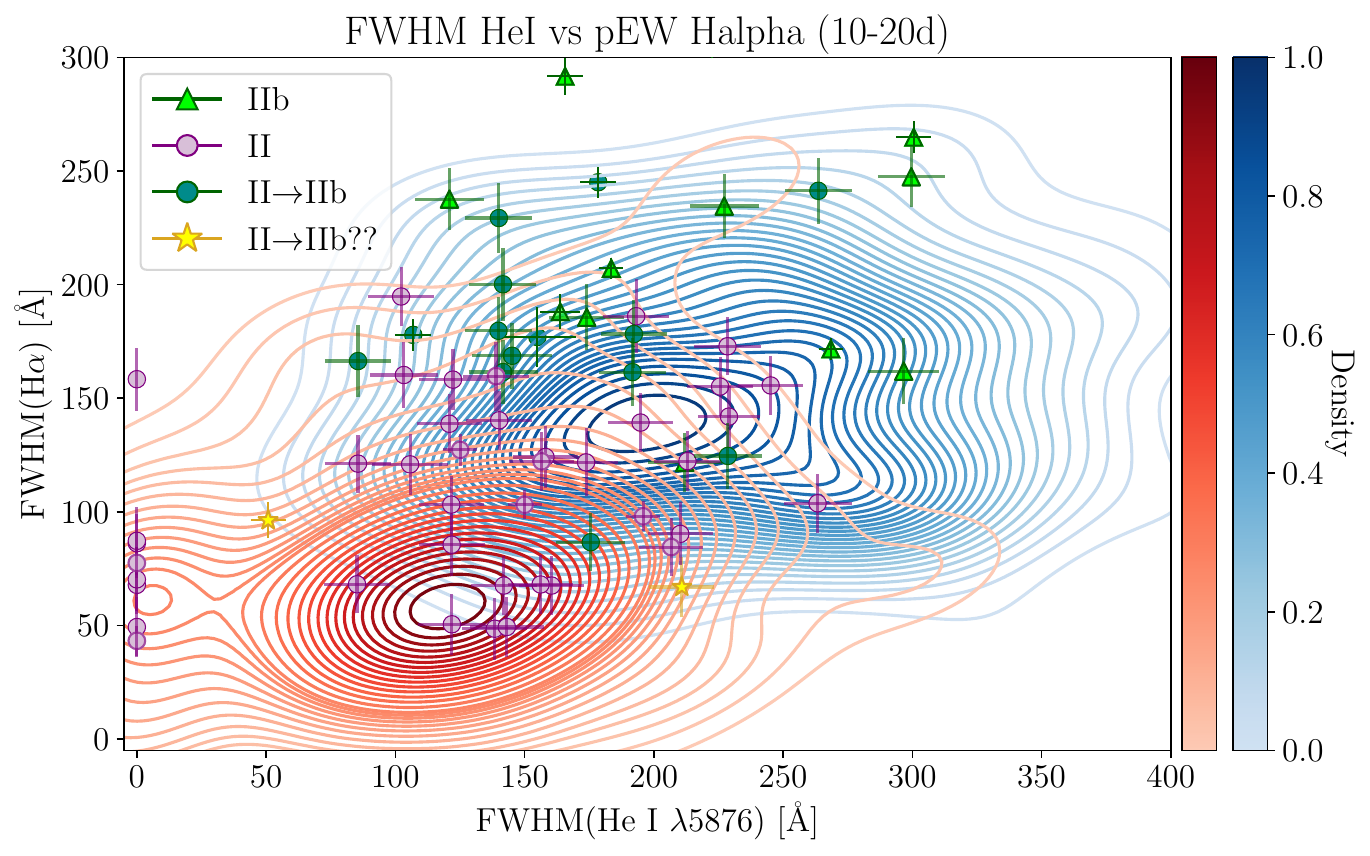}
\includegraphics[width=0.32\textwidth]{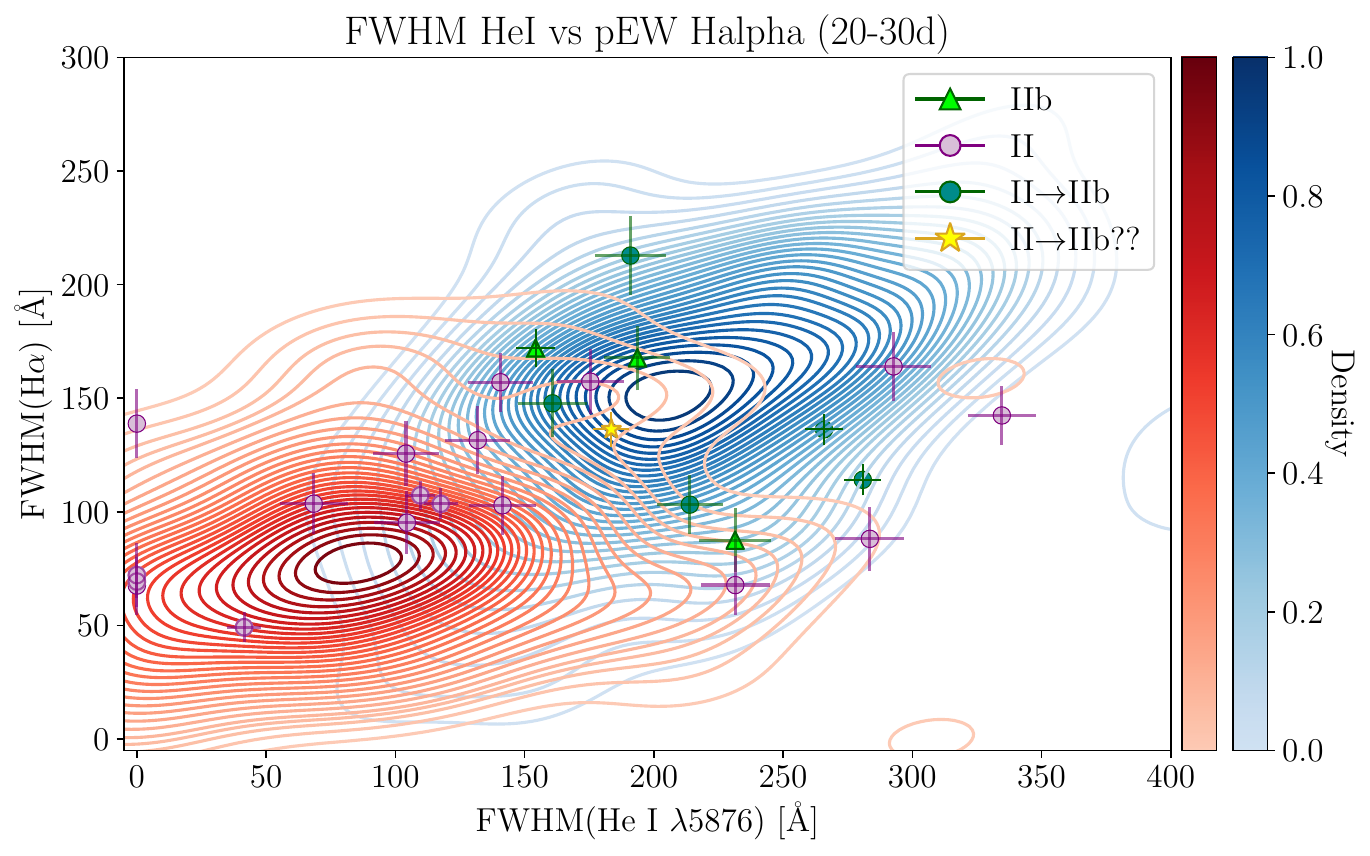}
\includegraphics[width=0.32\textwidth]{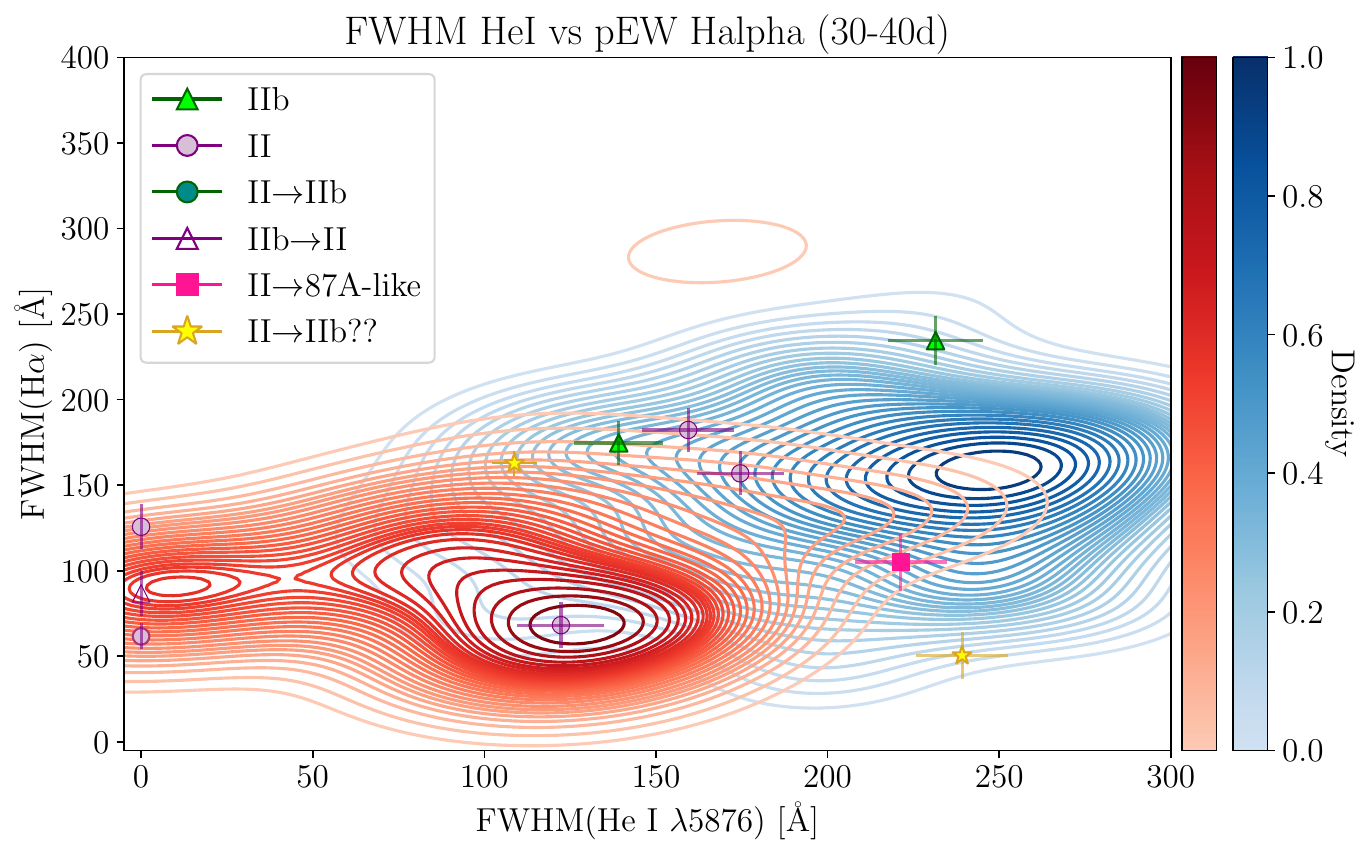}
\includegraphics[width=0.32\textwidth]{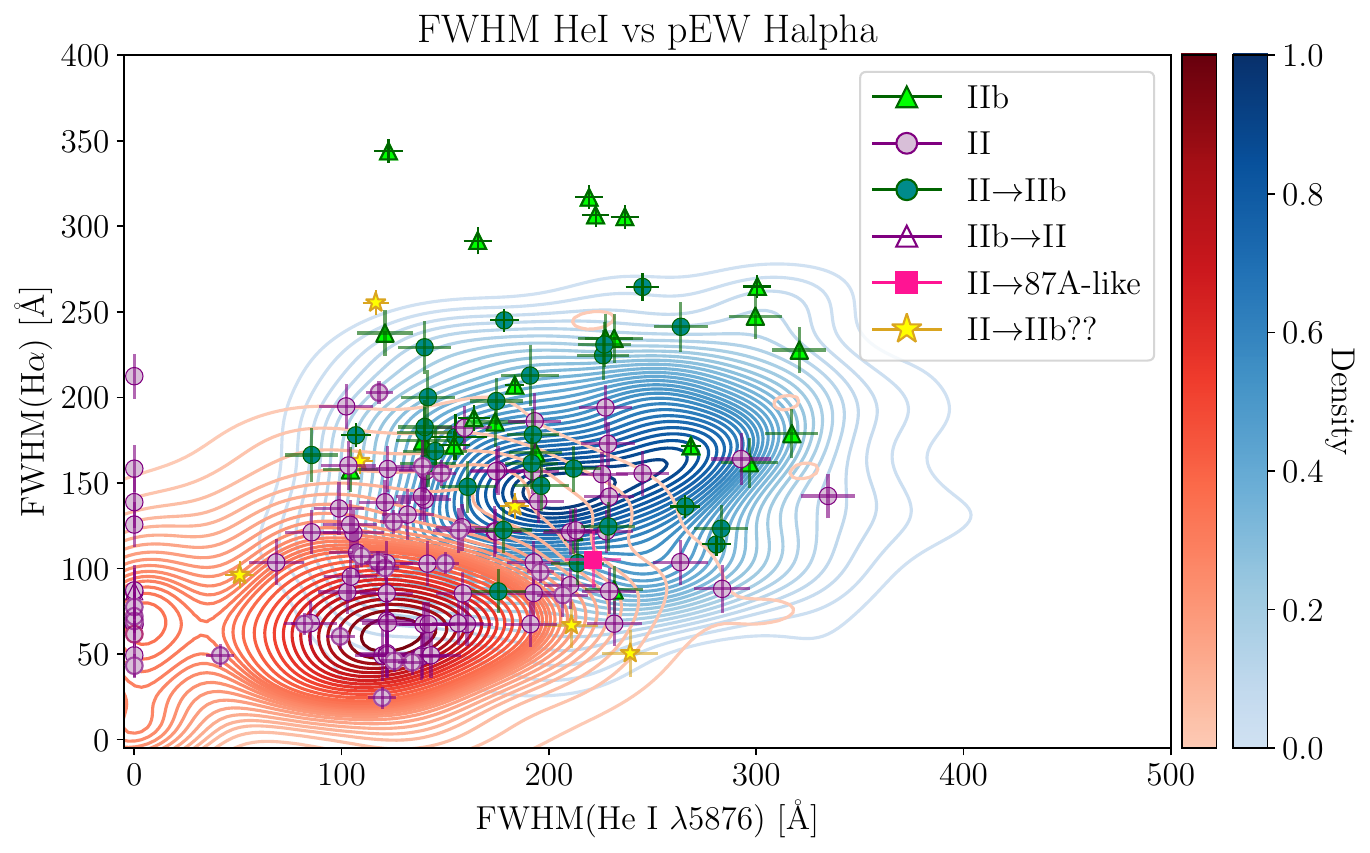}
\caption{The density contours of the FWHM along with the SEDM and the new classified SNe used as test sample, are shown. The comparison sample is represented by SNe~II as filled purple dots and SNe~IIb as filled green triangles. Dark green circles correspond to SNe initially classified as Type II but exhibiting SNe~IIb-like properties. Open purple triangles represent SNe initially classified as Type IIb but spectroscopically belonging to the SN~II class. Pink squares denote Type II SNe showing 87A-like characteristics. Yellow stars highlight Type II SNe that appear to exhibit SNe~IIb characteristics but lack sufficient data to confirm this classification.}
\label{fig:FWHM_CONTOURNS+SEDM}
\end{figure*}

\begin{figure*}
\centering
\includegraphics[width=\columnwidth]{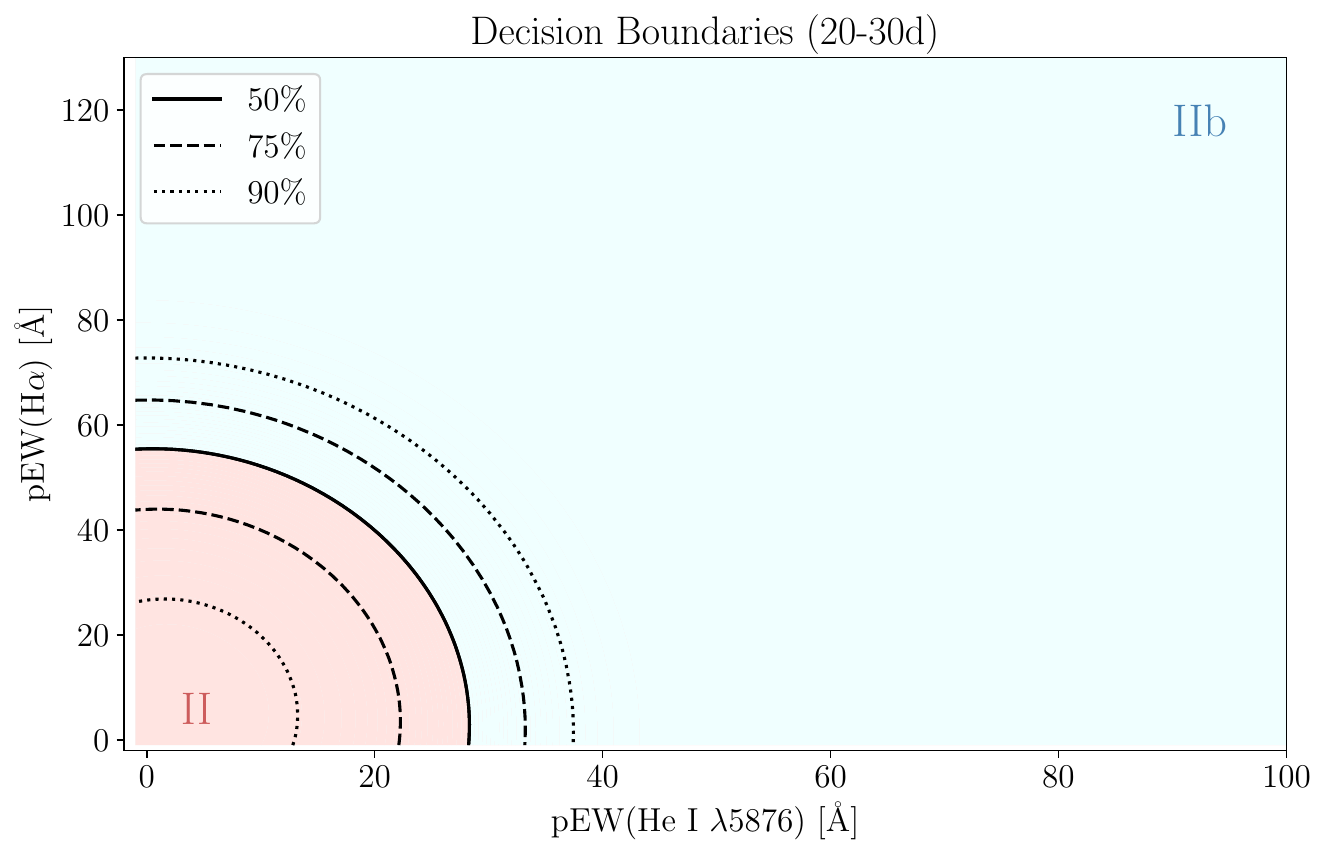}
\includegraphics[width=\columnwidth]{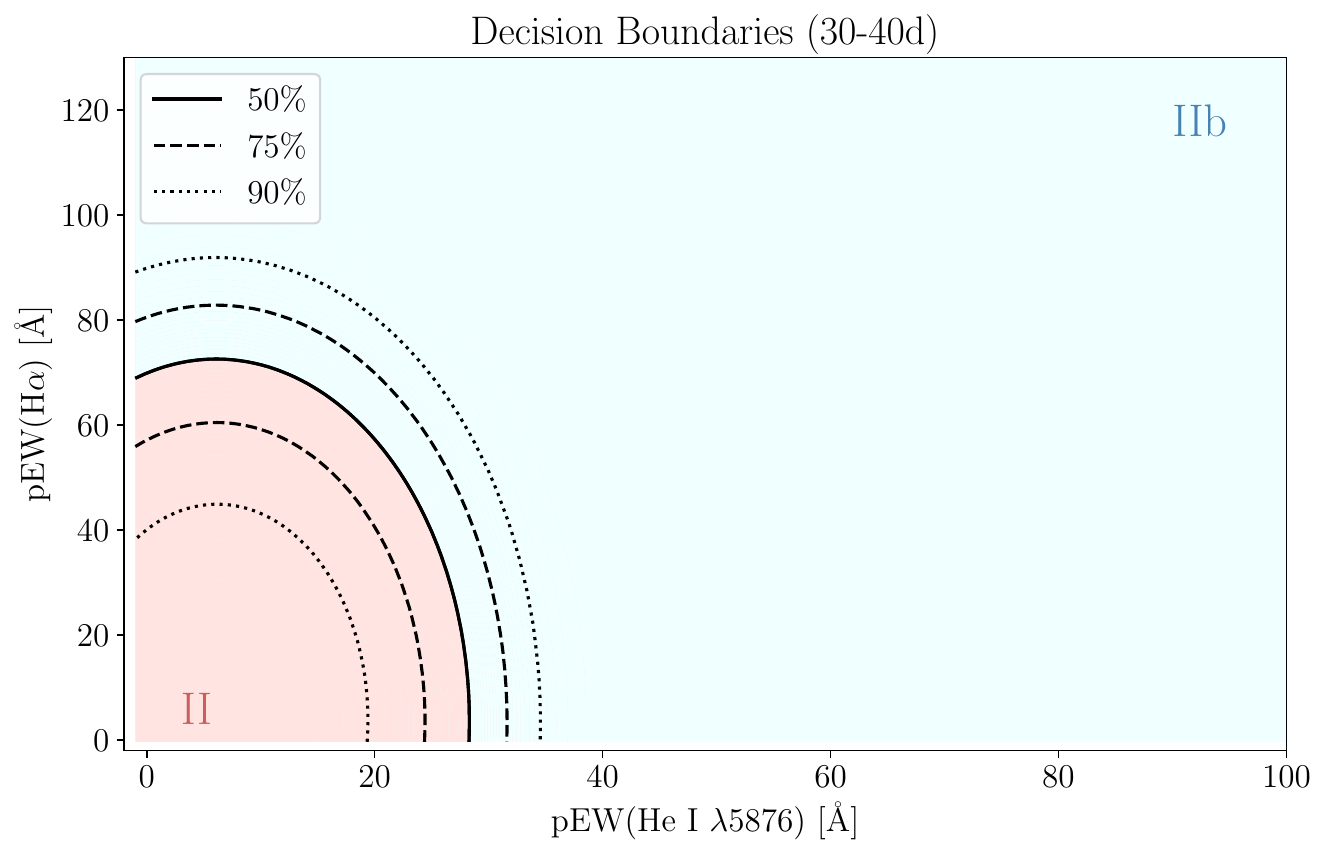}
\caption{The decision boundaries defined by the QDA based on the measures performed for the II and IIb pEW for the following time ranges 20-30 days and 30-40 days after the explosion. The red region corresponds to the type II SNe, and the blue region to the IIb SNe.}
\label{fig:QDA_pEWtimeranges}

\end{figure*}
\begin{figure*}
\centering
\includegraphics[width=0.32\textwidth]{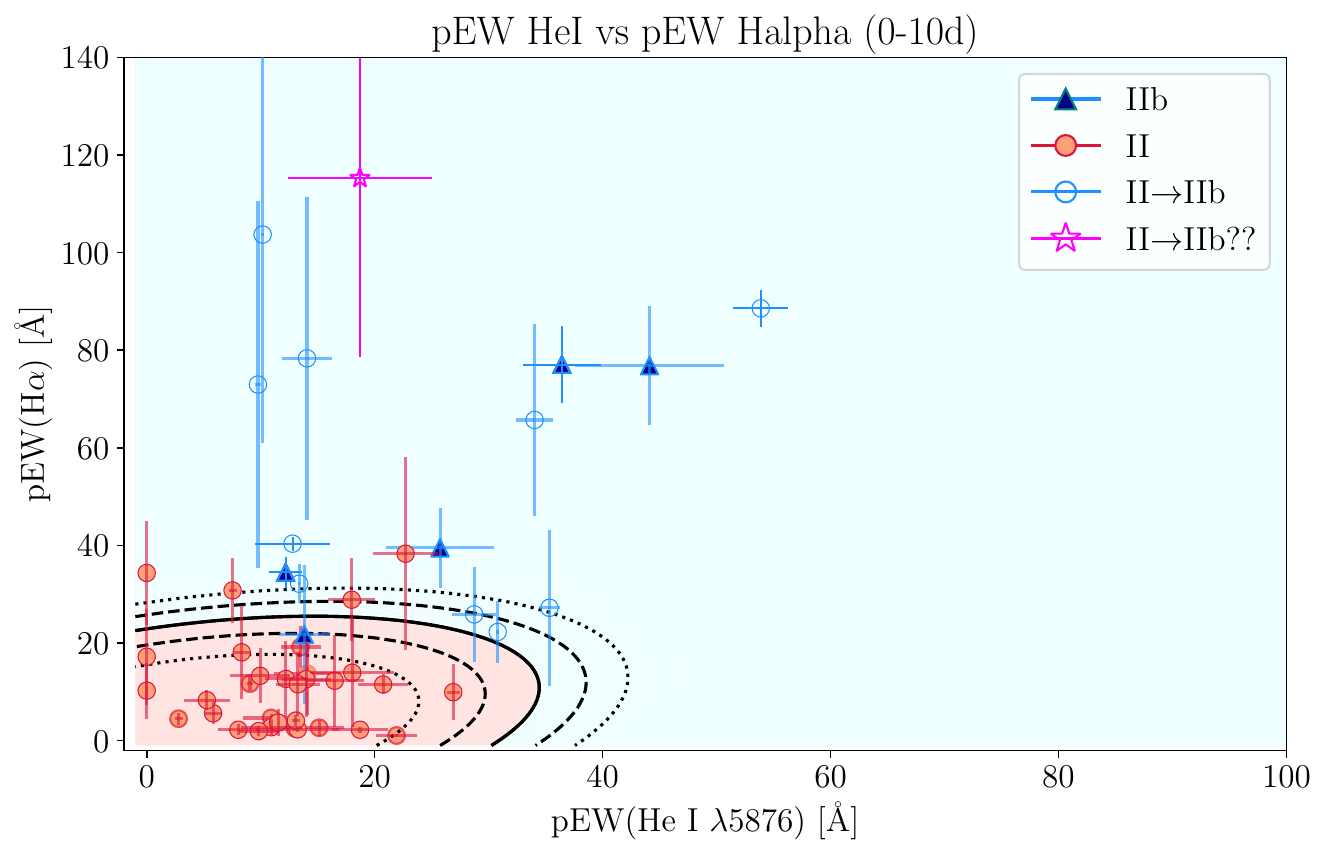}
\includegraphics[width=0.32\textwidth]{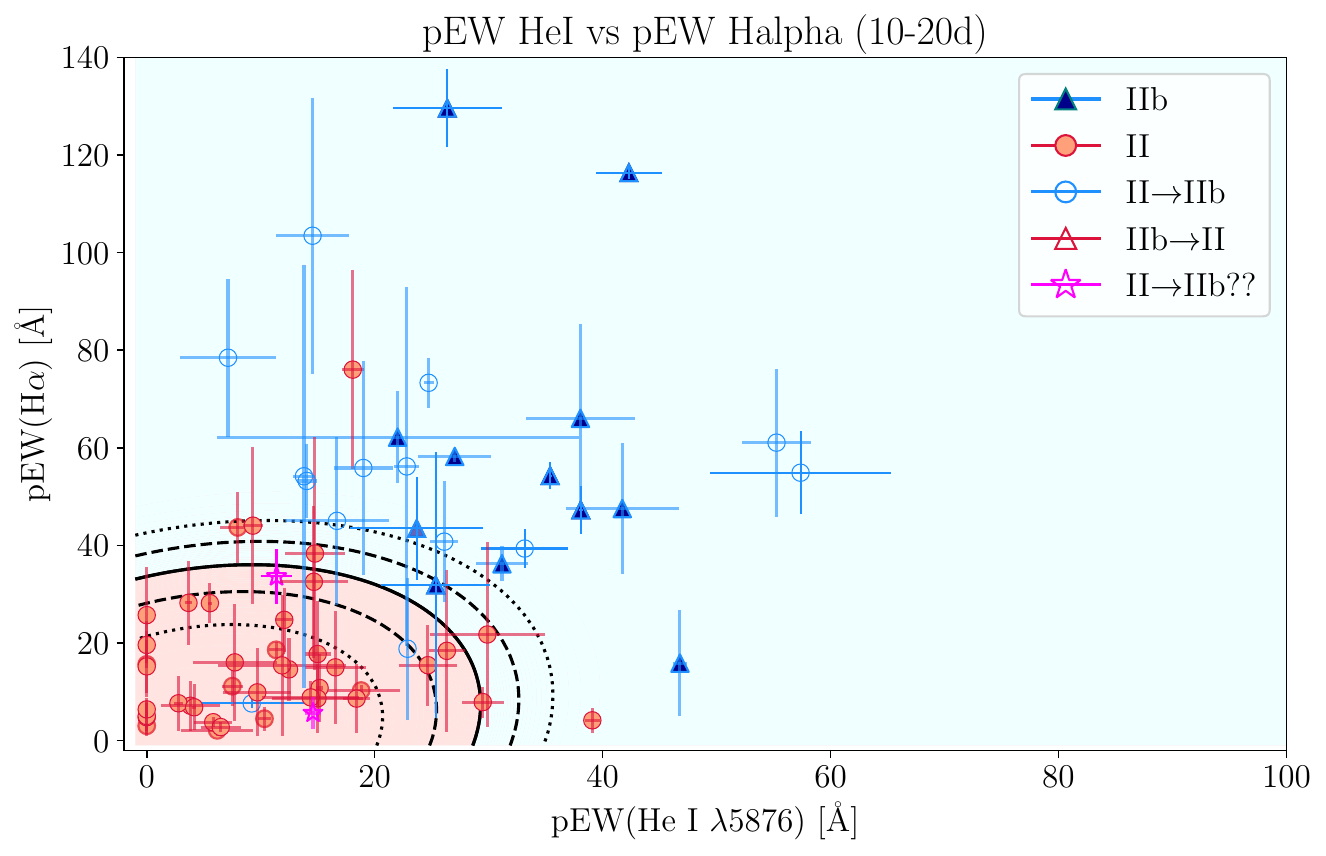}
\includegraphics[width=0.32\textwidth]{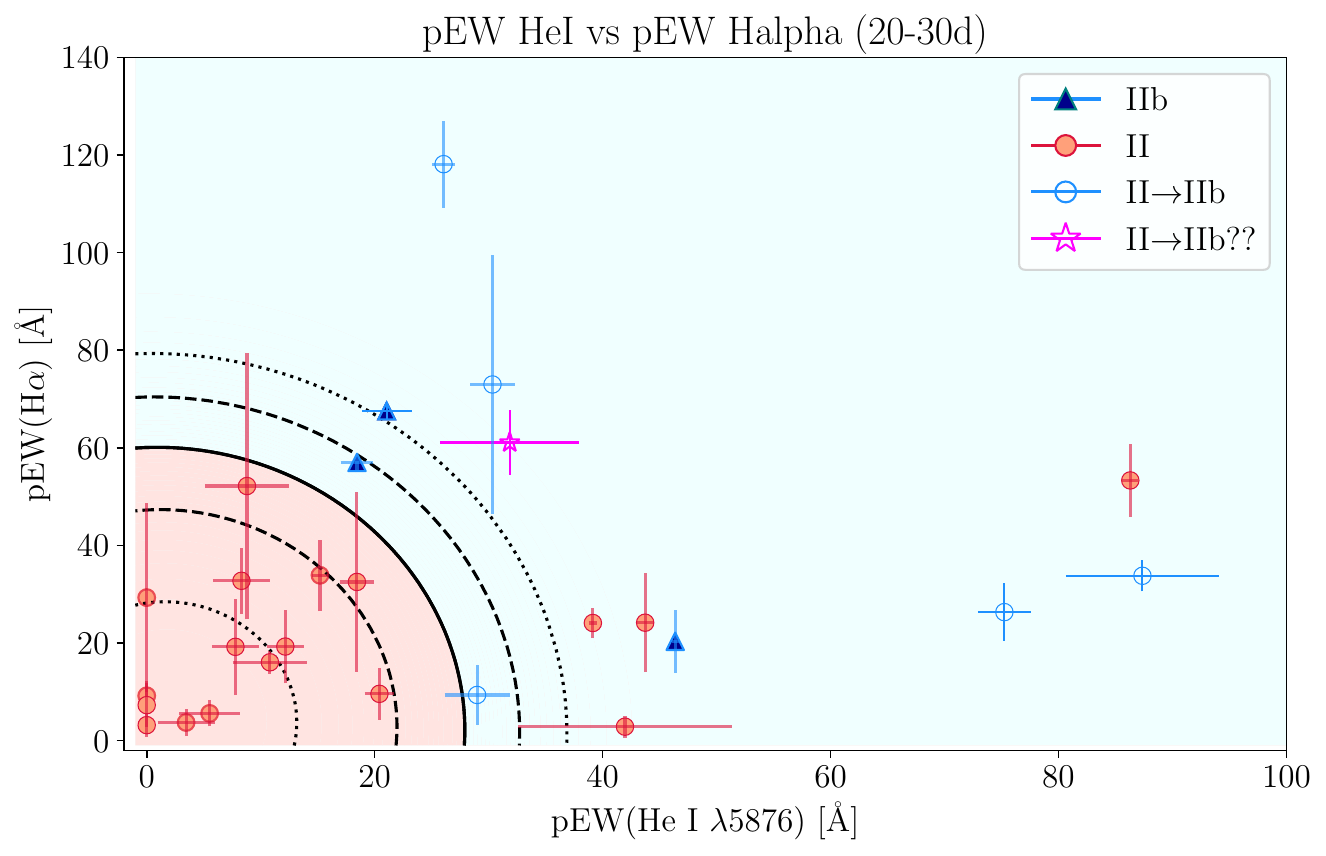}
\includegraphics[width=0.32\textwidth]{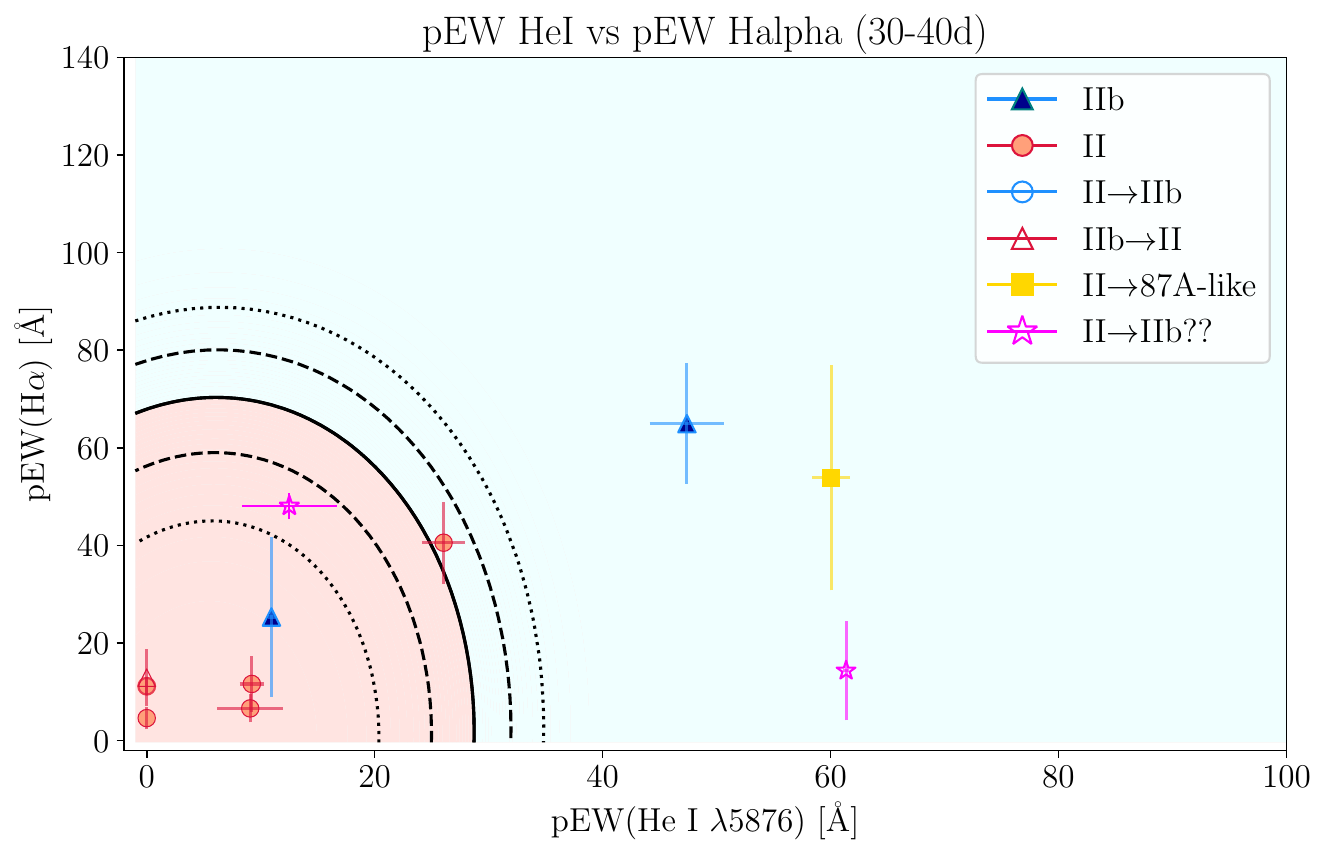}
\includegraphics[width=0.32\textwidth]{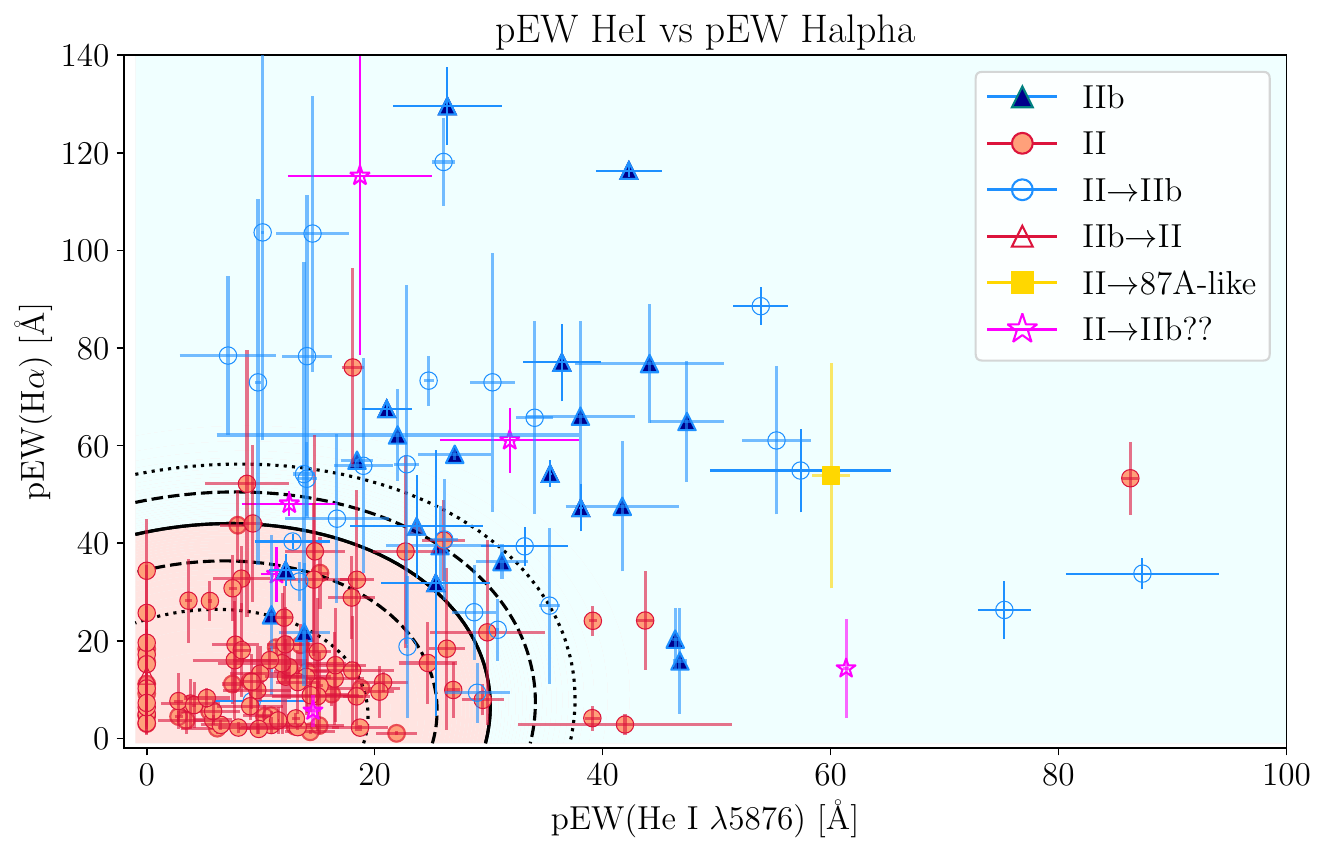}
\caption{The ranges defined by QDA based on the measures performed for the II and IIb pEW along with the SEDM and the new classified SNe. The closed red dots mark the type II data points, the closed blue triangles the type IIb and the yellow square a 87A-like. The open markers are those that have been reclassified based on our method. So the open blue circles belong to the SNe that were first classified as type II and then reclassified as IIb. The open red triangles are those first classified as IIb and then reclassified as II in this work. The purple open stars are SNe classified as II that seem to exhibit characteristics of IIb but could not be confirmed. In Table \ref{table:reclassification_SEDM} are presented plotted spectra with their old and new classifications. }
\label{fig:QDA+SEDM_timeranges}
\end{figure*}

\begin{figure*}
\centering
\includegraphics[width=0.32\textwidth]{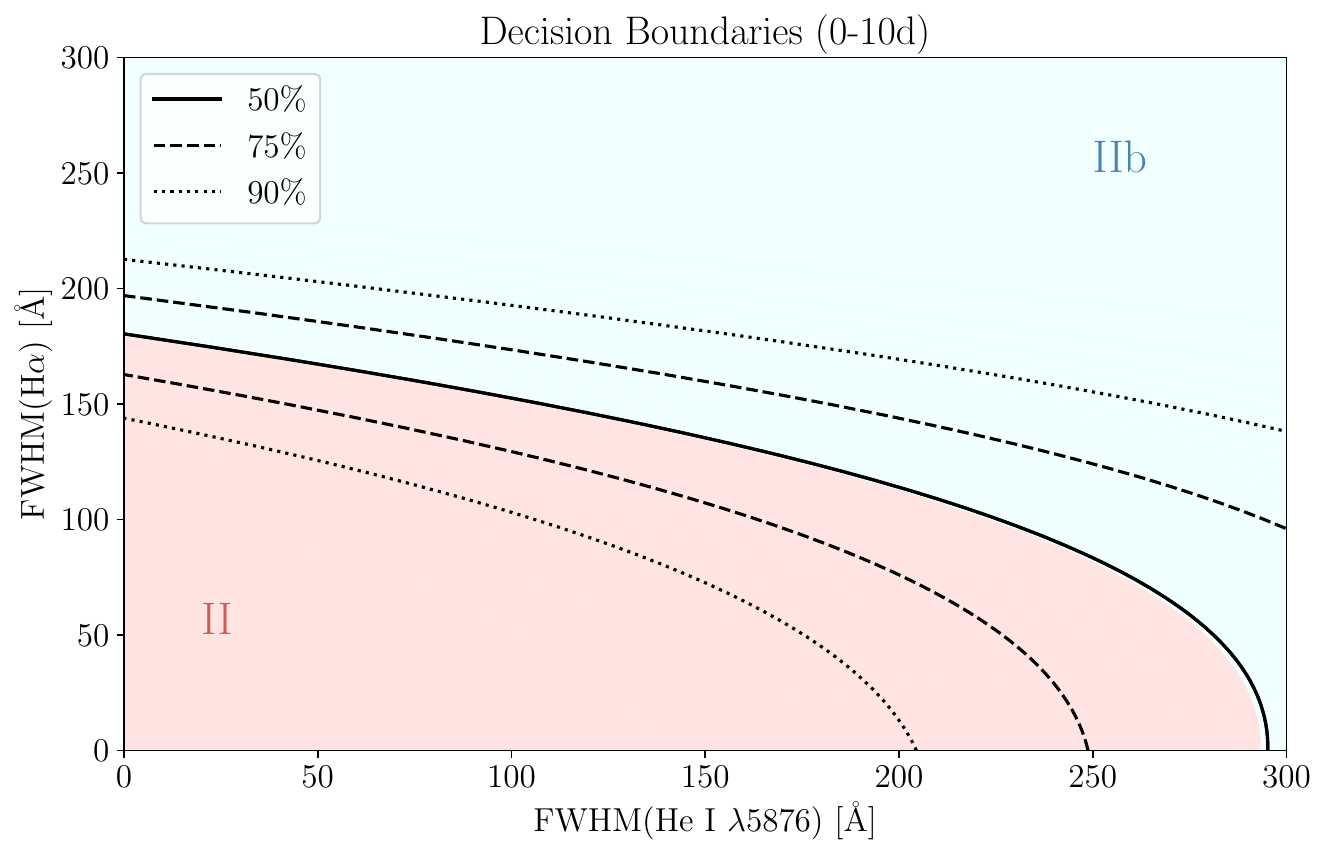}
\hspace{-0.3cm}
\includegraphics[width=0.32\textwidth]{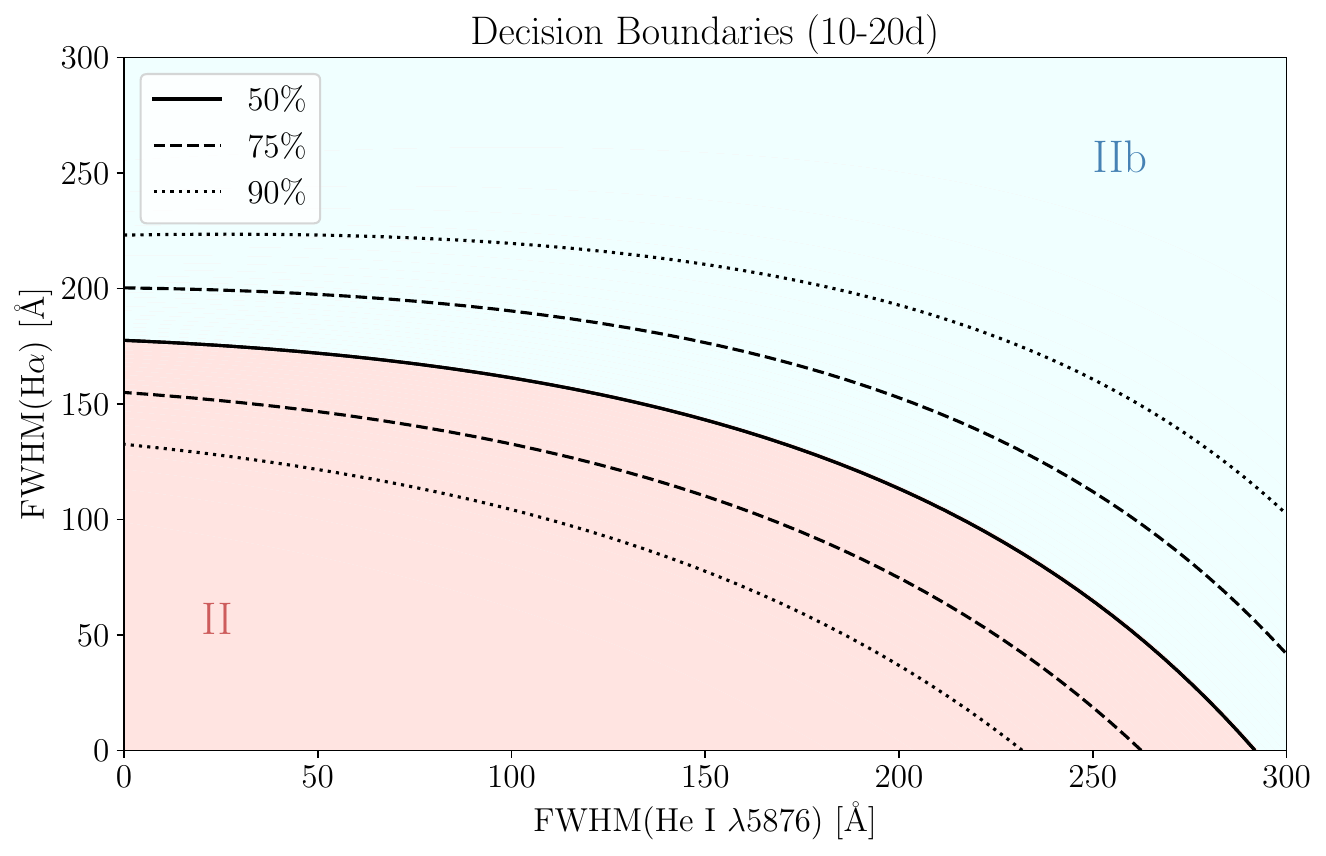}
\hspace{-0.3cm}
\includegraphics[width=0.32\textwidth]{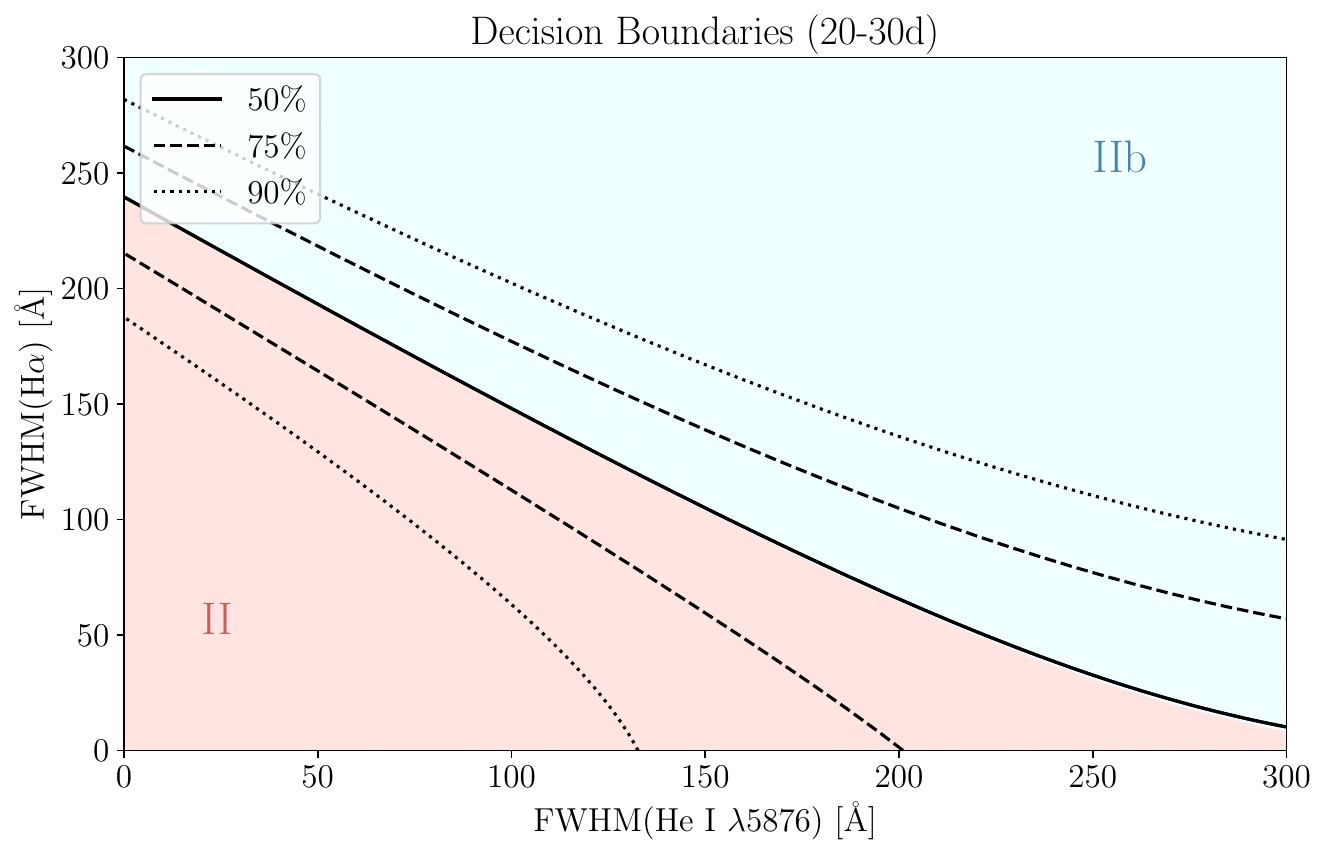}
\hspace{-0.3cm}
\includegraphics[width=0.32\textwidth]{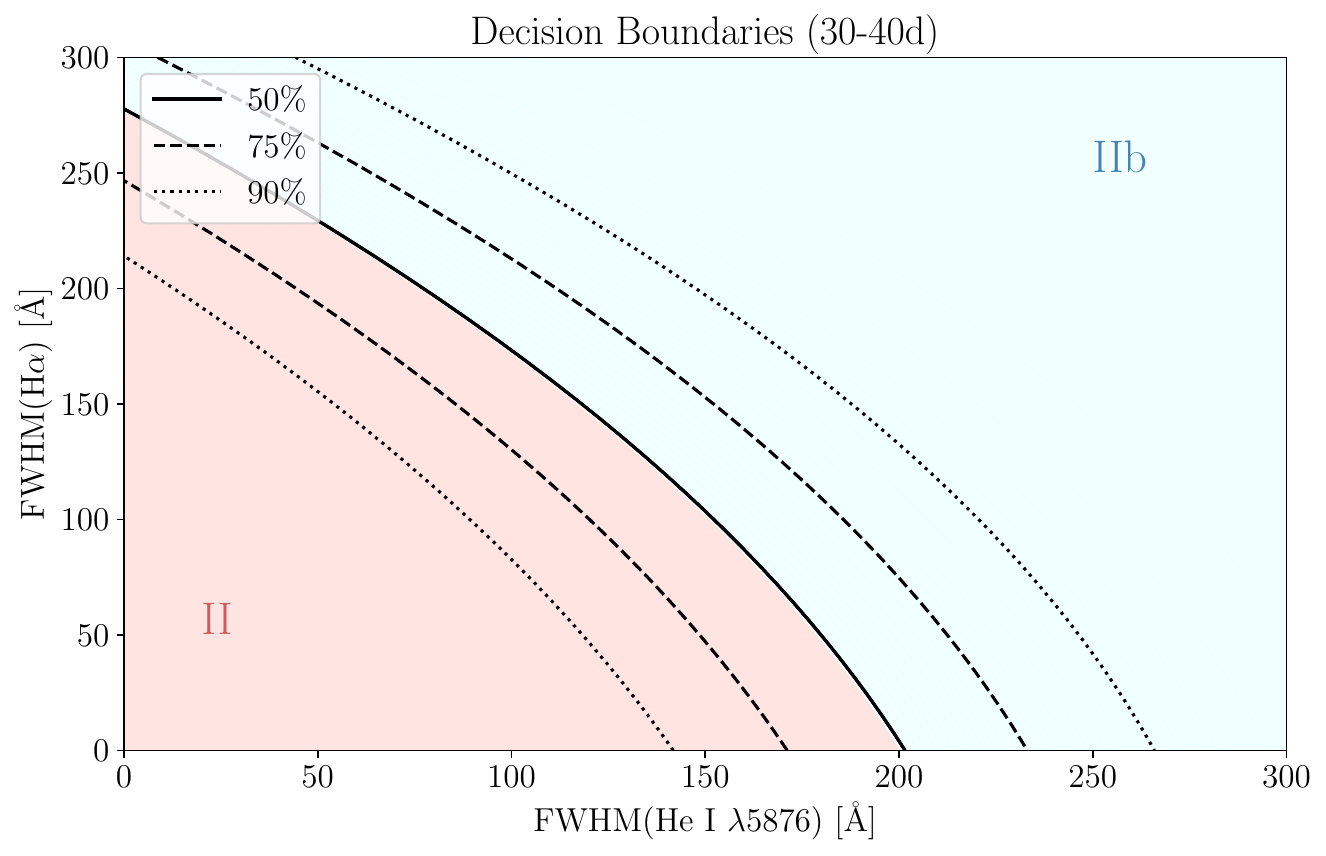}
\hspace{-0.3cm}
\includegraphics[width=0.32\textwidth]{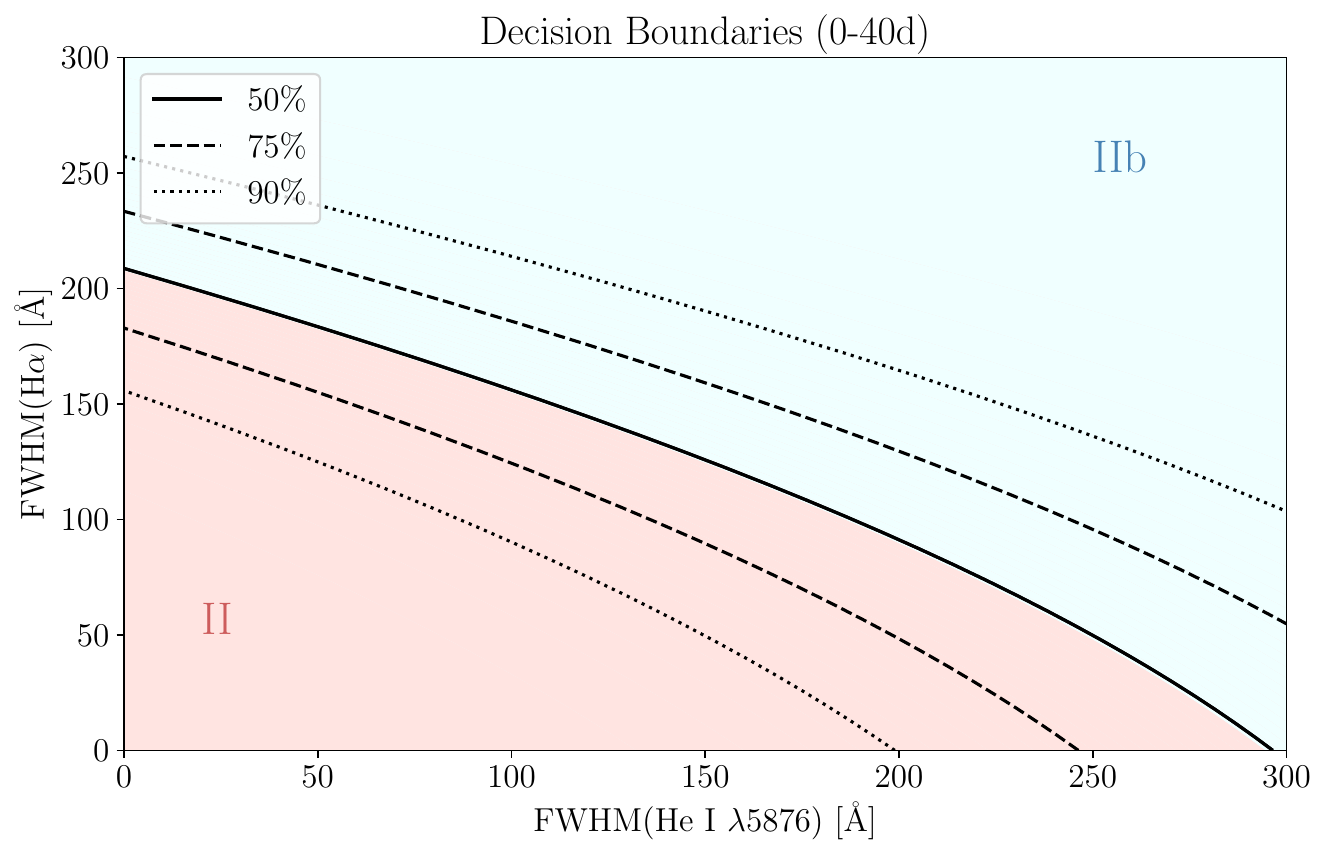}
\caption{Decision boundaries defined by the QDA classifier based on the measures performed for the FWHM of SNe~II and IIb at all the analysed time intervals. The red region corresponds to the SNe~II, and the blue region to SNe~IIb.}
\label{fig:QDA_FWHMtimeranges}
\end{figure*}

\begin{figure*}
\centering
\includegraphics[width=0.32\textwidth]{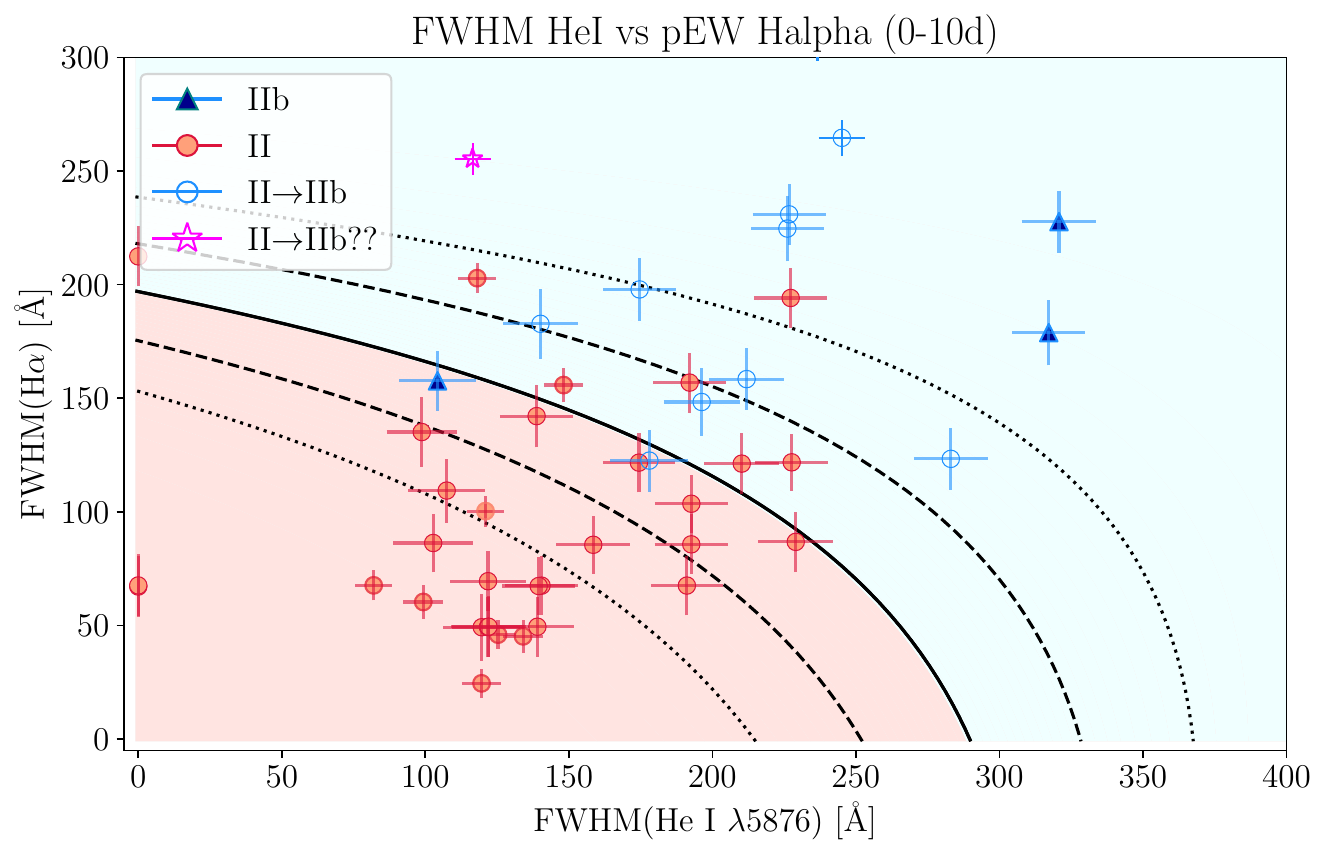}
\includegraphics[width=0.32\textwidth]{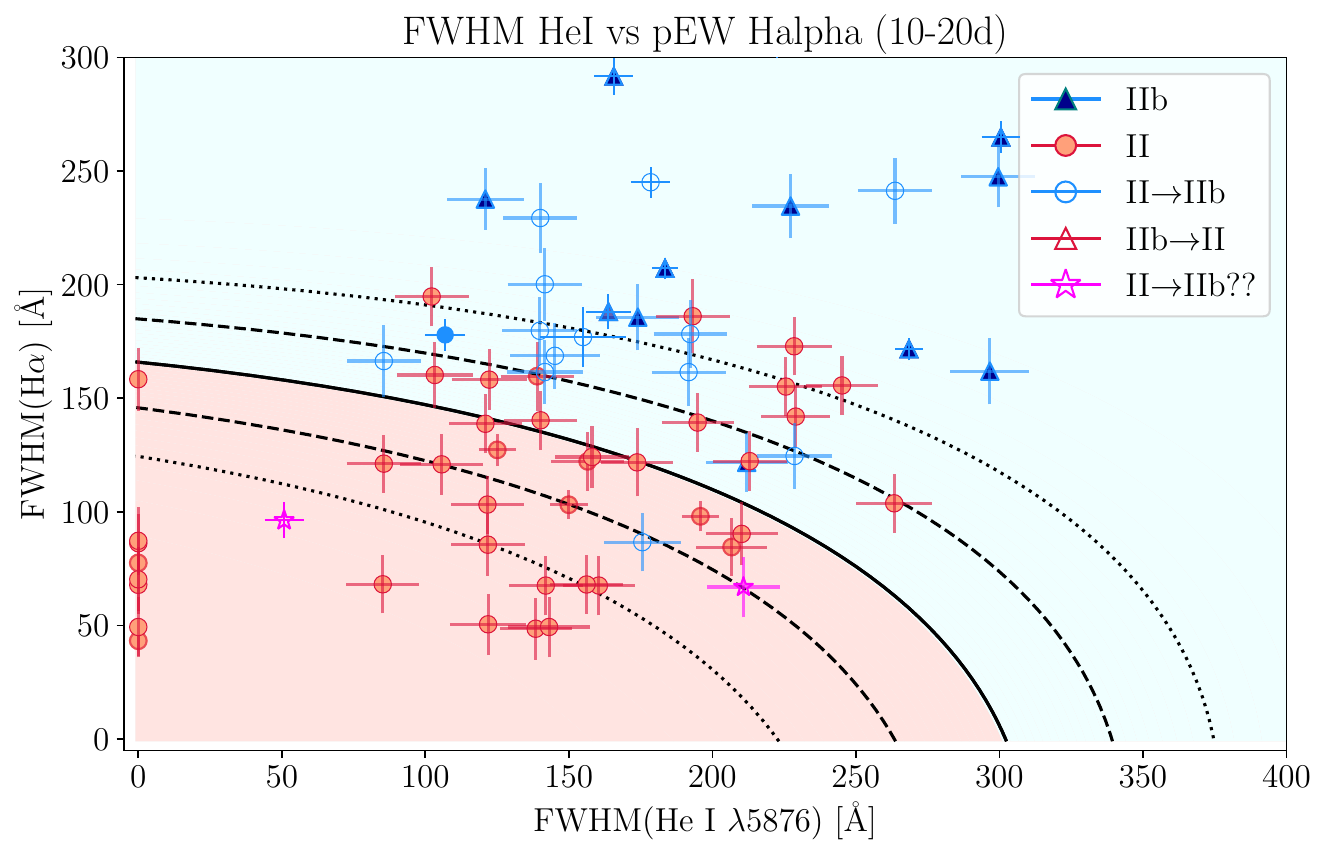}
\includegraphics[width=0.32\textwidth]{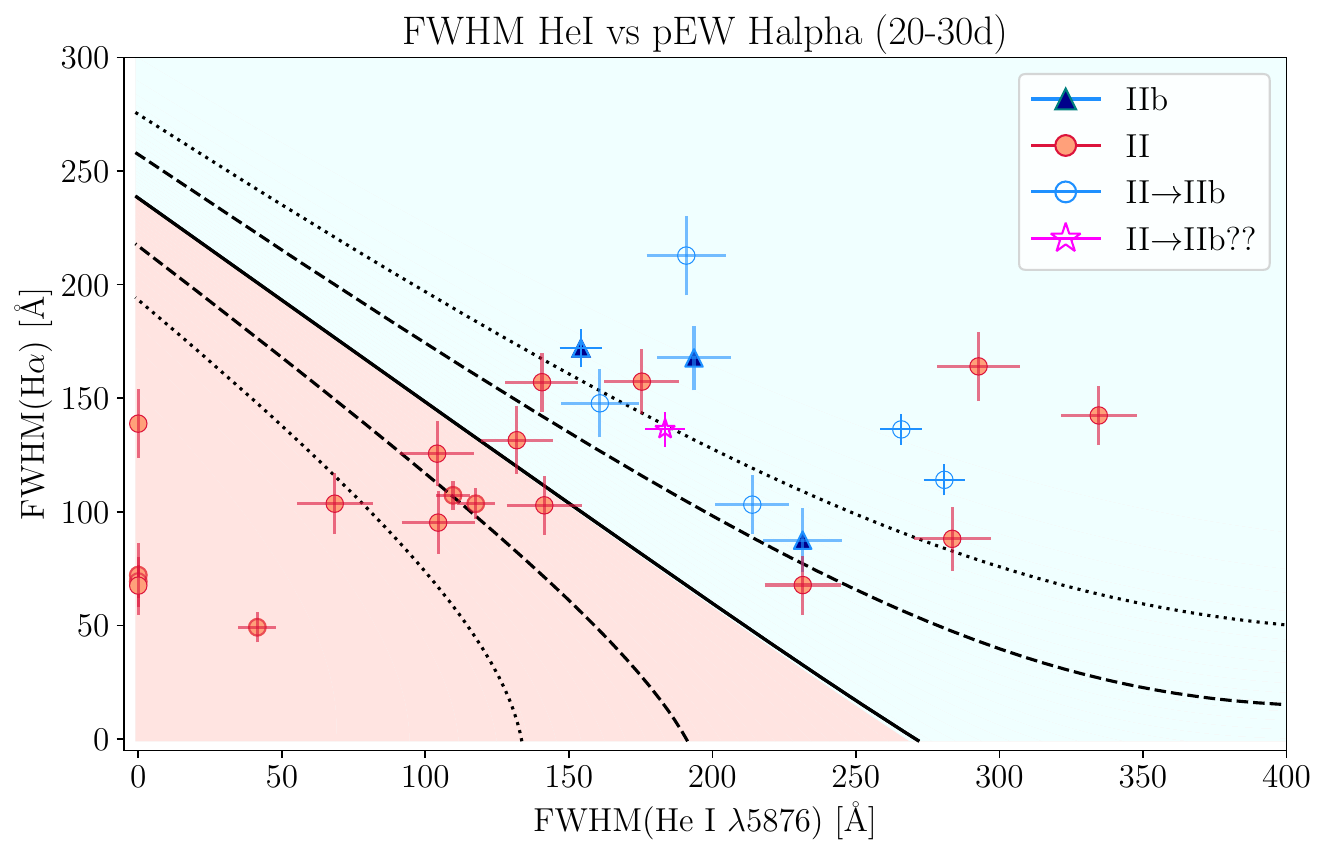}
\includegraphics[width=0.32\textwidth]{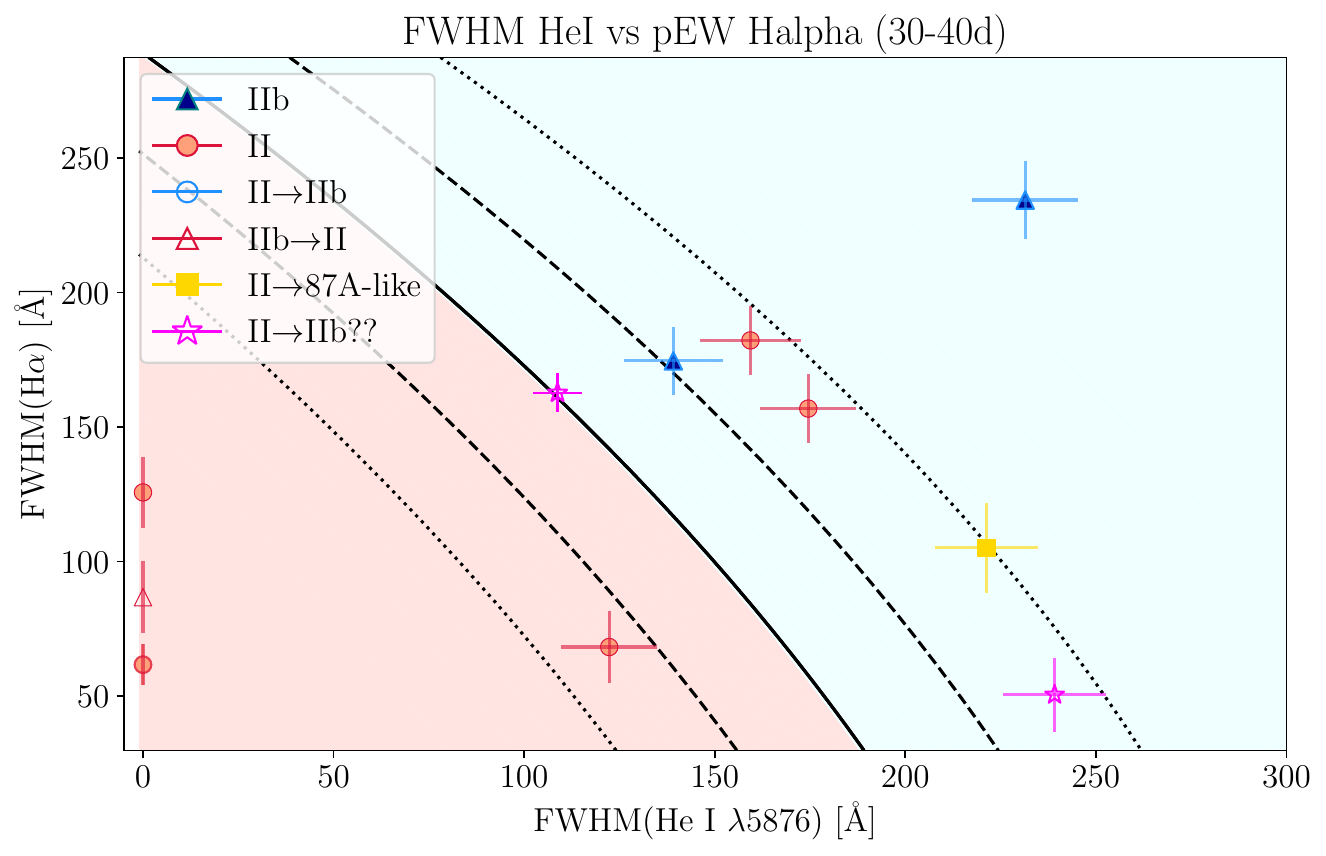}
\includegraphics[width=0.32\textwidth]{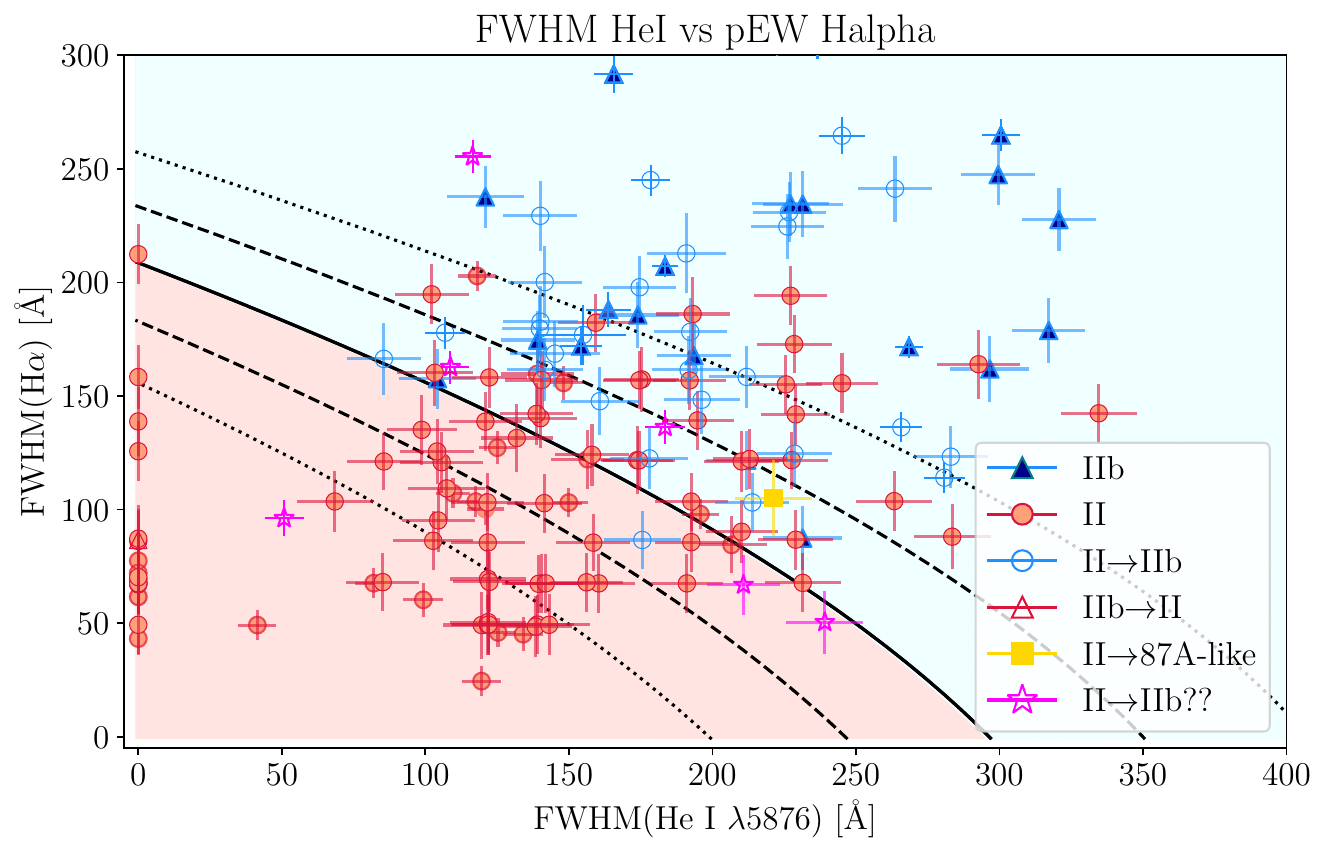}
\caption{The ranges defined by QDA based on the measures performed for the II and IIb FWHM, along with the SEDM and the new classified SNe. The closed red dots mark the type II data points, the closed blue triangles the type IIb and the yellow square a 87A-like. The open markers are those that have been reclassified based on our method. So the open blue circles belong to the SNe that were first classified as type II and then reclassified as IIb. The open red triangles are those first classified as IIb and then reclassified as II in this work. The purple open stars are SNe classified as II that seem to exhibit characteristics of IIb but could not be confirmed. In Table \ref{table:reclassification_SEDM} are presented plotted spectra with their old and new classifications. }
\label{fig:regiones_fwhm}
\end{figure*}

\begin{figure*}
\centering
\includegraphics[width=0.247\textwidth]{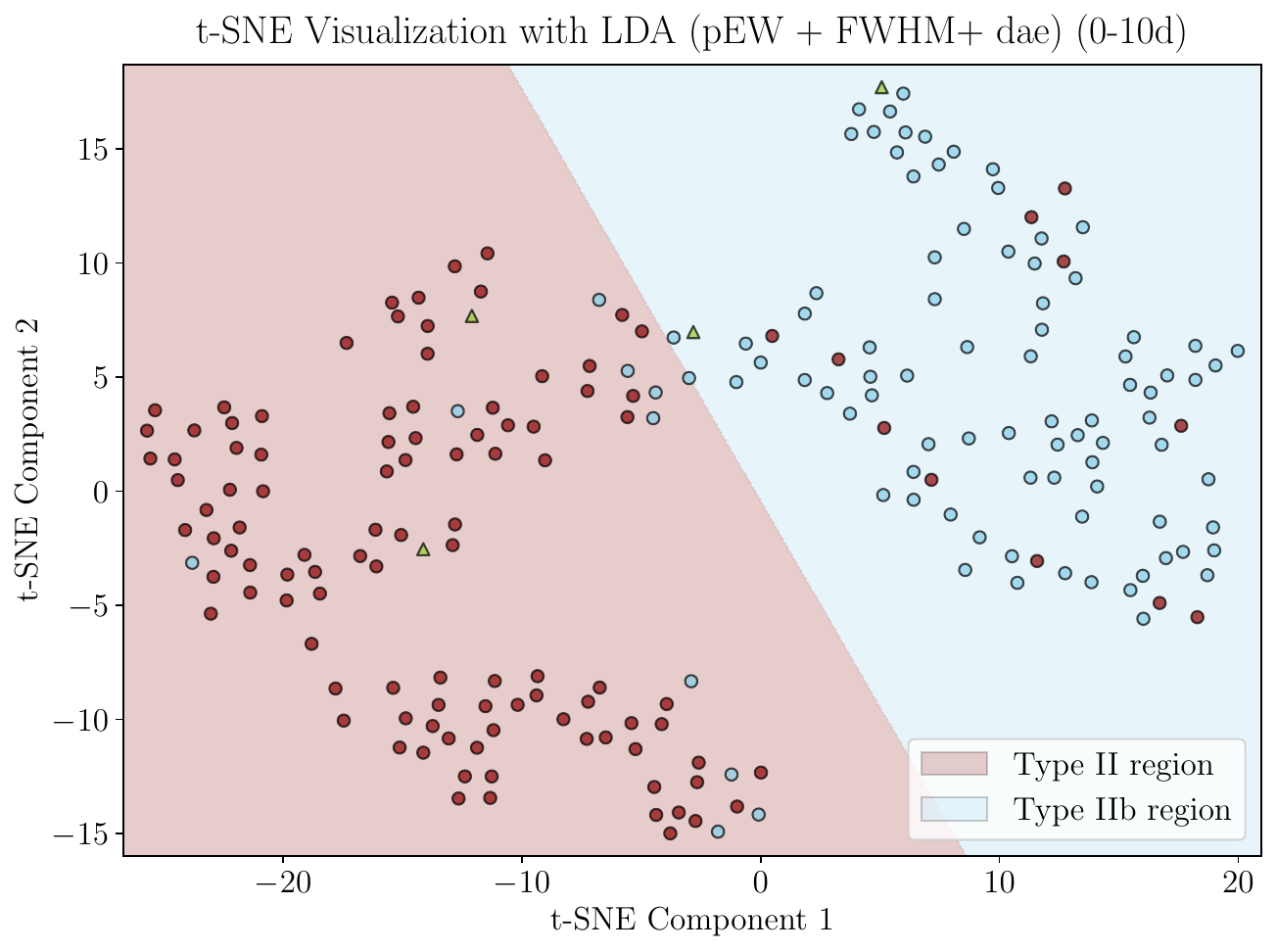}
\hspace{-0.15cm}
\includegraphics[width=0.247\textwidth]{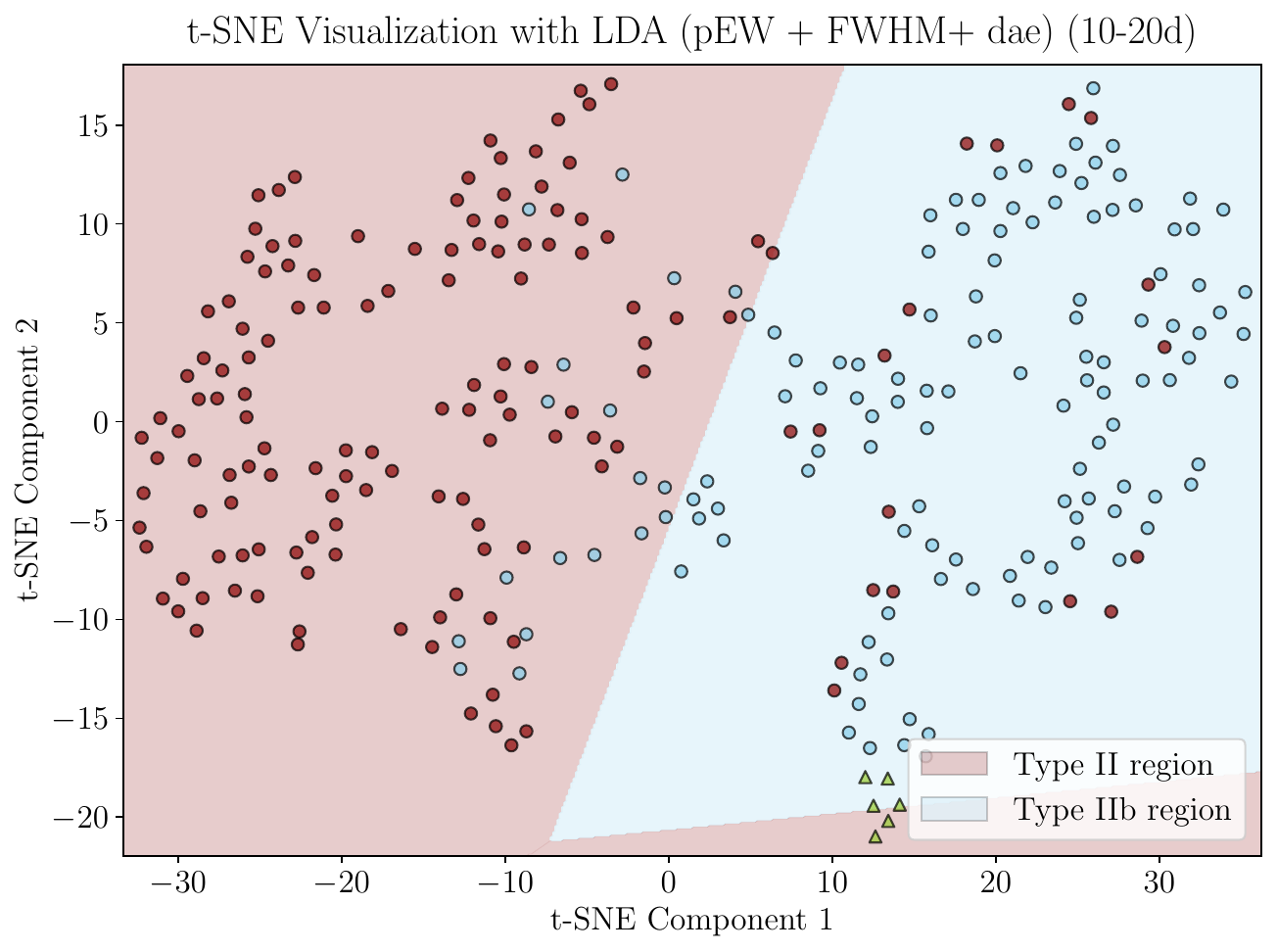}
\hspace{-0.15cm}
\includegraphics[width=0.247\textwidth]{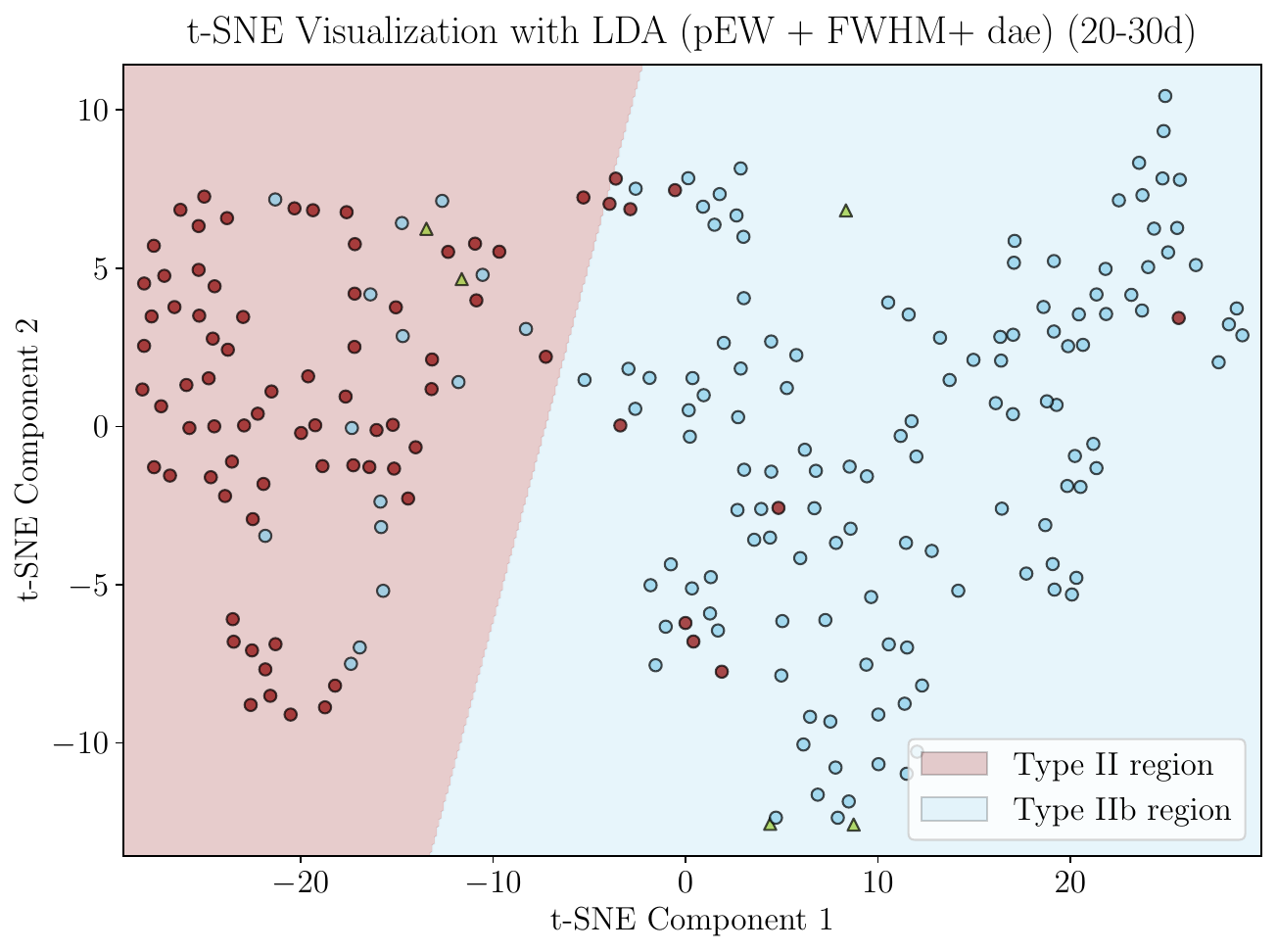}
\hspace{-0.15cm}
\includegraphics[width=0.247\textwidth]{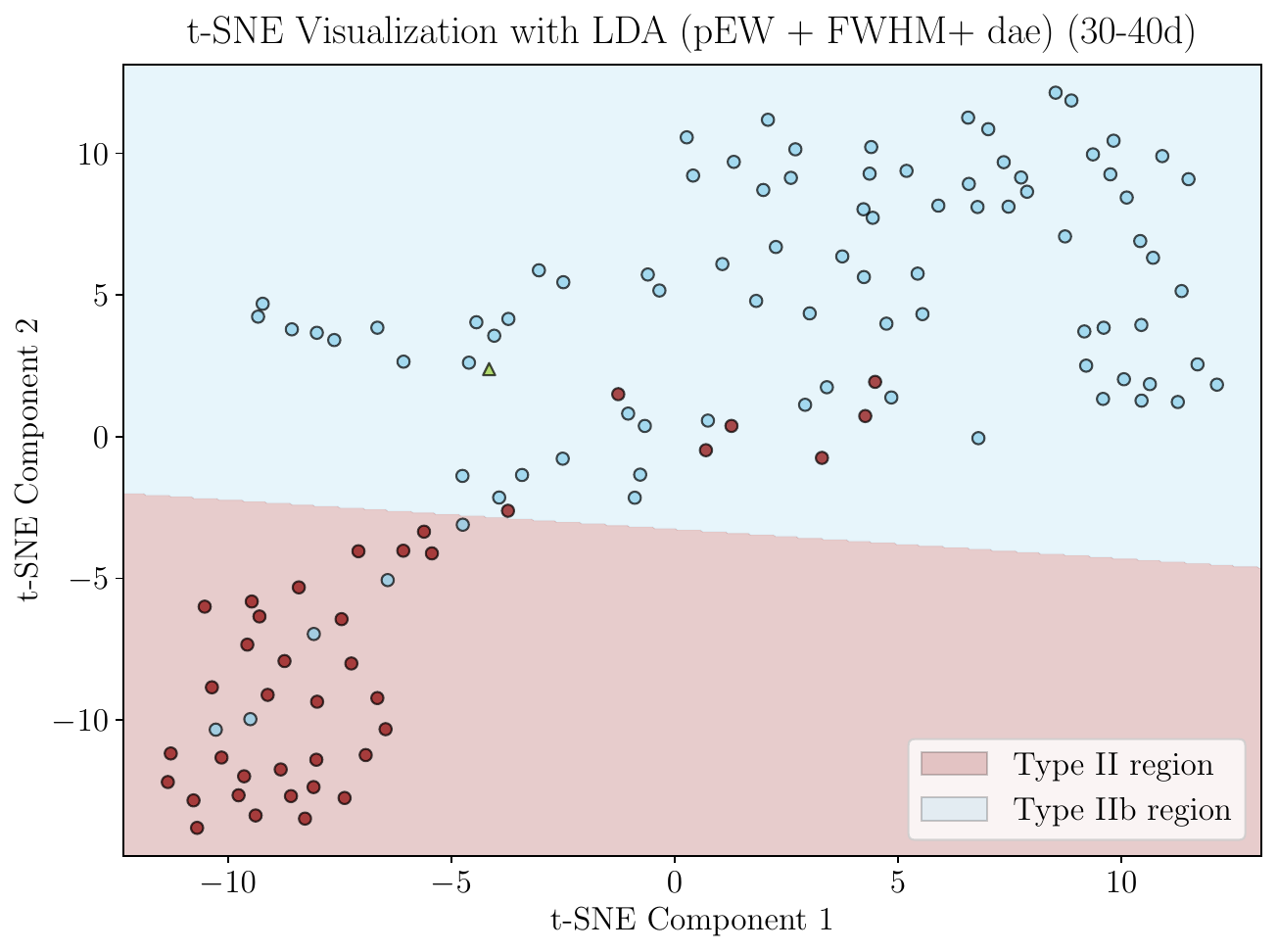}
\caption{The projection of t-SNE with a perplexity value of 20, along with a LDA analysis of the obtained 2D plot for all the analysed time ranges. }
\label{fig:t-SNE+LDA_timeranges}
\end{figure*}

\begin{figure*}
\centering
\includegraphics[width=0.26\textwidth]{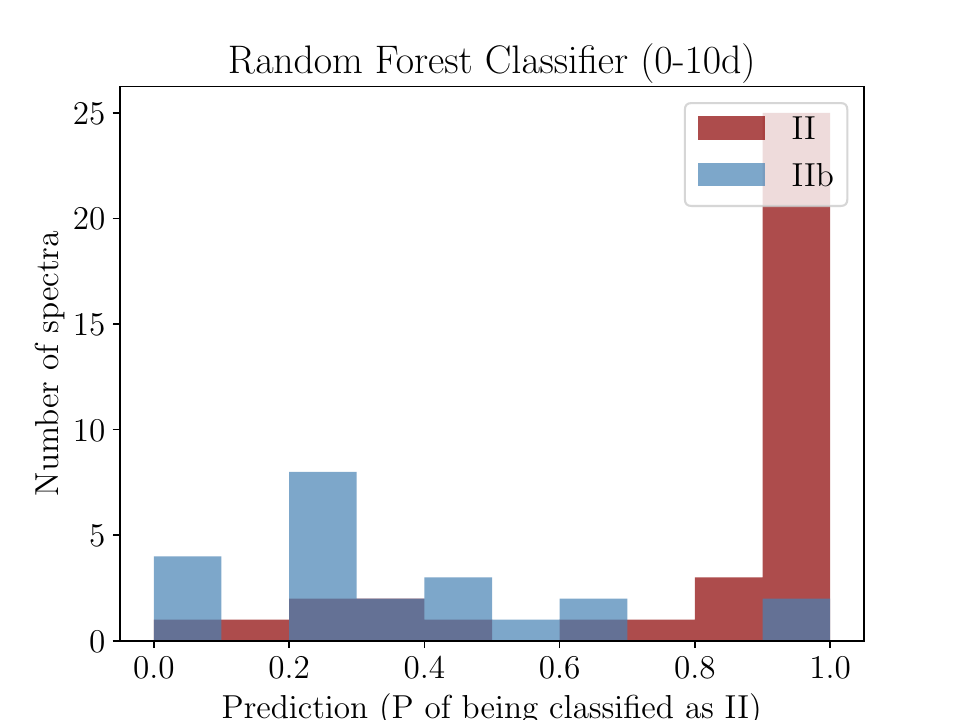}
\hspace{-0.5cm}
\includegraphics[width=0.26\textwidth]{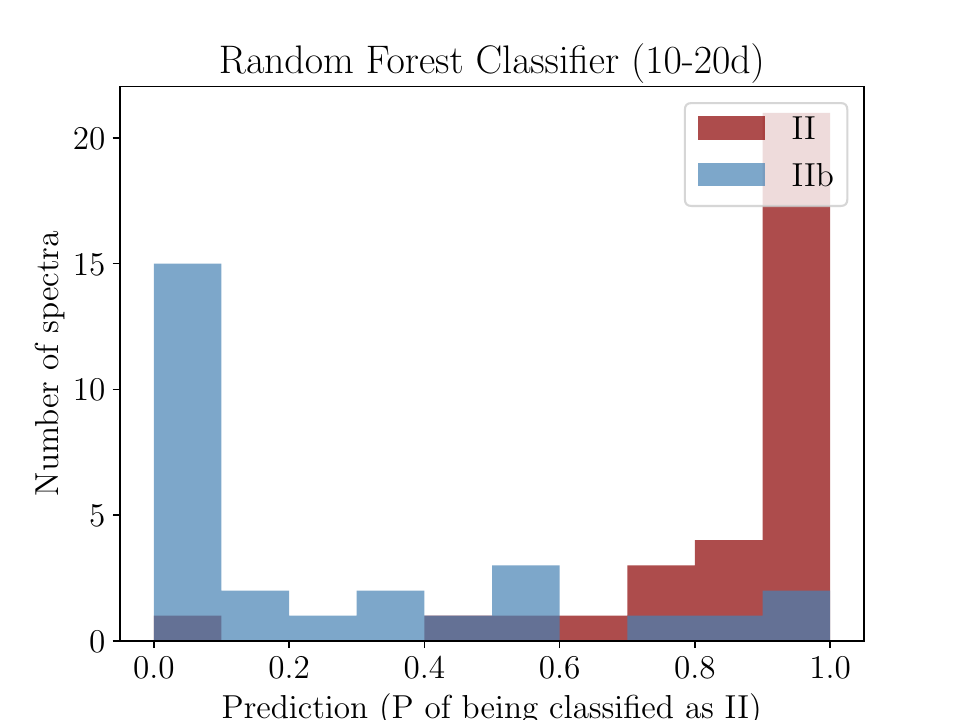}
\hspace{-0.5cm}
\includegraphics[width=0.26\textwidth]{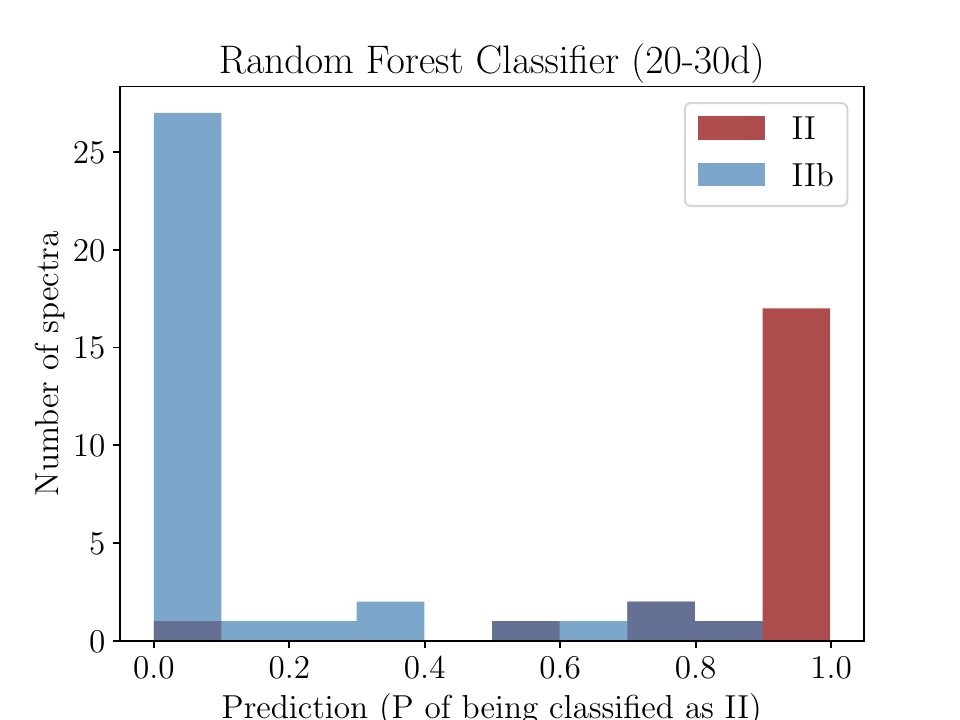}
\hspace{-0.5cm}
\includegraphics[width=0.26\textwidth]{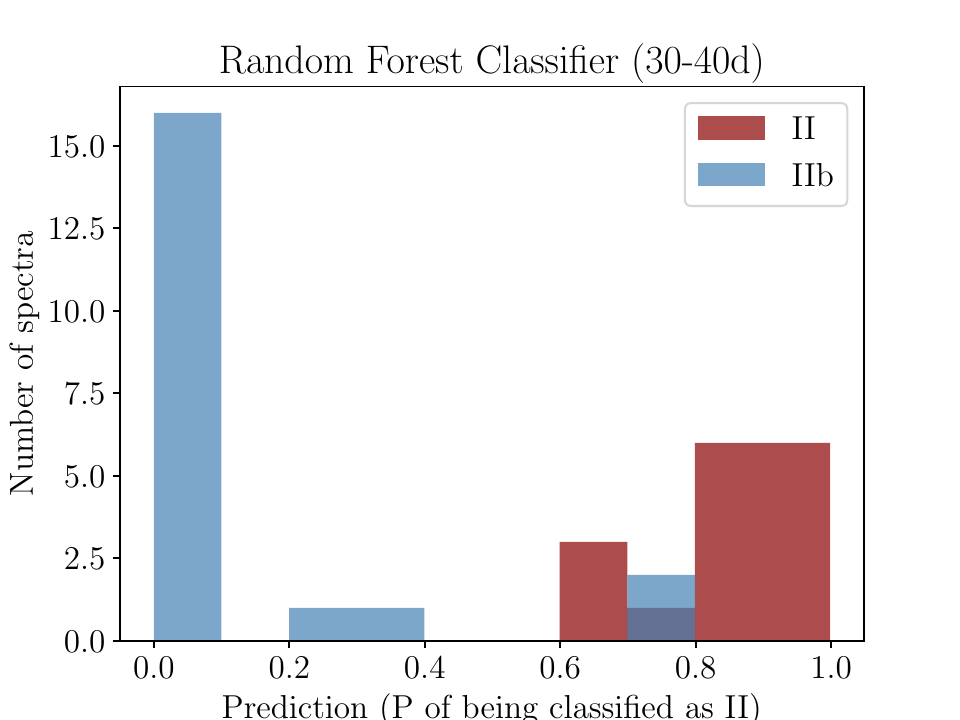}
\caption{Probabilities given by the Random Forest Classifier for supernova spectra, with the data split into four time ranges after the explosion: 0–10 days, 10–20 days, 20–30 days, and 30–40 days.  The classifier uses the pEW, FWHM, epoch, and type of each SN spectrum as input parameters,  with 60\% of the total sample used for training and the remaining 40\% for testing. The red and blue bins correspond to SNe~II and IIb, respectively.} 
\label{fig:RF_AZKENA}
\end{figure*}

\begin{figure*}
\centering
\includegraphics[width=0.45\textwidth]{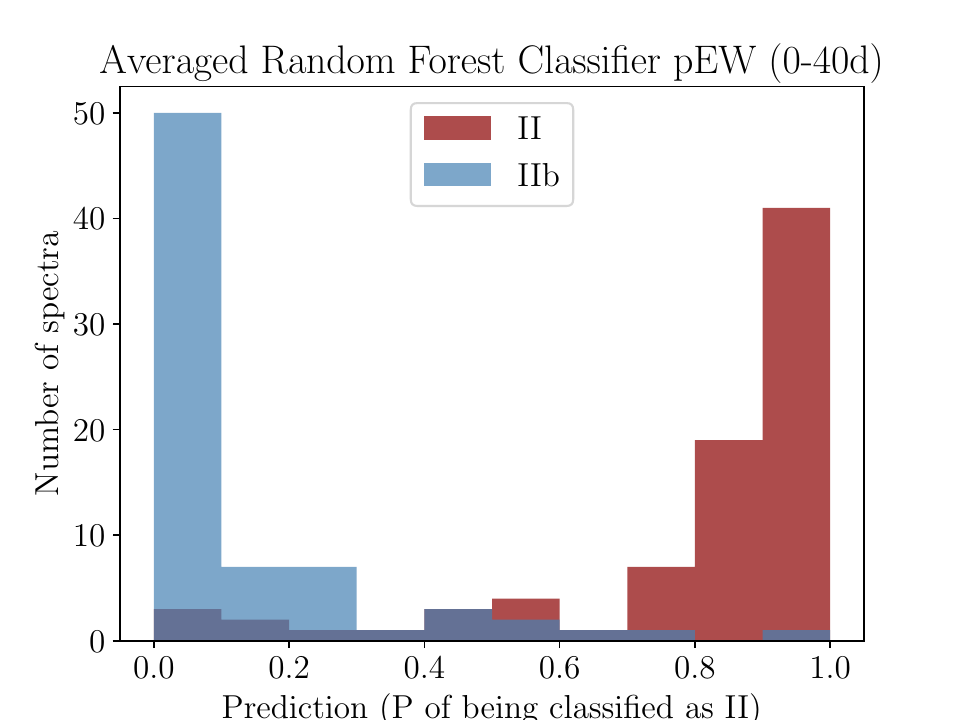}
\centering
\includegraphics[width=0.45\textwidth]{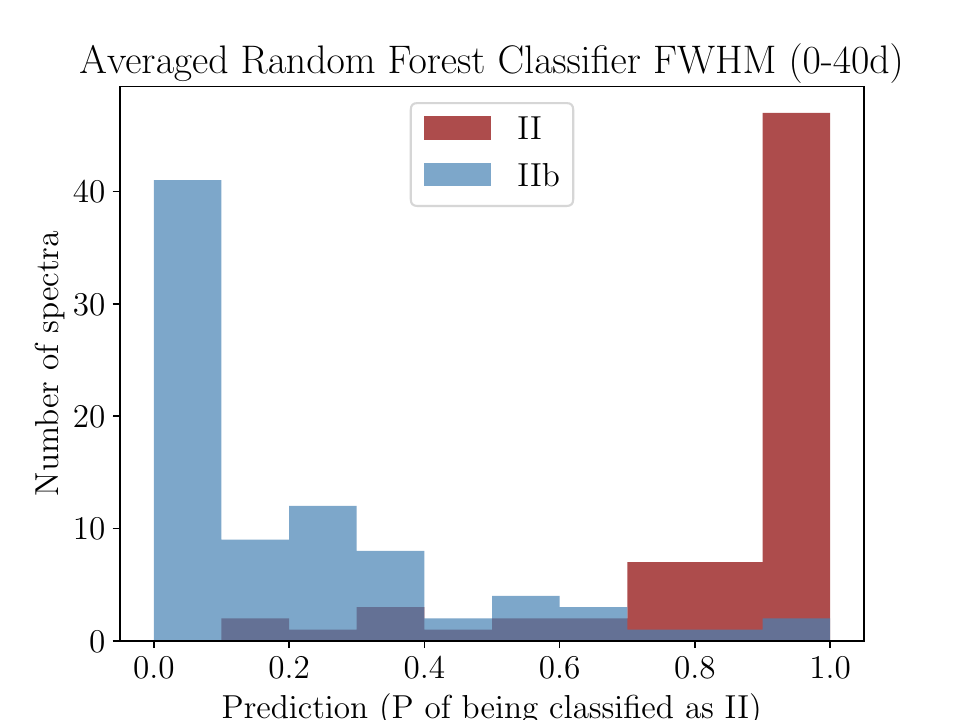}
\caption{\textbf{Left panel:} Probabilities given by the Random Forest Classifier for the whole sample, taking into account the pEW measurements for all the spectra within the first 40 days after the explosion the SN, using \%60 of the total sample as training sample and the rest as test sample. \textbf{Right panel:} Probabilities given by the Random Forest Classifier for the whole sample, taking into account only the FWHM measurements for all the spectra within the first 40 days after the explosion, using \%60 of the total sample as training sample and the rest as test sample. The red and blue bins correspond to SNe~II and IIb, respectively.} 
\label{fig:RF_timeranges2}
\end{figure*}

\clearpage
\onecolumn

\section{Tables}
\label{app:tables}

\clearpage
\centering
\begin{table}
\setlength{\tabcolsep}{3pt}
\caption{SNe with lower resolution data}
\label{table:SEDM_list_ONA}
\begin{tabular}{lc||lc}
\textbf{SNe} & \textbf{Classification} & \textbf{SNe} & \textbf{Classification} \\
\hline 
\hline
2016csv  & II  & 2020rmk*  & II  \\
2016hgm  & II  & 2020sy   & II  \\
2017aub  & II  & 2020tjd*  & IIb \\
2017fys  & II  & 2020uhm  & II  \\
2018aex  & II  & 2020urc  & IIb \\
2018amc  & II  & 2020uzx  & II  \\
2018dfc  & II  & 2020xcu  & II  \\
\end{tabular}
\begin{list}{}{}
\item The SNe marked with * are those that were reclassified (see Table \ref{table:reclassification_SEDM}). 
\end{list}
\end{table}

\clearpage
\begin{table}
\centering
\setlength{\tabcolsep}{8pt}
\caption{Reclassified SEDM spectra based on the regions we defined. } 
\label{table:reclassification_SEDM}
\begin{tabular}{lcccccc}
{\textbf{SNe}}& \textbf{Phase}         & {\textbf{Initial class.}} & {\textbf{This work}} & \multicolumn{3}{c}{\textbf{RF probabilities}} \\  
              & (days after explosion) &                           &                      & II        & IIb       & 87A-like     \\ 
\hline
\hline 
\rule{0pt}{2.5ex}
2018dfi       &  27          & II                               & IIb                                           & 0.28      & 0.72      & 0.00         \\
2018hmx       &  5           & II                               & IIb                                           & 0.31      & 0.69      & 0.00         \\
2018hqt       &  15          & II                               & \textbf{II}-\textgreater{}IIb? {(MC: 100.0\% II; 0.0\% IIb)}  & 0.93      & 0.05      & 0.02         \\
2018iby       &  15          & II                               & IIb                                           & 0.22      & 0.78      & 0.00         \\
2019cjx       &  20          & II                               & IIb                                           & 0.56      & 0.40      & 0.04         \\
2019hvg       &  5           & II                               & IIb                                           & 0.21      & 0.79      & 0.00         \\
\end{tabular}
\begin{list}{}{}
\item For objects denoted as "II->IIb?", the run MC simulation established the classification. 
\end{list}
\end{table}

\clearpage
\begin{table}
\begin{center}
\caption{New SN classifications.}
\label{table:new_classifications}
\setlength{\tabcolsep}{2pt}
\begin{tabular}{lcccccc}
{\textbf{SNe}} & {\textbf{Phase}} &{\textbf{Initial class.}} & {\textbf{This work}} & \multicolumn{3}{c} {\textbf{RF probabilities}} \\
                              &                 (days after explosion)                                  &             & &                          II            & IIb          & 87A-like        \\ 
\hline 
\hline
2013gg     &28         &            II                                     &        \textbf{II}-\textgreater{}IIb?  {(MC: 55.6\% II; 44.4\% IIb)}                                         &    0.0      &  1.0       &    0.00         \\

2015dj  & 11 & II & IIb & 0.66     & 0.34     &   0.00      \\
        & 21 & II & IIb & 0.03     & 0.97     &   0.00       \\
        & 24 & II & IIb & 0.00     & 1.00     &    0.00      \\


2019oxi  &32            & II                                                & \textbf{II}-\textgreater{}IIb?  {(MC: 83.0\% II; 17.0\% IIb)}                                                 &   0.60       & 0.40         & 0.0            \\
2024jxm   & 24          & II                                                & {II}                                                                                              &              &              &                 \\
\end{tabular}
\begin{list}{}{}
\item For objects denoted as "II->IIb?", the run MC simulation established the classification. For the SN 2015dj, as we have 3 spectra we included the obtained classification probabilities for the three of them.
\end{list}
\end{center}
\end{table}

\begin{table}
\small
\caption{Evolution of the mean value of pEW and FWHM for each type over the examined time ranges along with their standard deviation. }
\label{tab:main_values}
\setlength{\tabcolsep}{3pt}
\begin{tabular}{c|cccc|cccc}
\multicolumn{1}{l}{} & \multicolumn{4}{c|}{\textbf{FWHM (\AA)}}                                                  & \multicolumn{4}{c}{\textbf{pEW (\AA)}}                                                        \\
\hline
\multicolumn{1}{l}{} & \multicolumn{2}{c}{\textbf{\ha}}          & \multicolumn{2}{c|}{\textbf{HeI}}             & \multicolumn{2}{c}{\textbf{\ha}}          & \multicolumn{2}{c}{\textbf{HeI}}              \\
\multicolumn{1}{l}{} & \textit{\textbf{IIb}} & \textit{\textbf{II}} & \textit{\textbf{IIb}} & \textit{\textbf{II}} & \textit{\textbf{IIb}} & \textit{\textbf{II}} & \textit{\textbf{IIb}} & \textit{\textbf{II}} \\ \hline
\textbf{0-10d}       & $159.68 \pm 54.26$                 & $68.39 \pm 52.08$                & $222.41 \pm 71.13$               & $105.63 \pm 77.22$               & $39.41 \pm 30.51$                 & $6.10 \pm 7.83$                & $31.83 \pm 22.64$                & $9.36 \pm 9.93$                 \\
\textbf{10-20d}      & $151.51 \pm 45.48 $               & $74.28 \pm 48.54$                & $225.98 \pm 72.14$                & $116.24 \pm  68.02$              & $40.86\pm 22.61$                & $9.56 \pm 11.20$                 & $39.27 \pm 21.25$                & $9.64 \pm 8.44$                \\
\textbf{20-30d}      & $150.26 \pm 45.62$               & $77.32 \pm 36.51$                & $221.92 \pm 62.84$               & $87.42 \pm 69.05$               & $50.55 \pm 28.91$                & $16.73 \pm 16.88$               & $48.13 \pm 26.54$                & $8.13 \pm 10.02$                \\
\textbf{30-40d}      & $146.03 \pm 43.09$               & $95.90  \pm 41.46$              & $210.96 \pm 54.64$              & $97.34 \pm 58.96$              & $53.72  \pm 28.74$               & $24.98  \pm 19.19$              & $57.77 \pm 26.61$                & $10.76 \pm 7.95$             
\end{tabular}
\end{table}

\clearpage
\small
\setlength{\tabcolsep}{1pt}
\begin{longtable}{p{2.cm}p{1.3cm}p{1.1cm}p{2cm}p{4.2cm}p{5.2cm}}
\caption{Full SN sample.}
\label{tab:SNe} \\
\textbf{SN} & \textbf{Type} & \textbf{nº Spec} & \textbf{Redshift} & \textbf{Discovery reference} & \textbf{Other references} \\
\hline
\hline
1986L       & II            & 1                   & 0.00435    & \cite{gutierrez2017type}      &         \cite{gutierrez2017type}                                                                                \\
1987A       & 87A-like      & 10                  & 0.001067   & \cite{kunkel1987supernova}    & \cite{pun1995ultraviolet}                                                                                       \\
1987K       & IIb           & 4                   & 0.00268    & \cite{pollas1987supernovae}   & \cite{filippenko1988supernova}                                                               \\
1988A       & II            & 2                   & 0.005067   & \cite{gutierrez2017type}      & \cite{ruiz1990sn}                                                                                               \\
1990E       & II            & 2                   & 0.00414    & \cite{gutierrez2017type}      & \cite{schmidt1993photometric}                                                                                   \\
1992af      & II            & 2                   & 0.01847    & \cite{gutierrez2017type}      & \cite{gutierrez2017type}; \cite{della1992supernova}                                            \\
\end{longtable}


\label{lastpage}

\end{document}